\documentclass[apj]{emulateapj}
\usepackage{color}

\def\sun{\ifmmode\odot\else$\odot$\fi}

\input{psfig.sty}
\shorttitle{Torus and AGN properties of nearby 
  Seyfert galaxies}
\shortauthors{A. Alonso-Herrero et al.}

\begin{document}

\title{Torus and AGN properties of nearby 
  Seyfert galaxies: Results from fitting IR spectral energy
  distributions and spectroscopy}

\author{Almudena Alonso-Herrero\altaffilmark{1,2}, 
Cristina Ramos Almeida\altaffilmark{3}, 
Rachel Mason\altaffilmark{4},
Andr\'es Asensio Ramos\altaffilmark{5,6}, 
Patrick F. Roche\altaffilmark{7}, 
Nancy A. Levenson\altaffilmark{8}, 
Moshe Elitzur\altaffilmark{9},
Christopher Packham\altaffilmark{10}, 
Jos\'e Miguel  Rodr\'\i guez Espinosa\altaffilmark{5,6}, 
Stuart Young\altaffilmark{11}, 
Tanio D\' iaz-Santos\altaffilmark{12}, 
and 
Ana M. P\'erez Garc\' ia\altaffilmark{5,6}
} 

\altaffiltext{1}{Departamento de Astrof\'{\i}sica, Centro de 
Astrobiolog\'{\i}a, INTA-CSIC, Carretera de Torrej\'on a Ajalvir km 4,
Torrej\'on de Ardoz, 
E-28850 Madrid, Spain; E-mail: aalonso@cab.inta-csic.es}
\altaffiltext{2}{Associate Astronomer, Steward Observatory, University
of Arizona, Tucson, AZ 85721}
\altaffiltext{3}{University of Sheffield, Department of Physics \&
  Astronomy, Sheffield S3 7RH, UK}
\altaffiltext{4}{Gemini Observatory, Northern Operations Center, 670 North
A'ohoku Place, Hilo, HI 96720} 
\altaffiltext{5}{Instituto de Astrof\'{\i}sica de Canarias, E-38205 La
Laguna, Tenerife, Spain} 
\altaffiltext{6}{Departamento de Astrof\' isica, Universidad de La Laguna, E-38205, La Laguna, Tenerife, Spain}
\altaffiltext{7}{Astrophysics, Department of Physics, University of
  Oxford, DWB, Keble Road, Oxford OX1 3RH, UK} 
\altaffiltext{8}{Gemini Observatory, Casilla 603, La Serena, Chile}
\altaffiltext{9}{Department of Physics and Astronomy, University of
  Kentucky, Lexington, KY 40506-0055}
\altaffiltext{10}{Astronomy Department, University of Florida, 211 Bryant Space Science Center, P.O. Box 112055, Gainesville, FL 32611-2055}
\altaffiltext{11}{Department of Physics, Rochester Institute of
  Technology, 84 Lomb Memorial Drive, Rochester, NY 14623}
\altaffiltext{12}{Department of Physics, University of Crete, GR-71003, Heraklion, Greece}

\begin{abstract}
We used the \textit{CLUMPY} torus models  and a Bayesian approach 
to fit the infrared spectral energy
distributions (SEDs) and ground-based high-angular resolution
mid-infrared spectroscopy of 13 nearby Seyfert galaxies. This allowed
us to put tight constraints on 
torus model parameters such as the viewing angle $i$, the
radial thickness of the torus $Y$, the angular size of the cloud
distribution $\sigma_{\rm torus}$, and the average number of clouds along
radial equatorial rays $N_0$. We found that the viewing angle $i$ 
is not the only parameter controlling the classification of a galaxy
into a type 1 or a type 2. In principle type 2s could be viewed at any
viewing angle $i$ as long as there is one cloud along the line of sight. A
more relevant quantity for  clumpy media is 
the probability for an AGN photon to escape unabsorbed.
In our sample, 
type 1s have relatively high escape probabilities, 
$P_{\rm esc}\sim 12-44\%$, while 
type 2s, as expected,  tend to have very low escape
probabilities. Our fits also
confirmed that the tori of Seyfert galaxies are compact with  
torus model radii  in the range 
1-6\,pc. The scaling of the models to the data also provided the 
AGN bolometric luminosities 
$L_{\rm  bol}({\rm AGN})$, which were found to be 
in good agreement with estimates from the literature. 
When we combined our sample of Seyfert galaxies with a sample of PG
quasars from the literature to span a range of $L_{\rm bol}({\rm AGN}) \sim 
10^{43}-10^{47}\,{\rm erg \, s}^{-1}$, we  found plausible evidence of the
receding torus. That is, there is a 
tendency for the torus geometrical covering factor to be lower
($f_2\sim 0.1-0.3$) at high
AGN luminosities than at low AGN luminosities ($f_2\sim 0.9-1$ at
$\sim 10^{43-44}\,{\rm erg \,s}^{-1}$). This is because at low AGN luminosities
the tori appear to have wider angular sizes (larger $\sigma_{\rm
  torus}$) and more clouds along radial equatorial rays. We cannot,
however rule out the possibility that this is due to contamination by
extended dust structures not associated with the dusty torus at low
AGN luminosities, since most of 
these in our sample are hosted in highly inclined galaxies. 

\end{abstract}


\section{Introduction}

The unified model for  active galactic nuclei (AGN)  proposes the
ubiquitous presence of an obscuring torus around their nuclei, with
type 1  and type 2 AGN being intrinsically similar (Antonucci
1993). The central region 
of an AGN (including the Broad Emission Line Region, BLR) is obscured when 
viewed along directions close to the equatorial plane of the torus,
and it is then classified as a type 2 AGN. In type 1 AGN the viewing
angle is close to the polar direction of the torus, and thus we have a
direct view of the central engine. This model received
strong support from the fact that broad lines have been revealed in
the spectra of the polarized emission of a number of type 2 AGN
(e.g., Antonucci \& Miller 1985; Tran, Miller, \& Kay 1992). 

There is also indirect evidence of the
presence of the dusty torus advocated by 
the unified model. First is the similarity of Seyfert 1s and
  Seyfert 2s with respect to a given isotropic indicator of the AGN luminosity 
 such as,  [O\,{\sc
  iii}]$\lambda$5007,  infrared (IR), hard X-ray, and radio 
luminosities of  Seyferts (see e.g., Mulchaey et al. 1994;
Alonso-Herrero, Ward, \& 
Kotilainen 1997; Nagar et al. 1999). The UV and soft X-ray continua
of Seyfert 2s, on the other hand are 
underluminous relative to the type 1s, because they are not
transmitted through the torus. Second, the presence of 
“double-cone” morphology of the narrow line region (NLR) structures
(e.g., Pogge 1989; Wilson \& Tsvetanov 1996; Schmitt et al. 2003), 
with orientations similar to those of radio jets (Nagar et al. 1999
and references therein),  are 
interpreted as emission collimated by the torus.

The role of the extinction produced by the host galaxy cannot
be understated (Alonso-Herrero et al. 2003). For instance, for some
Seyfert 2s broad lines are identified in the near-IR  (e.g., Blanco,
Ward, \& Wright 1990;  Ruiz, Rieke, \& Schmidt 1994; Goodrich et al. 1994;  
Veilleux, Goodrich, \& Hill 1997; Ramos Almeida, P\'erez-Garc\'{\i}a,
\& Acosta-Pulido 2009), and counter-cones are seen in direct
and/or polarized near-IR light  (e.g., Packham et al. 1997 for NGC~1068; 
Maiolino et al. 2000 for Circinus). 
Finally,  there is the tendency of intermediate
types (Seyferts 1.8+1.9) to be hosted in edge-on galaxies (Maiolino \& Rieke
1995).

In recent years much progress has been made toward understanding the
properties of the molecular dusty torus. From the theoretical point of
view, torus models with smooth density distributions (e.g., Pier \&
Krolik 1993; Granato \& Danese 1994; Efstathiou \& Rowan-Robinson
1995) have been superseded 
by more complex clumpy dust distributions (e.g., Nenkova, Ivezic, \&
Elitzur 2002; H\"onig et al. 2006; Nenkova et al. 2008a,
2008b; Schartmann et al. 2008; H\"onig \&
Kishimoto 2010). The clumpy  models reproduce well 
the near-IR and mid-IR emission of different types of AGN 
(e.g., Mason et al. 2006, 2009; Nenkova et al. 2008b; 
Schartmann et al. 2008; Polletta et al. 2008; 
Mor, Netzer, \& Elitzur 2009; Thompson et
al. 2009;  Nikutta, Elitzur, \& Lacy
2009; Ramos Almeida et al. 2009 and 2011, RA09 
and RA11 hereafter; H\"onig et al. 2010; Lira et al. 2011), and 
overcome some of the difficulties faced   by smooth torus models in
fitting  IR data of AGN (see e.g., Alonso-Herrero et al. 2001, 2003).

From the observational point of view, mid-IR  interferometric
observations have revealed that the torus 
is relatively compact, with typical sizes of 
a few parsecs, geometrically thick, and with evidence of clumpiness
(e.g., Jaffe et al. 2004; Tristram et al. 2007, 2009; Burtscher et
al. 2009; Raban et al. 2009). Also, direct mid-IR imaging observations
of Circinus gave a torus size of $\le 4\,$pc (Packham et
al. 2005). In terms of their molecular gas, the tori contain $\sim
10^7\,{\rm M}_\odot$, extend for tens of parsecs (typically radii of 
30\,pc) and have gas column densities, as derived from near-IR
molecular hydrogen lines, ranging from 1 to $10\times
10^{23}\,{\rm cm}^{-2}$ (Davies et al. 2006; Hicks et al. 2009).

%{\rotate

\begin{table*}

\footnotesize
\caption{Properties of the sample.}
\begin{tabular}{lcccccccccccc}
\hline
\hline
Galaxy & z &d & Type & Ref & N$_{\rm H}$(X-ray)& Ref & $\Theta_{\rm
  cone}$ & Ref & $s_{12\mu{\rm m}}$ & Ref & $\log {\rm L}_{\rm bol}$ & Ref\\
       & &  Mpc&    &     & $10^{22}$cm$^{-2}$ & & deg & & pc & & erg s$^{-1}$\\
\hline 
Circinus &0.001448 & 4 & Sy2 & A1 & 430. & B1 & 80-90 & C1 & $0.4, 2.0$ & D1 &
43.6 & E1\\ 
IC~4329A &0.016054 & 65 &Sy1.2 & A2 & 0.61 &B2 & -- & -- & $<$10.8 & D2 & 45.0 & E2\\
IC~5063  &0.011348 & 46 &Sy2 & A3 & 21.78 & B2  &60 & C2 &-- &-- &  44.3-44.7 & E2\\ 
MCG~$-$5-23-16 & 0.008486 &34 &Sy2 (broad Pa$\beta$) &A4 &1.6 &B3 & -- &
--&
2.8 & D2 & 44.4 & E3\\
NGC~1068 &0.003793 &15 & Sy2 &A5 & $>1000.$ &B4 & 40, 65 & C3 &
$0.5\times 1.4$, 
$3\times 4$ & D3 & 45.0 & E4\\
NGC~2110 &0.007789 & 31 & Sy2 (broad Br$\gamma$) & A6 & 2.84 &B2 & 30
& C4 &-- &-- &
43.8-43.9 & E2\\ 
NGC~3227 &0.003859 & 17 & Sy1.5 & A5 &1.74  &B2 &60-70 & C5 &-- & -- & 43.2-43.5 & E2\\ 
NGC~4151 &0.003319 & 13 & Sy1.5 & A5& 6.9 &B5 & 67, 75 &C3 & 2 & D4 & 43.7 & E4\\
NGC~5506 &0.006181 & 25 & NLSy1 (broad Pa$\beta$) & A4, A7 &2.78 &B2 &90-100 & C6 & -- & D2  & 44.1-44.3 & E2\\
NGC~7172 &0.008683 & 35 & Sy2 & A2  & 8.19 & B2 &-- &-- &-- &-- & 43.7-43.8 & E2\\
NGC~7469 &0.016317 & 66 & Sy1 & A5& 0.05 & B6 &-- &-- & 10.5 & D2 & 45.0-45.1 & E2\\
NGC~7582 &0.005254 & 21 & Sy2 (broad Br$\gamma$) & A6 & 5.,$\simeq 100.$ & B7 &86,
120& C3  & -- & D2 &
43.3 & E2\\
NGC~7674 &0.028924 &118 & Sy2 (broad Pa$\beta$) & A8 &$>1000.$ &B8
&--& C2 &-- &-- & $\sim 45$
& E5\\ 
\hline
\end{tabular}
Notes.--- Distances are for $H_0=75\,{\rm km\,s}^{-1} {\rm Mpc}^{-1}$
and for the nearby objects are taken from RA09 and RA11.
References.--- A1. Oliva et al. (1994). A2: V\'eron-Cetty \& V\'eron
(2006). A3: Colina, Sparks, \& Macchetto (1991). A4: Blanco et  al. (1990). 
A5: Osterbrock \& Martel (1993)  and references therein.
A6: Reunanen et al. (2003). A7: Nagar et al. (2002). A8: Ruiz et al. (1994).
B1: Matt et al. (1999).
B2: Vasudevan et al. (2010).
B3: Perola et al. (2002).
B4: Risaliti, Maiolino, \& Salvati (1999).
B5: Beckmann et al. (2005).
B6: Guainazzi et al. (1994).
B7: Bianchi et al. (2009).
B8: Matt et al. (2000).
C1: Maiolino et al. (2000).
C2: Schmitt et al. (2003).
C3: Wilson \& Tsvetanov (1994).
C4: Pogge (1989).
C5: Mundell et al. (1995).
C6: Wilson et al. (1985), Maiolino et al. (1994).
D1: Tristram et al. (2007). D2: Tristram et al. (2009). 
 D3: Raban et al. (2009). D4: Burtscher et al. (2009). E1:  Moorwood et
al. (1996). E2: Vasudevan et al. (2010). 
E3: From the $2-10\,$keV flux of Weaver \& Reynolds (1998) 
and applying a bolometric
correction of 20.
E4: Woo \& Urry (2002) and
references therein. E5: Estimated from the scattered
$2-10\,$keV luminosity  by Malaguti et al. (1998).
\end{table*}
  
\smallskip

This is the third paper in a series using high angular resolution
IR observations, the  \textit{CLUMPY} torus models of Nenkova et
al. (2008a,b), 
and a Bayesian approach for fitting the data 
(Asensio Ramos \& Ramos Almeida
2009) to derive the torus and AGN properties.  In the first two
papers of the series (RA09 and RA11)  we fitted the IR photometric 
spectral energy 
distributions (SEDs) of a sample of nearby Seyfert 1 and Seyfert 2
galaxies. We constrained several
torus  model parameters, namely its angular width and the average number of
clouds along radial equatorial rays,
as well as the viewing angle to the torus.  We found, in
clear contrast with the simplest unified model predictions, that type 2s
appear to have tori with wider cloud distributions 
(the torus angular width $\sigma_{\rm torus}$ parameter, see Figure~1,  and
Section~3.1) and more clumps than
those of type 1s. This may suggest that some of
the properties of the tori of type 1 and type 2 AGNs are
intrinsically different.  For a sample of nearby active galaxies 
H\"onig et al. (2010)
  constrained the number of clouds along equatorial rays and
  their distribution using high-angular resolution
 ground-based mid-IR spectroscopic
  observations, while fixing the other torus model parameters. 
They found however, no differences
in the number and distribution of clouds  between type 1 and type 
  2 AGN.

In this paper we expand on our previous work
by combining photometric 
SEDs with high angular resolution ($\sim 0.3-0.4\arcsec$) mid-IR
spectroscopic observations of a sample of 13 nearby Seyfert
galaxies. For the first time in this work we fit these data for a
sizeable sample of Seyferts to  put tighter constraints on  torus
 model
parameters. This paper is organized as follows. Section~2 presents the new
observations and the data compiled from the literature. Section~3
describes the clumpy dusty torus models and the modelling
technique. A discussion of the fits and inferred torus parameters 
is presented in Section~4. Section~5 discusses the properties of the
torus and AGN. Finally
our conclusions 
are given in Section~6. 

\begin{figure}[h]

\includegraphics[width=10cm]{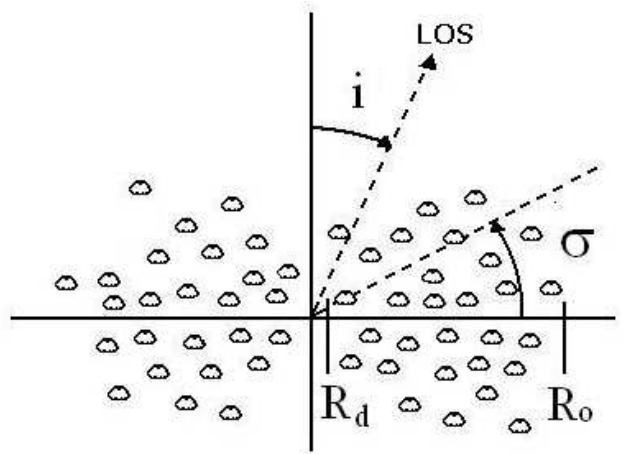}

\caption{Sketch of the  \textit{CLUMPY} models of Nenkova et al. (2008a,b).
The  radial torus thickness $Y$ is defined as the ratio between outer radius 
($R_{\rm o}$) and the dust sublimation radius ($R_{\rm d}$). All the clouds 
have the same $\tau_{V}$, and $\sigma_{\rm torus}$  
characterizes the width of the angular distribution of clouds. The number of
cloud encounters is a function of the viewing angle $i$, the width of
the angular distribution of the clouds $\sigma_{\rm
  torus}$, and
the mean number of clouds along radial equatorial rays $N_0$ (see Equation~2). 
\label{clumpy_scheme}}
\end{figure}

\section{Observations}

\subsection{The sample}
Our sample is composed of 13 nearby  Seyfert galaxies at a mean distance of
38\,Mpc (median of 31\,Mpc). 
We selected the galaxies to have high ($<0.8\arcsec$) 
angular resolution near-IR ($1-5\,\mu$m) imaging, 
and mid-IR ($8-20\,\mu$m) imaging and spectroscopy in the literature
(Sections~2.3 and 2.4).  Although the sample is not complete and
  may not be representative, the high-angular resolution IR data
  ensure that we can isolate the IR emission arising from the torus
  surrounding the AGN. Currently such high angular resolution imaging and
  spectroscopy 
can only be attained from the ground using 8-10\,m-class
telescopes (typically $\sim 0.3-0.4\arcsec$ at $10\,\mu$m). 
The properties of this sample relevant to this work, and the corresponding
references  are summarized in Table~1. The sample includes type 1 and
type 2 Seyfert galaxies, as well as Seyfert 2 galaxies with a 
 broad component detected in the near-IR. The bolometric 
luminosities  of the AGN, $L_{\rm bol}$,  are taken from the works of
Woo \& Urry (2002) and Vasudevan et al. (2010), or estimated from the
hard X-ray luminosities using a typical bolometric correction of 20
(Elvis et al. 1994). The AGN bolometric luminosities of our sample 
span almost two
orders of magnitude, from $\sim 10^{43}$ to $10^{45}\,{\rm erg \,
  s}^{-1}$. 
In terms of their X-ray neutral hydrogen column 
density, $N_{\rm H}$(X-ray), the sample contains both Compton thin
objects and  Compton thick galaxies. We finally list in Table~1 two
observational properties related to the torus, namely 
the opening angle of the ionization cones
$\Theta_{\rm cone}$, and the torus size $s_{12\mu{\rm m}}$ (
  FWHM) as derived from the modelling of
mid-IR interferometric observations. These two parameters 
will be compared with the fitted torus model 
parameters in Section~5.1.

As explained in the Introduction, in  this paper in addition to the IR
photometric SEDs we will
fit ground-based mid-IR ($\sim 8-13\,\mu$m) spectroscopy, including the
$9.7\,\mu$m silicate feature. Shi et al. (2006) found that the $9.7\,\mu$m
silicate features measured from {\it Spitzer}/IRS spectroscopy of a
large sample of AGN vary from emission to 
absorption with increasing neutral hydrogen column densities. These
authors interpreted this result with a scenario where the obscuring material
is located in two different physical scales ($0.1-10$\,pc disk and a
large scale disk extending up to 100\,pc) with the dust distributed in clouds of
different properties. Although the mid-IR ground-based spectroscopy 
in this work (see Sections~2.2 and 2.4) probes much smaller physical
scales than those probed by the {\it Spitzer}/IRS data,  
it is of interest to place our sample of galaxies in the context
of larger samples. In Figure~2 we show the apparent strength of the $9.7\,\mu$m
silicate feature (${\rm S}_{9.7}$) as a function of the X-ray hydrogen column
density for Seyfert 1s, Seyfert 2s, PG quasars, and 2MASS quasars
adapted from the work of Shi et al. (2006). The strengths of
the silicate feature in this figure
are mostly measured from {\it Spitzer}/IRS spectroscopy\footnote{The
  typical extraction apertures of the Shi et al. (2006) IRS short-low
  (SL) spectra  were 6pixels or 10.8\arcsec.} and thus
correspond©∂ to large physical sizes (at the mean distance
  of our sample $\sim 2\,$kpc). For NGC~4151, NGC~7469 and NGC~3227
$S_{9.7}$ were also measured from {\it Spitzer}/IRS spectroscopy
(Thompson et al. 2009), and for NGC~7582, NGC~7674 and IC~5063 from
ground-based data (H\"onig et al. 2010, and Section~2.4). 
 From this figure it is clear that our
relatively small sample probes well the observed ranges for Seyfert 1s
and Seyfert 2s. We will come back to this issue in Section~5.

\begin{figure}

\includegraphics[width=8.5cm,angle=-90]{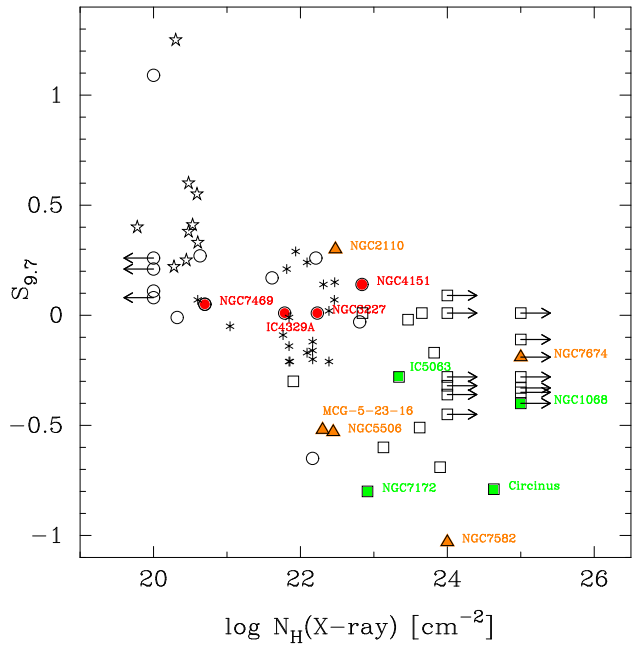}
\caption{Apparent strength of the $9.7\,\mu$m silicate feature vs. the X-ray
  hydrogen column density, adapted from the work of Shi et
  al. (2006). For the strength of the silicate feature, positive
  values mean the feature is observed in emission and negative values
  in absorption. Star-like symbols are PG quasars, asterisks are 2MASS
  quasars, circles are Seyfert 1s, and squares are Seyfert 2s. The
  filled symbols indicate the galaxies in our sample. The triangles
  are those Seyfert 2s in our sample with broad lines detected in the
  near-IR (see Table~1).  The X-ray column densities are from Shi et
  al. (2006) except for the 
galaxies in our sample, which are taken from the references given in
Table~1. The strength of the silicate features are measured 
from {\it Spitzer}/IRS data and ground-based data (see Section~2.1). 
}

\vspace{0.5cm}
\end{figure}

\subsection{New observations}
$N$-band spectroscopy of NGC~4151 was obtained with Michelle (Glasse
et al. 1997) on the
Gemini North telescope on 20070318 (Program ID GN-2006B-Q-18). The
standard mid-IR chop-nod technique was employed for the observations,
and the chop and nod distances were 15\arcsec. The LowN grating and 2-pixel
(0.36\arcsec) slit were used, giving $R\sim200$. The slit was orientated at 60
degrees E of N, along the extended mid-IR emission discovered by
Radomski et al. (2003). Two blocks of spectroscopy were obtained, each
of 450 sec on-source (in the guided chop beam), and the galaxy nucleus was
reacquired in between. Telluric standard stars were observed before
and after the NGC~4151 spectroscopy,  with the same observational
setup. Michelle data files contain planes consisting of the difference image
for each chopped pair for each nod. The chopped pairs were examined
for anomalously high background or electronic noise, but none needed
to be excluded from the final data set. The chopped difference images
were then combined until a single file was obtained for each
spectroscopy block, and this file for the two blocks then averaged
together. The resulting spectrum was extracted in a 2-pixel aperture,
wavelength-calibrated using telluric lines, divided by the standard
star, and multiplied by a black-body spectrum. 
%[If more detail is
%needed, I used the first, brighter standard star, Psi Uma, K1III,
%T=4610 K] 

$Q$-band ($\sim 20\,\mu$m) spectroscopy of NGC~1068 was obtained with
Michelle on 20090923 (Program ID GN-2009B-Q-58), using the same
chop-nod technique. The 3-pixel (0.54\arcsec) slit and lowQ grating were
used, for $R\sim100$, with the slit orientated along the ionisation cones
(20 degrees E of N).  The galaxy nucleus was observed for a total of
1800 sec on-source, and telluric standard stars were observed before
and after NGC~1068. 
The data reduction procedure was similar to that used for the NGC~4151
$N$-band spectroscopy, including extraction of the spectrum in a 2-pixel
(0.36\arcsec) aperture. 
%[Tom Geballe helped me with the telluric cancellation, and I extracted
%the spectrum in a 2-pix aperture to compare with his results. I don't
%think 2 pixels is really appropriate in the Q band, but your scaling
%to the photometric points probably means this is not very
%important. It may be appropriate to acknowledge Tom for his help with
%the data reduction.] 
However, cancellation of the strong and variable $Q$-band telluric lines
was poor in the initial, ratioed spectrum. This was improved by adding
a sloping baseline to the standard star spectra before dividing\footnote{
i.e., fitting a continuum, multiplying by a few
per cent, adding that back to the star spectrum, and dividing by that
instead.}. The effect on the spectral slope of the resulting NGC~1068
spectrum is small.  The flux calibration was done with the telluric
standard stars. The  
resulting flux densities are probably affected by slit losses, and
therefore in Section~2.4 we recalibrate the spectra with
  photometric observations. Finally the $Q$-band spectrum used for the
modelling was rebinned using a 3 pixel box.

%{\rotate
\begin{table*}

\footnotesize
\caption{Wavelength coverage of the observations and references}
\begin{tabular}{lcccccc}
\hline
\hline
Galaxy & near-IR SED & Ref. & mid-IR SED & Ref & Spectroscopy & Ref\\
\hline 
Circinus & $JHKK'L'M'$  & A1 & 8.7, $18.3\,\mu$m & B1
& $N$-band MIDI 0.60\arcsec & C1 \\
%         &            &                    & 11.8, $18.7\,\mu$m&
%         B2\\
IC~4329A & $HK$ & A2 &8.99, $11.88\,\mu$m & B2 & $N$-band VISIR 0.75\arcsec & C2 \\
         & $L$  & A3\\
IC~5063 & $H$ & A4 & 8.7, $18.3\,\mu$m 
& B1 & $N$-band T-ReCS 0.67\arcsec & C3\\ 
        & $K$  & A5 &10.5, 11.3, 11.9, $12.3\,\mu$m    &
        B2\\  
MCG~$-$5-23-16 & $JHKL'M$ & A2 & 8.59, 8.99, $11.88\,\mu$m  & B2 &
$N$-band VISIR 0.75\arcsec & C2\\ 
             &         &   & $18.72\,\mu$m & B3\\
NGC~1068 & $JHK$          & A2     & 8.8,
$18.3\,\mu$m & B4 & $N$-band Michelle 0.36\arcsec & C4\\
             &$LM$               & A6    &  &   & $Q$-band  Michelle
             0.54\arcsec & This work\\
NGC~2110 & $L'$  &A7  & $11.2\,\mu$m & B5  &
$N$-band Michelle 0.36\arcsec  & C5 \\ 
         &    &                    & 8.6, 9.0, $11.9\,\mu$m    
& B2\\
         &    &                     &  $20\,\mu$m & B6\\
NGC~3227 & $HK$ & A8 & $11.3\,\mu$m & B1 & $N$-band VISIR 0.75\arcsec & C2\\
         & $L$    &A3  & 8.99, $11.88\,\mu$m & B2\\
NGC~4151 & $JHK$ & A8 & 10.8, $18.2\,\mu$m & B1
& $N$-band Michelle 0.36\arcsec & This work \\
         & $LM$    &  A3\\
NGC~5506 & $JHKL'$ & A1 & 11.3, $18.1\,\mu$m &
B1 & $N$-band T-ReCS 0.36\arcsec  & C6 \\
         & M & A3 & \\
NGC~7172 & $HKL'M$     & A2           & $10.36\,\mu$m & B1 & 
$N$-band T-ReCS 0.36\arcsec  & C6 \\
         &      &           & $12.27\,\mu$m & B7\\
NGC~7469 & $JHK$ & A2 & 8.7, $18.3\,\mu$m & B8 &
$N$-band VISIR 0.75\arcsec  & C2\\
         & $L'$    & A1   & 10.49, 11.25, 11.88, 12.27, $13\,\mu$m& B2\\
NGC~7582 & $HKLM$ & A1 & 10.8, $18.2\,\mu$m & B1 &
$N$-band VISIR 0.75\arcsec  & C2 \\
         &      &   & 8.6, 9.0, $11.9\,\mu$m & B2 & \\
NGC~7674 & $JHKL'M$ & A2 & $12.81\,\mu$m & B2 & $N$-band VISIR 0.75\arcsec
& C2\\
\hline
\end{tabular}

References. A1: Prieto et al. (2010). A2: Alonso-Herrero et
al. (2001). A3: Ward et al. (1987), the photometry is used as an upper
limit. A4: Quillen et al. (2001).
A5: Kulkarni et al. (1998). A6: Marco \& Alloin (2000). 
A7: Alonso-Herrero et al. (1998), the photometry is an upper limit.
A8: Alonso-Herrero et al. (2003).
%A8: Quillen et al. (2001). 
B1: RA09. B2: H\"onig et al. (2010).
B3: Reunanen, Prieto, \& Siebenmorgen (2010).
B4: Tomono et al. (2001), for the 0.4\arcsec \, fluxes. B5: Mason et al. (2009).
B6:  Lawrence et al. (1985). B7: Horst et al. (2008).
B8: RA11.
C1: Tristram et al. (2007). C2: H\"onig et al. (2010). 
C3: Young et al. (2007). C4: Mason et al. (2006).
C5: Mason et al. (2009). C6: Roche et al. (2007).
\end{table*}
%}  
\smallskip

\subsection{Published unresolved nuclear fluxes}
As discussed at length in our previous papers (Alonso-Herrero et
al. 2001, 2003; RA09 and RA11) high
angular resolution observations are required to isolate the
emission
associated with the torus, and with the direct view
of the AGN in type 1s as
well.  At the distances of our galaxies and the current angular
resolutions of the near and mid-IR imaging and spectroscopic observations 
the torus emission appears unresolved. 
In the near-IR up to $\lambda \sim 2\,\mu$m, extended stellar
emission arising in the host galaxy 
contaminates and even dominates the nuclear fluxes of type 2 Seyferts 
(Alonso-Herrero, Ward, \& Kotilainen 1996; Videla et al. 2011) and is
not negligible 
in type 1 Seyferts 
(Kotilainen et al. 1992). At longer wavelengths ($\lambda > 3\,\mu$m)
contamination by stellar photospheric emission is greatly
reduced. However, any extended nuclear emission not directly 
related to the dusty torus,
such as dusty clouds in the NLR and the 
coronal line region (e.g., Bock et al. 2000; Alloin et
al. 2000; Radomski et al. 2003; Packham
et al. 2005; Mason et al. 2006; Roche et al. 2006; 
Reunanen et al. 2010) and/or dust heated by  young massive stars
(Siebenmorgen, Kr\"ugel, \& Spoon 2004; 
Alonso-Herrero et al. 2006; Mason et al. 2007; Reunanen et al. 2010),  
needs to be removed. 

To isolate as much as possible the emission
of the torus (and the AGN when seen directly)  we compiled
high-angular resolution  near-IR and mid-IR  fluxes from the
literature with estimates of the unresolved emission when available. 
These unresolved fluxes are the result of removing, using
  various methods, the underlying
 near-IR stellar emission and the mid-IR emission produced by star
formation. The  compiled photometry includes near-IR ground-based
and {\it HST}/NICMOS 
observations (with the NIC2 camera, angular resolutions
$0.15-0.2$\arcsec) and ground-based mid-IR measurements with angular
resolutions   of $\le0.8\arcsec$. The only exception are the $L$-band
measurements of Ward et al. (1987) for type 1 Seyferts (Table~1), but
we use them as upper limits. 
All the mid-IR photometric points have angular
resolutions in the range $0.3-0.5$\arcsec \, to match the resolution and
slit widths of the mid-IR spectroscopic data (see Section~2.4). When
possible, for a given galaxy 
we tried to match the angular resolution of the photometric
points to make sure we are modelling similar physical scales.  
Table~2 lists for each galaxy in our 
sample the wavelengths and references of the photometric data used to construct
their SEDs. 
Finally, we used the {\it Spitzer}/IRS $30\,\mu$m continuum 
fluxes of Deo et al. (2009) for those galaxies in our sample without
nuclear star 
formation as upper limits in our fits.  

Based on discussions in the papers listed in Table~2 and comparisons
between different works we use the following errors for our
analysis. For near-IR ground-based data, except for the NACO data (see
below) of Prieto et
al. (2010),  we use for the $J$-band 30\%, for the $H$ and $K$ bands 25\%,
and for the 
$L$-band 20\%. These include the photometric error,
the background subtraction uncertainty and the uncertainty from
estimating the unresolved flux. For the last one we note that the
stellar emission contribution within a given aperture decreases with 
increasing wavelength and thus the stellar contribution 
has a minimum in the $L$-band 
(see e.g., Kotilainen et al. 1992 and
Alonso-Herrero et al. 1996). The $M$-band fluxes are always considered as 
upper limits because estimating the unresolved component was not
possible. For the NACO AO observations of Prieto et al. (2010), which
were measured through
0.1-0.2\arcsec \, apertures, the smaller 
contamination by stellar emission when compared to natural seeing
observations results in lower
uncertainties. We therefore use 20\% in J, and 15\% in 
the $HKLM$ NACO bands. Finally, for the NICMOS data, which have
very stable photometric calibration and PSFs, we use 20\% in the $J$-band,
and $10-20$\% in the $H$ and $K$ bands (unless otherwise specified in the
corresponding references). The estimated NICMOS 
uncertainties are based on the comparison of the unresolved fluxes reported
for the same galaxies in different works (Alonso-Herrero et al. 2001;
Quillen et al. 2001; Gallimore \& Matthews 2003, and 
Kishimoto et al. 2007). For the $N$ and $Q$ band measurements we use 15\%
and 25\% errors, respectively to account for the photometric
calibration and unresolved component uncertainties (see details in RA09). 

We finally address the issue of possible variability of the
  near-IR fluxes and the simultaneity of the SEDs. 
Nine out of the 13 galaxies in our sample have been
  reported to show variability in the near-IR: NGC~7674 (Quillen et
  al. 2000), NGC~1068, IC~4329A, NGC~2110, MCG~$-$5-23-16, NGC~5506,
  and NGC~7469 (Glass 2004), NGC~3227 (Suganuma et al. 2006), and
  NGC~4151 (Koshida et al. 2009). There
are no reports on mid-IR variability of our sources. 
The typical variability  in the near-IR
 is $\simeq 40$\% on average, with amplitude variations around the
 median of between 
    $10-30\%$ for our sample. For all these galaxies the 
  $1-2.2\,\mu$m data, when available,
 were taken simultaneously. The $LM$ fluxes in some cases were not
 simultaneous with the shorter wavelength fluxes. However, in most cases the
 $LM$ fluxes are taken as
 upper limits as there was no estimate of the unresolved
 emission. Glass (2004)  showed that the $L$-band variability 
is typically less than 20\% around the median flux, which is within
the photometric and unresolved emission uncertainties.  We 
can therefore assume that variability in the near-IR does not affect the
compiled SED within the above discussed uncertainties of the
unresolved measurements.

\subsection{Published ground-based mid-IR spectroscopy}
The published mid-IR spectroscopy used in this work (see Table~2) 
was obtained with four different instruments:
\begin{itemize}
\item
(i) The Thermal-Region
Camera Spectrograph (T-ReCS; Telesco et al. 1998) on the Gemini-South
Telescope. The T-ReCS data were obtained with the low-resolution
grating, which provides a spectral resolution of $R\sim 100$. The slit
widths were  0.67\arcsec \, (Young et al. 2007) and
0.36\arcsec \, (Roche et al. 2006, 2007). 
\item
(ii) The Michelle instrument
(Glasse et al. 1997) on the Gemini-North telescope. The Michelle
observations (Mason et al. 2006; 2009 and Section~2.2) 
were obtained with a 0.36\arcsec \, slit width 
and a spectral resolution of $R\sim 200$ in the $N$-band, and  a
  0.54\arcsec \, slit and $R\sim
100$ in the $Q$-band. 
\item
(iii) VISIR, the mid-IR 
imager and spectrograph mounted on the 8.2m UT3 telescope at the
ESO/Paranal observatory in Chile. The VISIR observations (H\"onig et
al. 2010) were obtained in low-spectral resolution mode ($R \sim
300)$ with a slit width of 0.75\arcsec.  
\item
(iv) MIDI (Leinert et al. 2003), the mid-IR
interferometer at the VLTI at the ESO/Paranal observatory. The MIDI
observations (Tristram et al. 2007) were taken with a 0.6\arcsec \,
width slit and $R\sim 30$. 
The data used here are a "total flux spectrum", where 4
spectra are obtained for this, two 
for each telescope of the interferometer and 2 for each "window" in
MIDI. The spectrum was extracted in a 6 pixel wide mask or
0.516\arcsec,  and calibrated 
individually for the windows and telescopes. 
\end{itemize}

Because the VISIR ground-based mid-IR spectra have slightly higher
spectral resolution than the other spectra, we rebinned them to contain
approximately 150 spectral points. The MIDI, T-ReCS and Michelle
spectra contain between 100 and 200 spectral points. We did not
attempt to remove any emission lines (e.g., [S\,{\sc
  iv}]$10.51\,\mu$m) or broad features (e.g., PAH features) 
in the spectra. 

\begin{figure}
\includegraphics[width=7cm,angle=-90]{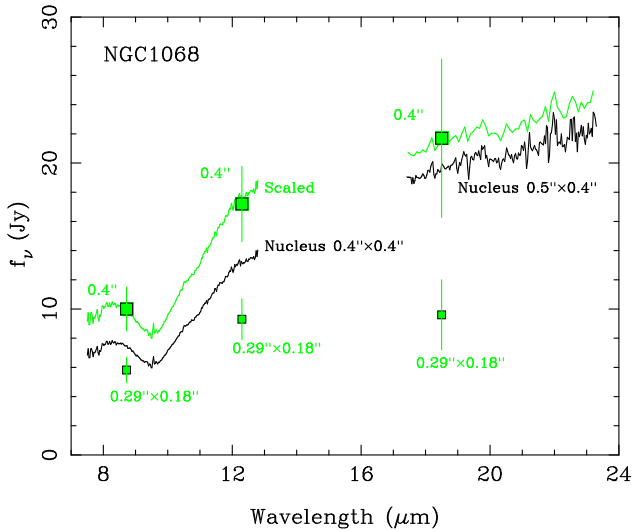}
\caption{The black lines are the 
original flux-calibrated 
Michelle $N$ and $Q$-band spectra of NGC~1068 (see Mason et al. 2006
and Section~2.1, respectively). The large and
small square symbols are the photometry of Tomono et al. (2001) for a
circular 0.4\arcsec-diameter aperture on the original data, and for a 
$0.29\arcsec \times 0.15\arcsec$ aperture on
the deconvolved data, respectively. The green lines are the $N$ and
$Q$-band spectra
scaled to the 0.4\arcsec \, photometric points at 8.7 and $18.5\,\mu$m
photometric points.  After the scaling, the $Q$-band spectrum was
  also rebinned (see text for details).}
\vspace{0.5cm}
\end{figure}

The final step was to  scale
the spectra to the photometric points. This was mostly necessary for the
spectra observed with the narrowest slits (see Table~2) because of
slit losses. As an
example, in Figure~3 we show the original $N$ and $Q$-band  spectra
(Mason et al. 2006 and Section~2.1, respectively) of NGC~1068
together with the mid-IR photometric points of 
Tomono et al. (2001) for a circular
aperture of 0.4\arcsec-diameter, and the photometry through
a rectangular $0.29\arcsec \times 0.15\arcsec$ aperture after
deconvolution of the data. To match the angular resolutions of the
imaging and spectroscopic data, the NGC~1068 $N$ and $Q$-band 
spectra were scaled to their corresponding
$0.4\arcsec$ \, photometric points.  Finally, for all the $N$-band spectra 
in our sample we added in quadrature the
intrinsic error of the spectra and the 
15\% error of the imaging data (see Section~2.3) used for scaling the
data, for each spectroscopic point. For the $Q$-band spectrum of
NGC~1068 the associated error for scaling the spectrum is 25\%. These
errors were added for the fitting.
\smallskip

\section{Modeling of the data}

\subsection{Clumpy Torus Models}
In this work we use an interpolated version (see Section~3.3) of 
the clumpy dusty torus models of Nenkova et al. (2002,
2008a,b) including the corrections for the previously 
erroneous AGN scaling factor
(see the erratum by Nenkova et al. 2010).  
The \textit{CLUMPY} models are described by the six parameters
listed in Table~3 that we explain in the following (see also
Figure~1). An AGN with a bolometric luminosity $L_{\rm bol}({\rm AGN})$ is 
surrounded by a torus of dusty clouds and all the clouds have the
same optical depth $\tau_V$, which is defined in the optical $V$-band. 
The torus clouds are located between the inner radius of the torus
$R_{\rm d}$ and the outer radius of the torus $R_{\rm o}$, and the
torus radial thickness is defined as $Y = R_{\rm o}/R_{\rm d}$. The inner
radius of the torus is set by the dust sublimation temperature
($T_{\rm sub} \approx 1500$ K), 

\begin{equation}
 R_{\rm d} = 0.4~\left( \frac{1500~{\rm K}}{T_{\rm sub}}\right)^{2.6}
 \left({\frac{L_{\rm bol}({\rm AGN})}{10^{45}\,\mathrm{erg 
  ~s^{-1}}}}\right)^{0.5} {\rm pc}.   
\end{equation}

The angular distribution of the clouds is assumed to have a smooth
boundary and it is described as a Gaussian with a width
parameter $\sigma_{\rm torus}$. The radial distribution is a declining
power law with index $q$ ($\propto r^{-q}$). 
The mean number of clouds along  radial equatorial rays is
$N_0$. The number of clouds along the line of sight (LOS) at a viewing angle
$i$ (measured from the polar direction, see Figure~1) is,

\begin{equation}
N_{LOS}(i) =
N_0~e^{-(90-i)^2/\sigma_{\rm torus}^2}.
\end{equation}

In a clumpy dust distribution the classification of an object as a
type 1 or type 2 AGN is not truly a matter of  the viewing angle but
 of the probability for direct view of the AGN (see Figure~1, and also Elitzur
2008). The probability for an AGN-produced
 photon to escape through the torus along a
viewing angle $i$  when all the clouds are optically 
thick ($\tau_V>1$) is,

\begin{equation}
P_{\rm esc} \simeq e^{(-N_{\rm LOS})}.
\end{equation}

\begin{table}
\caption{Axial ratios, foreground extinction measurements and strength
  of the silicate feature}
\begin{tabular}{lccccc}

\hline
Galaxy & $b/a$ & $A_V$(frg) & Ref & $\rm{S}_{9.7}$ & Ref\\
       & & (mag)\\
\hline
Circinus & 0.44 & 9 & A1, A2 & $-$1.8/$-$2.4 & B1\\
IC~4329A & 0.28 &- &- & $-0.02$ &B2 \\
IC~5063  & 0.68 & 7 & A3     & $-$0.3 & B2\\
MCG~$-$5-23-16 &0.46 &$>6$ & A4 & $-$0.3 & B2\\
NGC~1068 & 0.85 &- &-& $-0.4$ & B3 \\
NGC~2110 & 0.74 & 5 & A2   & $0.0$3& B2\\
NGC~3227 & 0.68 & -& - & 0.01 & B4\\
NGC~4151 & 0.71 & - & - & 0.14 & B4\\
NGC~5506 & 0.30 & $\ge 11$ & A5 & $-$1.1/$-$1.4 & B1\\
NGC~7172 & 0.46 & - & - &  $-$3.2/$-$3.2& B1\\
NGC~7469 & 0.72 & - & - & 0.05 & B4\\
NGC~7582 & 0.42 & 8, 13 & A6 & $-$1.0 & B2\\
NGC~7674 & 0.91 & $\sim 3-5$ & A7  & $-$0.2 & B2\\
\hline
\end{tabular}

Notes.--- The axial ratios are the ratio of to minor to major axis of
the host galaxies and are taken from 
de Vaucouleurs et al. (1991), except for that of NGC~7172 that is from
Jarrett et al. (2003).
 S$_{9.7}$ are the observed  strengths of the silicate feature,
with positive numbers indicating that the feature is in emission, and 
negative numbers the feature is
in absorption. For the Roche et al. (2007) galaxies, the two measurements
correspond to fits to the feature done using two different silicate grain
profiles.  References. A1: Maiolino et al. (2000). A2: Storchi-Bergmann et
al. (1999). A3: Heisler \& Robertis (1999).
A4: Veilleux et al. (1997). A5: Goodrich et al. (1994). 
A6: Winge et al. (2000). A7: Riffel, Rodr\'{\i}guez-Ardila, \&
Pastoriza (2006). B1: Roche et 
al. (2007). B2: H\"onig et al. (2010). B3: Mason et al. (2006).
B4: Thompson et al. (2009).
\end{table}
\smallskip

In the models the radiative transfer equations are solved for each
clump and thus the solutions depend mainly on
the location of each clump within the torus, its optical depth, and
the chosen dust composition. We adopt a dust extinction profile
corresponding to the OHMc dust (i.e., the standard cold oxygen-rich
ISM dust of Ossenkopf et al. 1992). The total torus emission is
calculated by integrating the source function of the total number of
clumps convolved with the radiation propagation probability along the
torus (Nenkova et al. 2002). For Type 1 and intermediate type Seyferts
where there is an unobscured view of the AGN, it is also possible to include 
its contribution to the resulting SED. The AGN continuum emission in
these models is characterized with a piecewise power law distribution
(see Nenkova et al. 2008a for details).

 We finally note that there is some evidence of the presence of an
extra hot dust component, originating very close to the AGN. 
Reverberation mapping (see discussion by Kishimoto et al. 2007) and near-IR
interferometric observations (e.g., Kishimoto et al. 2009 and
references therein) of Seyfert galaxies  
have shown that the sublimation radii appear to be smaller than
expected from Equation~1 (similar to the Barvainis 1987 relation).
However in this work we  take the simplest approach for fitting the data
assuming that all the near-IR emission is originating in the
¨classical¨ torus, with no additional hot dust components.

\subsection{Foreground dust components}
In this section we examine the possibility of having contributions to 
the observed IR SEDs and mid-IR spectra of Seyfert galaxies from dust
emission  and absorption unrelated to the AGN. There are several
pieces of evidence we can look
into, namely the inclination of the galaxies, and the presence
of nuclear dust lanes and spatially resolved variations of the
silicate feature. 

Approximately half of the galaxies in our sample have
minor-to-major axis ratios of $b/a<0.5$ (inclined host galaxies, see Table~4),
and as suggested by Deo et al. (2009), it is likely that 
dust in the host galaxy
disk can contribute significantly to the observed silicate
absorption and the long-wavelength continuum.  
The only exception appears to be IC~4329A, which is a highly inclined
system, but does not show silicate absorption. Deo et al. (2009)
concluded for this galaxy  that it is likely that our line
of sight does not intersect any dense clouds in the host galaxy. 

\smallskip

\begin{table}
\caption{Parameters of the \textit{CLUMPY} Torus Models}
\begin{tabular}{lll}
\hline
\hline
Parameter & Symbol& Interval\\
\hline
Torus radial thickness & $Y$ & [5, 30]\\
Torus angular width  & $\sigma_{\rm torus}$ & [$15\arcdeg$, $70\arcdeg$]\\
Number of clouds along an equatorial ray & $N_0$ & [1, 15]\\
Index of the radial density profile & $q$ & [0, 3]\\
Viewing angle & $i$ & [$0\arcdeg$, $90\arcdeg$]\\
Optical depth per single cloud & $\tau_V$ &  [5, 150]\\
\hline
\end{tabular}

Notes.--- Torus radial thickness: $Y=R_{\rm o}/R_{\rm d}$ where $R_{\rm o}$ is the
outer radius and $R_{\rm d}$ is the inner radius (Equation~1). 
The cloud distribution
between $R_{\rm d}$ and $R_{\rm o}$ is parametrized as $r^{-q}$.
\end{table}
\bigskip

Using color maps  
Martini et al. (2003) showed that
in general the dust in the nuclear regions of Seyfert galaxies is on
physical scales not associated with those of the dusty   
torus. It is then likely that this dust is
 located in front of the nucleus (see also Regan \& Mulchaey   1999)
 and arises from  the galactic ISM. In our sample, 
Circinus, IC~5063, NGC~5506, NGC~7582, NGC~2110,
NGC~7172, and possibly NGC~7674 show dust features in the nuclear 
(central $\sim 1-2\,\arcsec$) regions (see e.g., Quillen et al. 1999; 
Maiolino et al. 2000; Martini et al. 2003).

Some of the galaxies in our sample show variations
of the $9.7\,\mu$m silicate feature on scales of $1-2$\arcsec \, indicating
the presence of extended dust components (Mason et al. 2006; Roche et
al. 2006, 2007: Colling, Roche, \& Mason 2009). Similarly, the spatially-resolved mid-IR
     polarimetric observations of NGC~1068 (Packham et al. 2007) can
     be explained with a geometrically, and
optically thick torus  surrounded by a larger, more diffuse
structure, associated with the dusty central regions of the
host galaxy. 

Roche et al. (2007) showed that 
    NGC~5506 and NGC~7172, both with prominent nuclear dust lanes
    (see e.g., Malkan et al. 1998),
have similar absorbing columns as derived
     from X-ray observations and the $9.7\,\mu$m features, for
       the latter just by using a foreground dust screen model. This
     suggests that some of the extinction measured from the $9.7\,\mu$m
     silicate feature might arise in the galactic ISM (see also the
     discussion in Section~2.1,
     and Figure~2). In other words, in these two galaxies 
if the silicate feature were to come only from the torus, it would be
shallower (at it is filled in 
       by emission from warm dust also in the torus).
These findings would also be consistent with the work of 
Guainazzi, Matt, \& Perola (2005) who showed that the presence of dust
lanes on scales of 100\,pc in Compton-thin Seyfert 2s is
correlated with the X-ray obscuration, mostly in the X-ray column
density range $\sim 10^{23}-10^{24}\,{\rm cm}^{-2}$. This was
interpreted by these authors as
due to the larger covering fraction of the gas in the dust lanes,
rather than the pc-scale dusty torus.

\setcounter{figure}{3}
\begin{figure*}
\includegraphics[width=7cm,angle=-90]{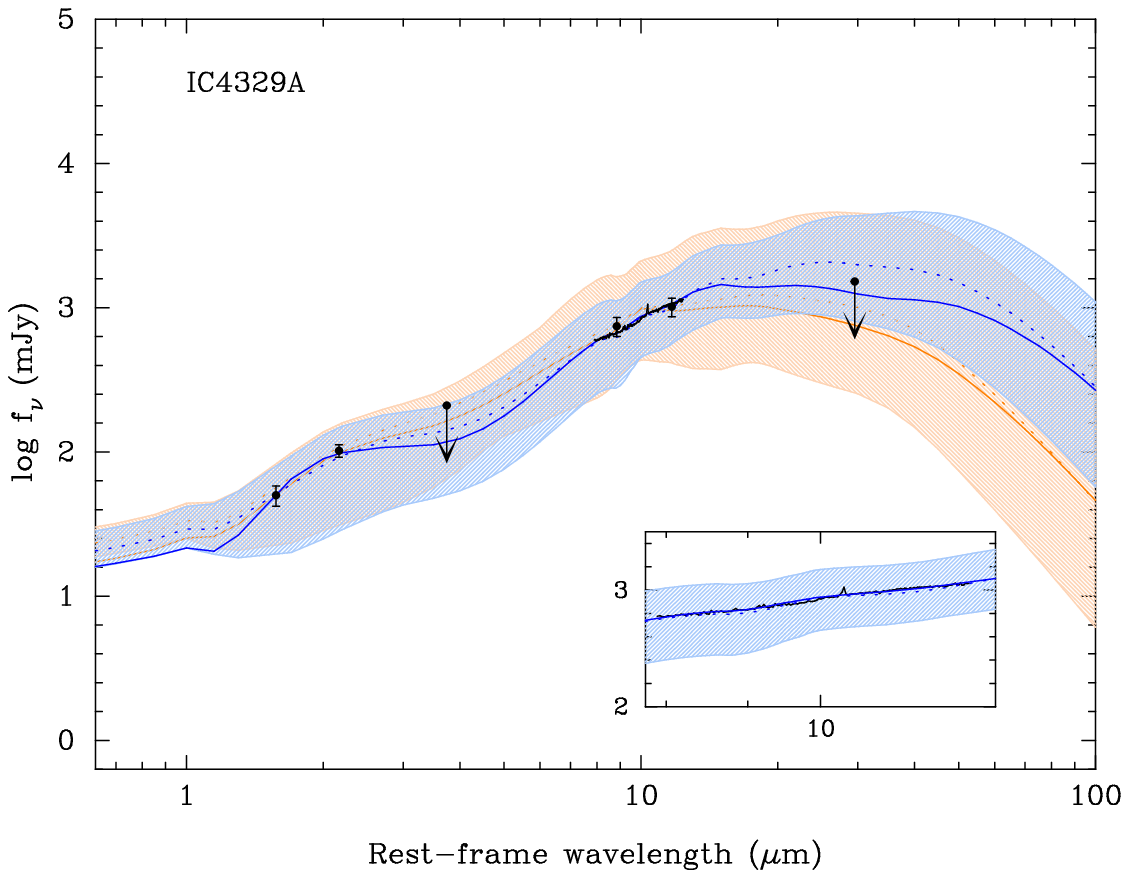}
\includegraphics[width=7cm,angle=-90]{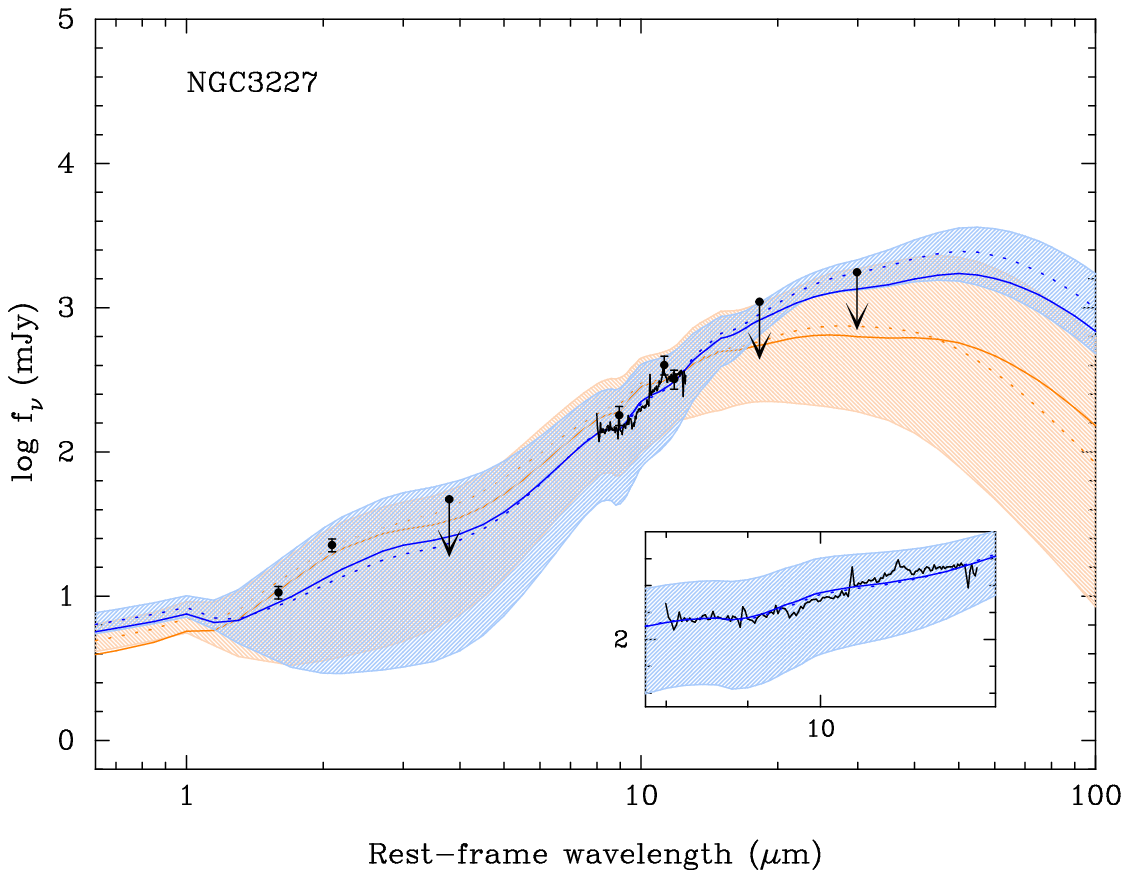}

\vspace{0.5cm}
\includegraphics[width=7cm,angle=-90]{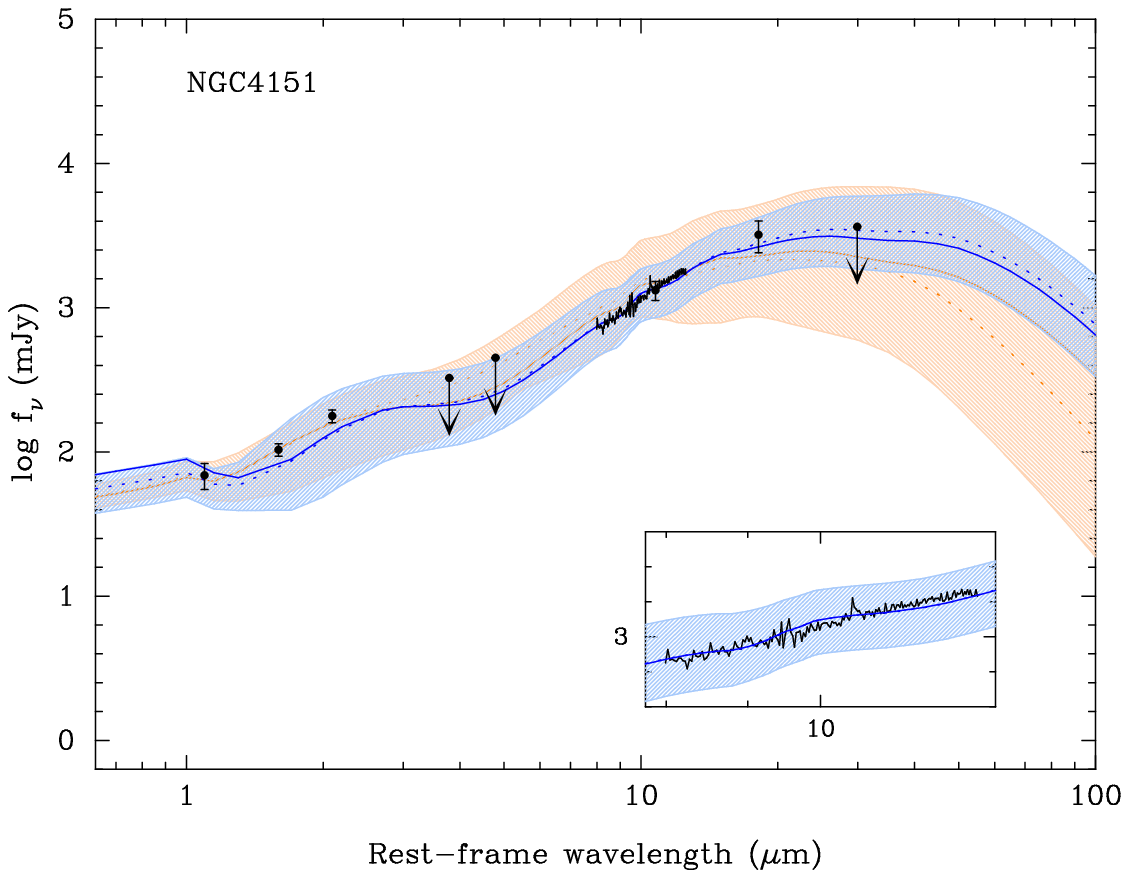}
\includegraphics[width=7cm,angle=-90]{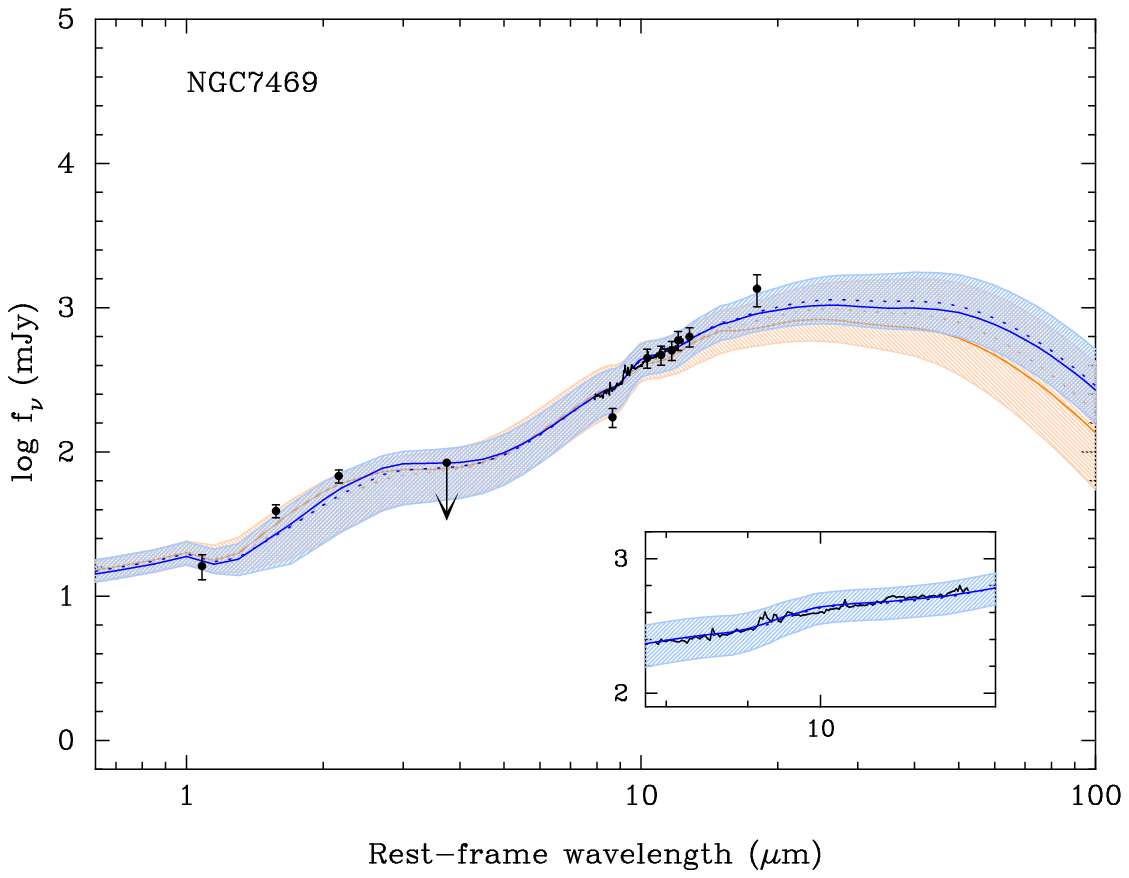}

\caption{Model fits for pure Seyfert 1 galaxies with 
torus + AGN emission. 
The filled dots are the SED photometric data and the black line
  is the mid-IR spectrum. The fits of
  the torus emission to the IR  SED data points alone
 are shown in orange, while the fits to  the IR SED and spectroscopy
  are shown in blue. The solid blue and orange lines are the models
  described by the combination of parameters that maximizes their
  probability (MAP) distributions, while the dashed lines correspond
  to the model  
  computed with the median value of the probability distribution of
  each parameter. The shaded areas indicate the range of models
  compatible with a 68\% confidence interval.  The inset 
shows in detail the fit to the $N$-band 
spectroscopy.   We did not restrict any of the torus model 
parameters. No
foreground extinction was used for the fits.}

\end{figure*}

\setcounter{figure}{4}
\begin{figure*}
\includegraphics[width=7.cm,angle=-90]{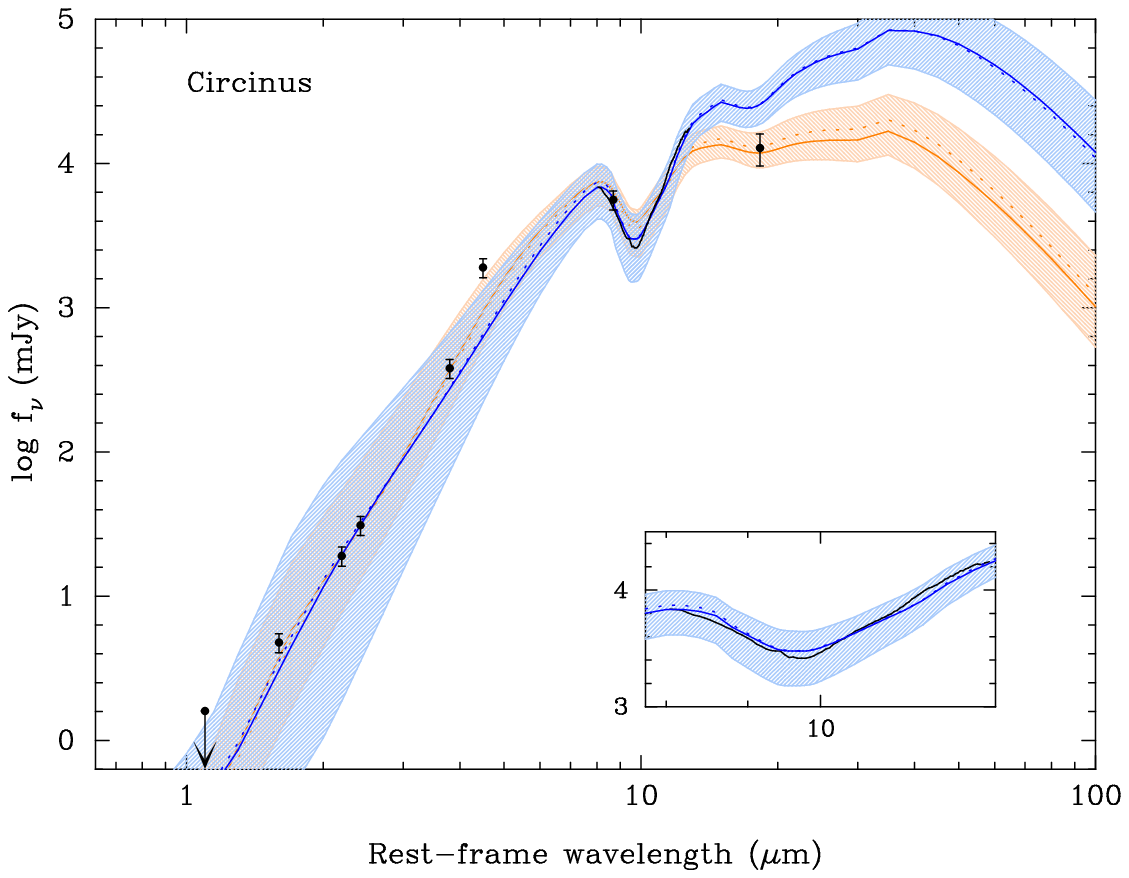}
\includegraphics[width=7.cm,angle=-90]{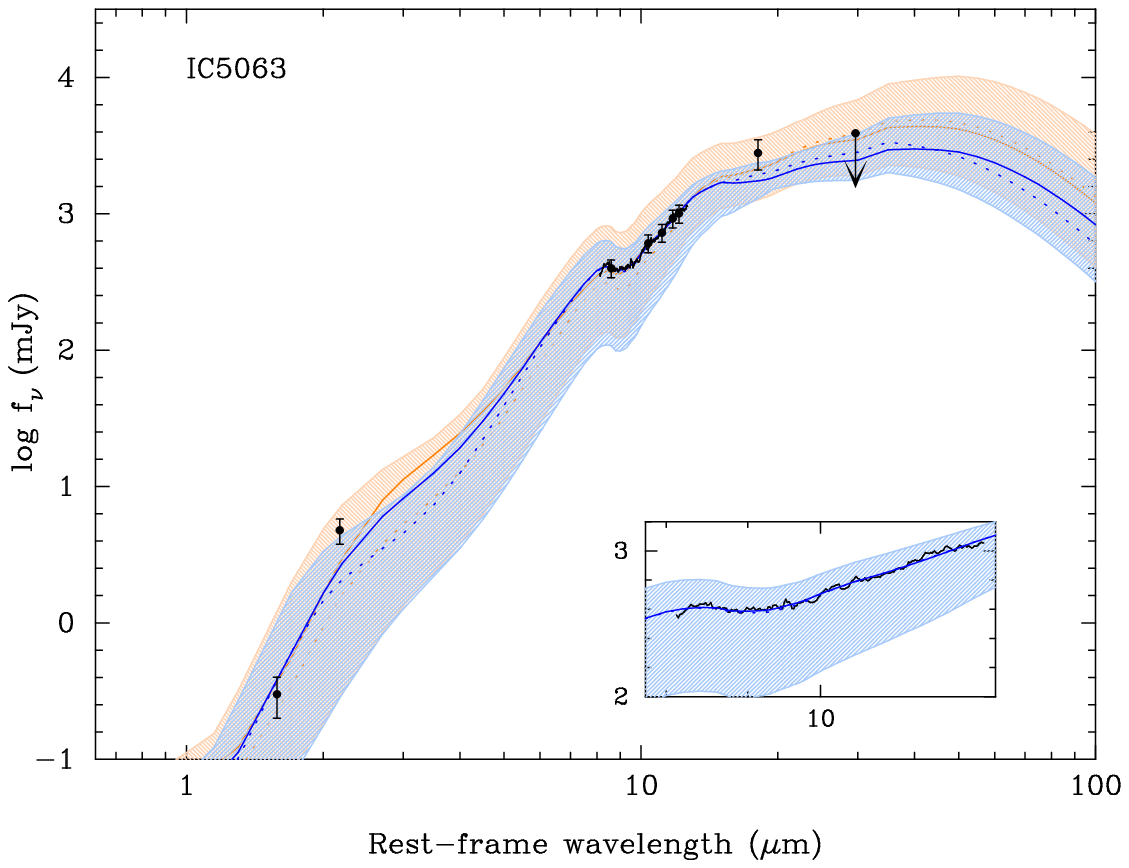}

\vspace{0.5cm}
\includegraphics[width=7.cm,angle=-90]{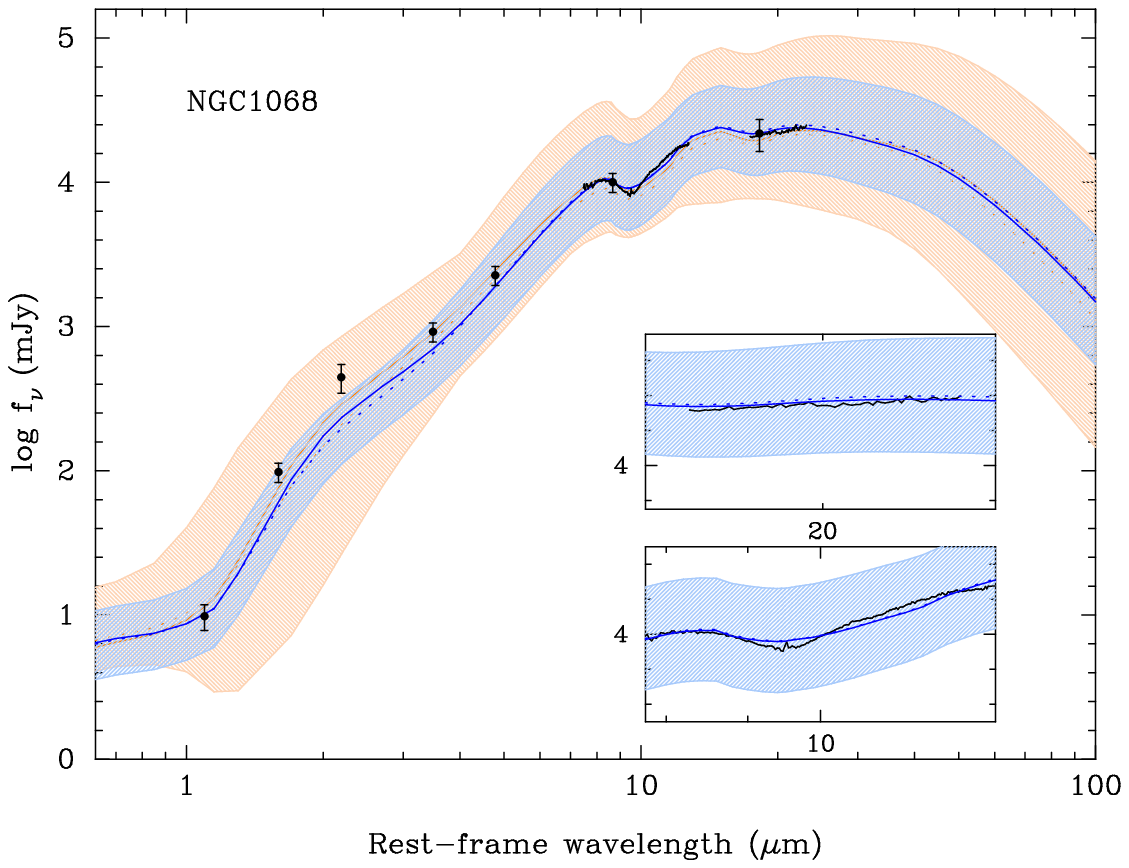}
\includegraphics[width=7.cm,angle=-90]{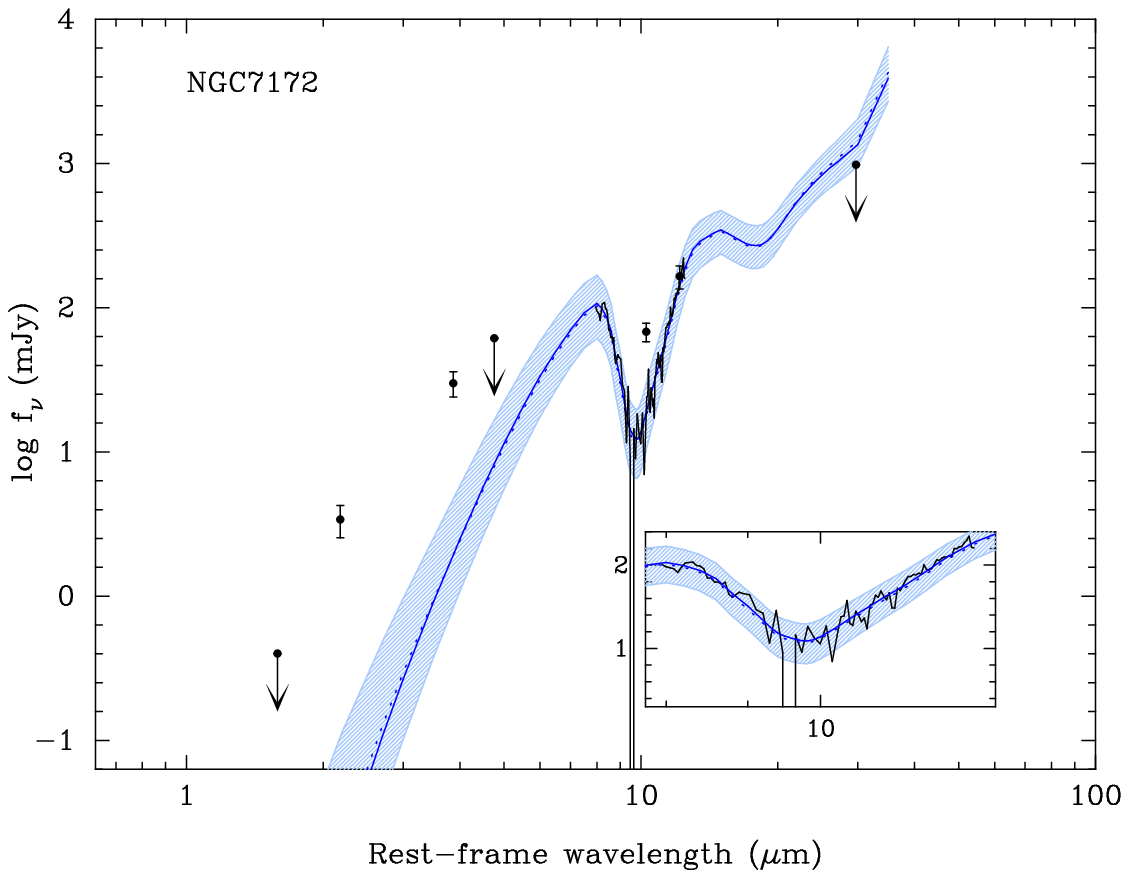}

\caption{Model fits for pure Seyfert 2 galaxies with 
 torus emission alone. Symbols and lines are as in Figure~4. {\it
   Upper left panel:} Circinus. 
We fixed the 
foreground extinction to $A_V({\rm frg})=9\,$mag (Maiolino et
al. 2000). We restricted the viewing angle to the range
$i=60-90\arcdeg$. {\it Upper right panel:} IC~5063. 
We did not restrict any of the torus  model parameters.  
The foreground extinction was fixed to A$_V{\rm (frg)}=7\,$mag based
on results by Heisler \& de Robertis (1999). {\it Lower left panel:}
NGC~1068. We restricted the viewing angle to the range
$i=60-90\arcdeg$ (see Section~3.3). For NGC~1068 the second inset
shows in detail the fit to the $Q$-band spectroscopy. 
{\it Lower right panel:}
NGC~7172. Only the SED+spectroscopy fit is shown due to the limited number of
photometric points in the near-IR. 
We did not restrict any of the torus model
  parameters. The foreground extinction was fixed to 
$A_V({\rm frg})=40\,$mag.
}
\end{figure*}

From a theoretical point of view, clumpy dusty models cannot produce
very deep silicate features ($S_{9.7}<-1)$ (see Nenkova et
al. 2008b; Elitzur 2008;  Sirocky et al. 2008), while observations
show that many Seyfert galaxies have relatively
deep  silicate features  (see Shi et al. 2006; Hao et
al. 2007; Deo et al. 2009). In our sample, this  includes 
Circinus, NGC~5506, NGC~7582, NGC~7172 (see Table~4 and
Figure~2). By analogy with those ULIRGs optically
classified as type-2 AGN and with deep silicate features modelled by 
Sirocky et al. (2008), the deep silicate
features of some Seyfert 2 galaxies in our sample 
may be explained by additional obscuration by cold foreground dust.

\setcounter{figure}{5}
\begin{figure*}
\includegraphics[width=7.cm,angle=-90]{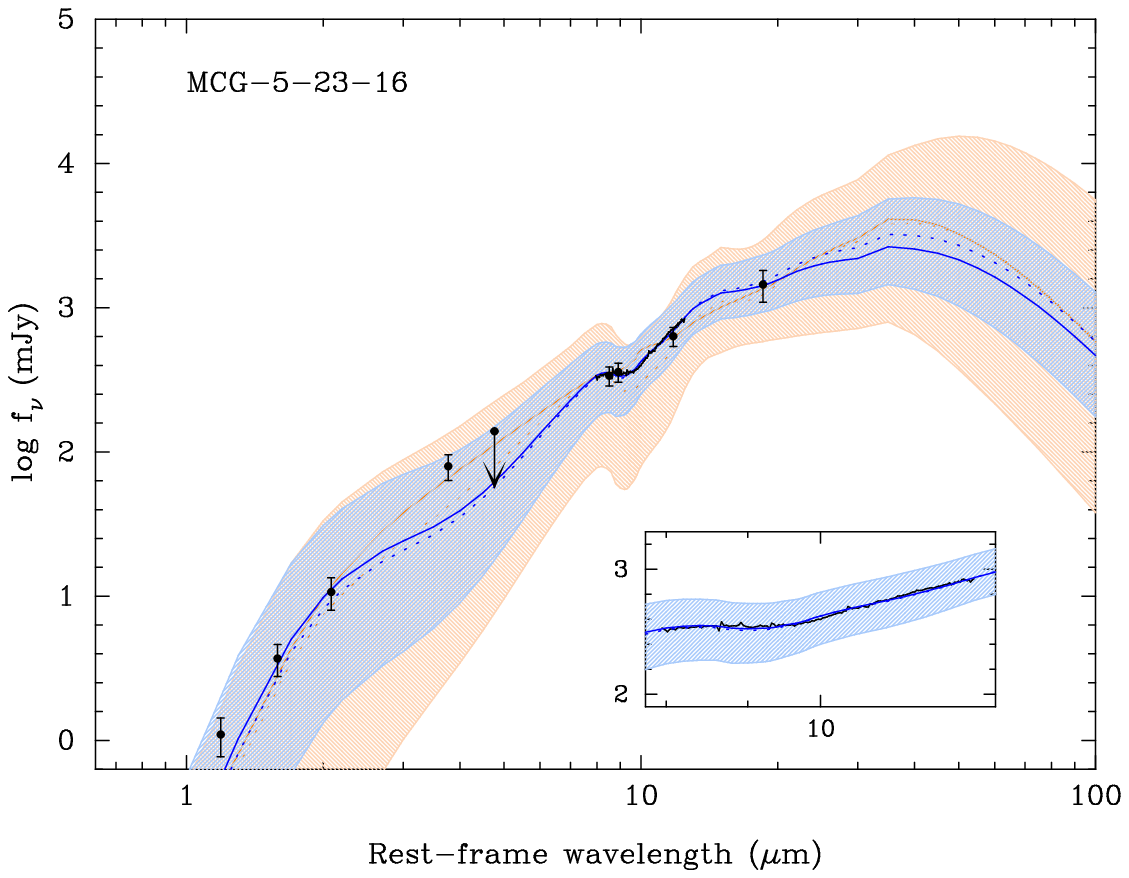}
\includegraphics[width=7.cm,angle=-90]{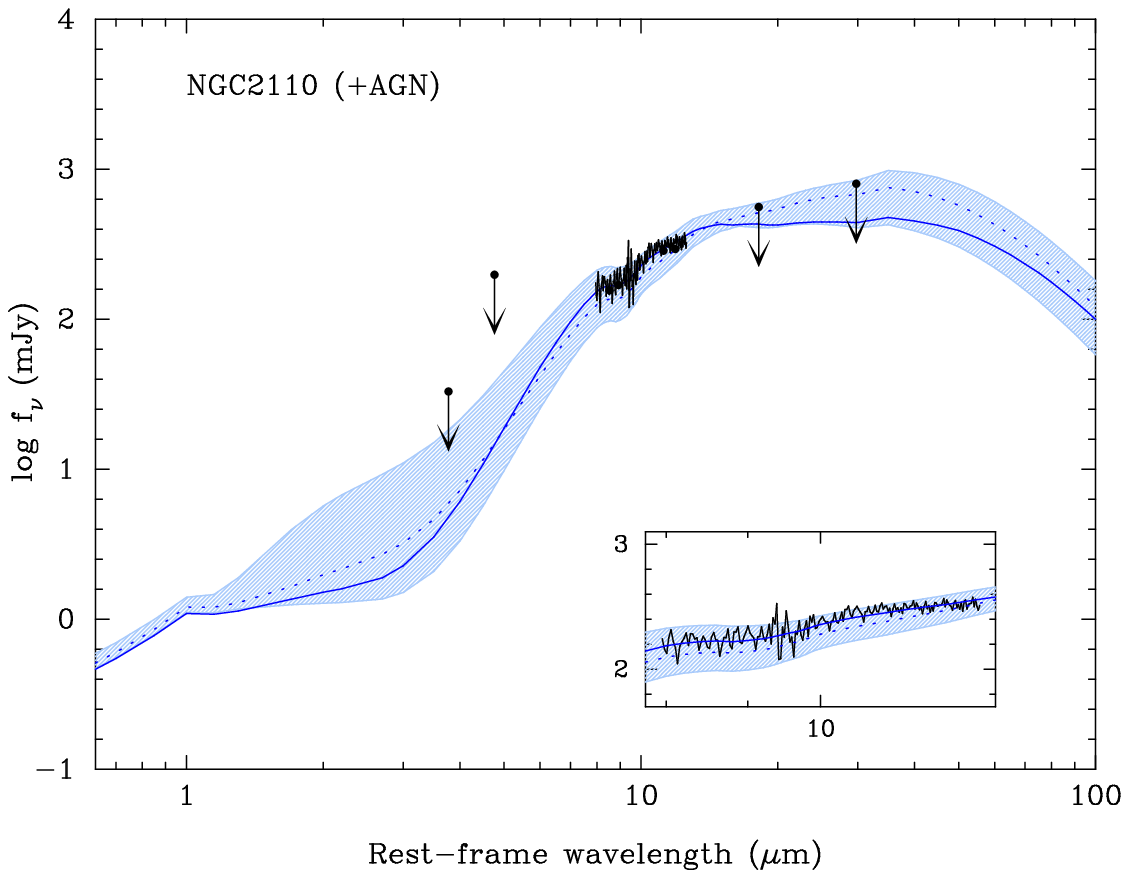}

\vspace{0.5cm}
\includegraphics[width=7.cm,angle=-90]{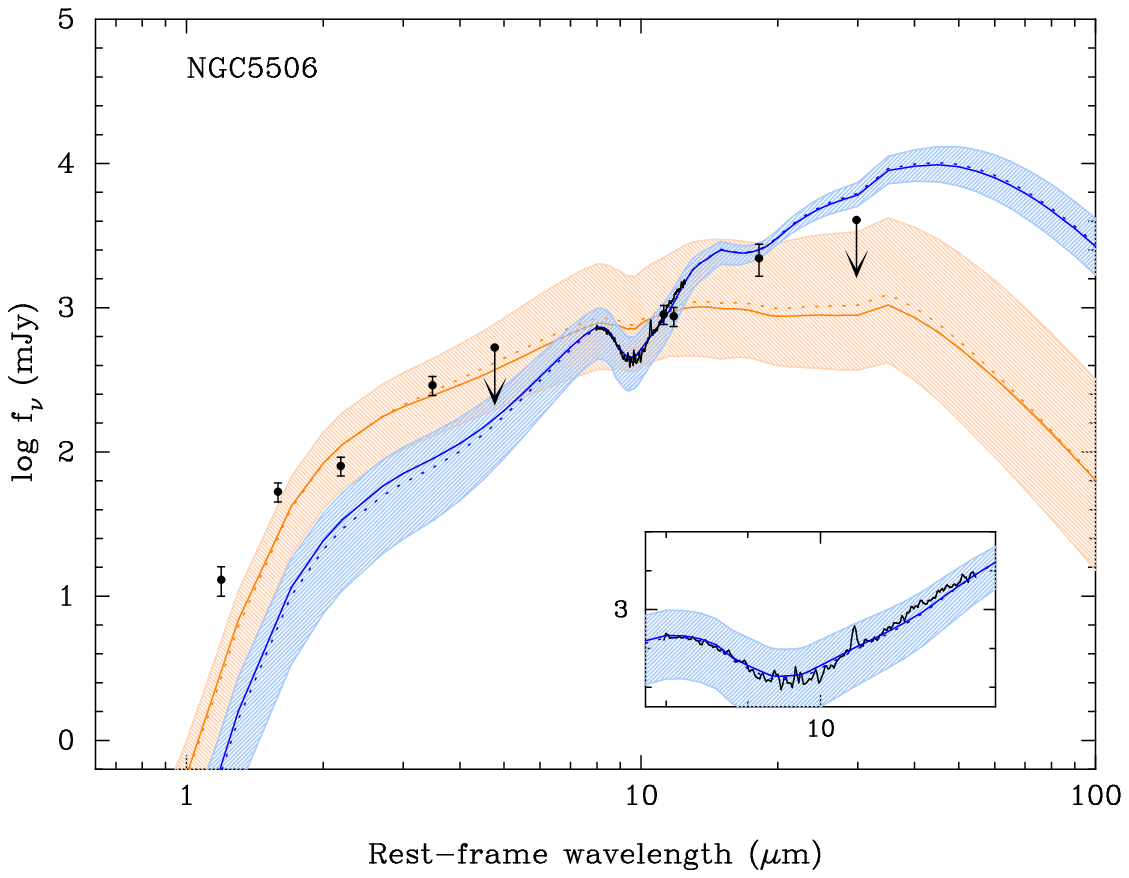}
\includegraphics[width=7.cm,angle=-90]{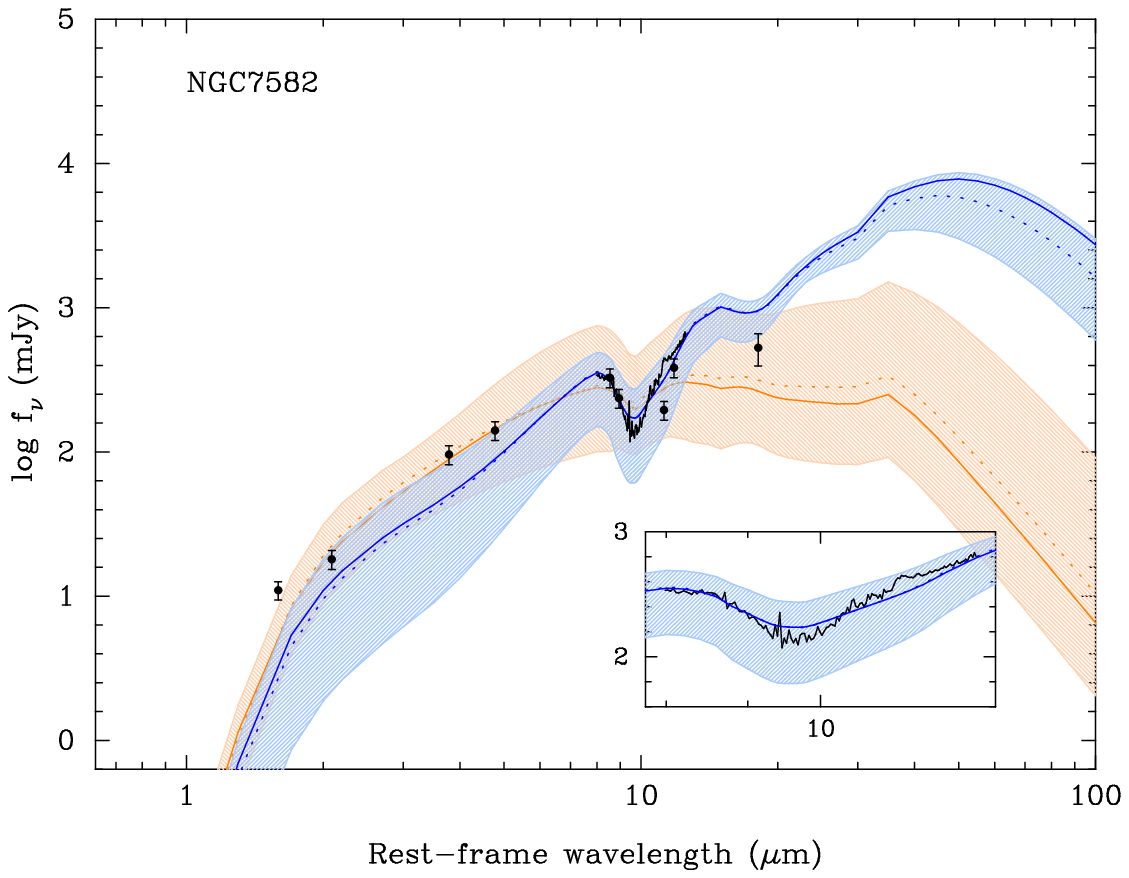}

\caption{(a) Model fits for Seyfert 2 galaxies with broad lines detected
  in the near-IR. {\it Upper left panel:} MCG~$-$5-23-16. Model fits
  are for torus emission alone. All symbols 
  and lines are as 
  in Figure~4. The viewing angle was restricted to a Gaussian distribution
  centered at $i=55\arcdeg$ \, with a 10\arcdeg \, width (see Section~3.3).  
The foreground extinction was fixed to A$_V{\rm (frg)}=7\,$mag based
on results by Veilleux et al. (1997).  {\it Upper right panel:}
NGC~2110. Model fits are for torus + AGN emission. 
The viewing angle was restricted to a Gaussian distribution
  centered at $i=40\arcdeg$ \, with a 10\arcdeg \, width (see
  Section~3.3).  
Only the SED+spectroscopy fit is shown due to the limited number of
photometric points in the near-IR.
The foreground extinction was fixed to 
$A_V({\rm frg})=5\,$mag (Storchi-Bergman et al. 1999).
{\it Lower left panel:} NGC~5506. Model fits are for only for torus
  emission (see Section~4.2). The viewing angle was restricted to a
  Gaussian distribution 
  centered at $i=40\arcdeg$ \, with a 10\arcdeg \, width (see
  Section~3.3). The foreground extinction was fixed to  
$A_V({\rm frg})=11\,$mag (Goodrich et al. 1994). {\it Lower right
  panel:} NGC~7582. Model fits are for 
torus emission alone. We did not restrict any of the torus
  model parameters. The foreground extinction was fixed to 
$A_V({\rm frg})=13\,$mag (Winge et al. 2000).}
\end{figure*}

\setcounter{figure}{5}
\begin{figure}
\includegraphics[width=7.cm,angle=-90]{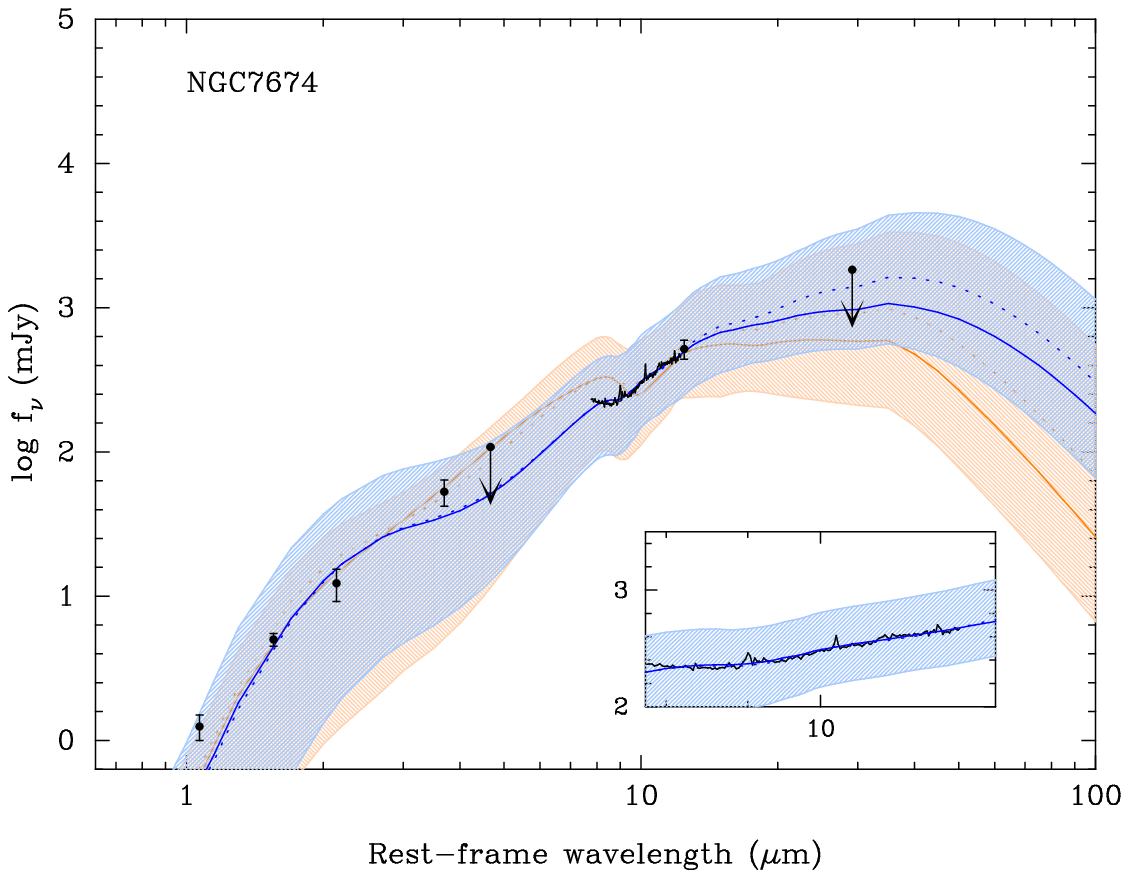}
\caption{(b) Continued. NGC~7674. Model fits are for 
torus emission alone. We did not restrict any of the torus
 model  parameters. The foreground extinction was fixed to 
$A_V({\rm frg})=5\,$mag.}
\end{figure}

\begin{figure*}
\hspace{2cm}
\includegraphics[width=10.cm,angle=-90]{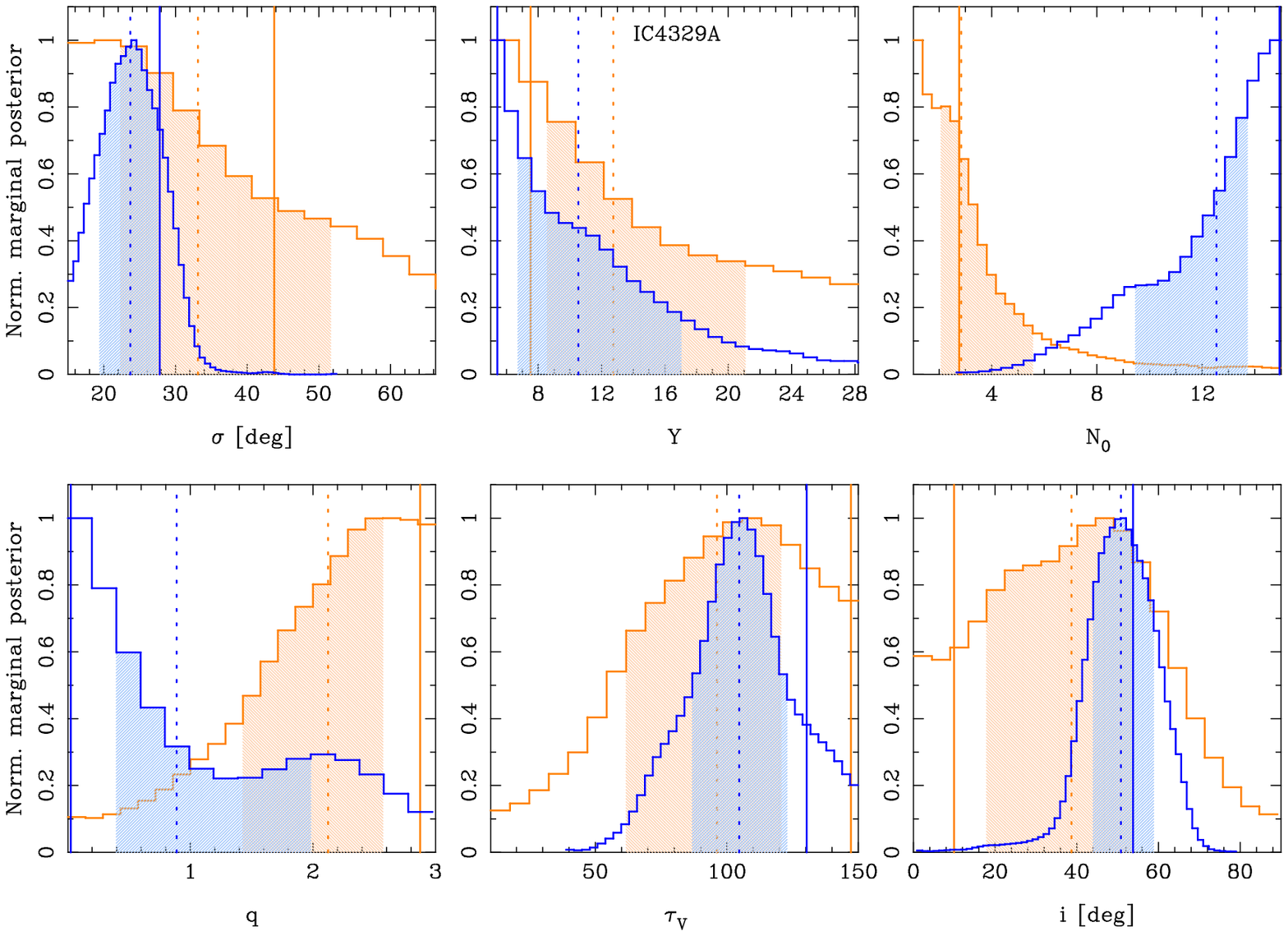}

\caption{Marginal posterior distributions of the free parameters that
  describe the   \textit{CLUMPY} models resulting from fitting
  the data for   IC~4329A  The
  SED alone fits are shown in orange and the SED+spectroscopy fits are
  shown in blue. The
  solid and dashed lines indicate the MAP and median values of the
  distributions, and the shaded areas are the $\pm 1\sigma$ values.
We did not
  use a foreground extinction. The marginal posterior distributions for the
  rest of the sample are shown in Appendix~A.}
\end{figure*}

For simplicity we will
     consider the extended dust component  as a purely foreground
     medium producing only absorption. We note however, 
 that high angular resolution mid-IR spectroscopic
     observations of the ionization cone dust of NGC~1068 show evidence of
     components of both absorption and emission. 
We use the IR extinction curve
     of Chiar \& Tielens (2006) of the local ISM in the wavelength
     range $\sim 1-35\,\mu$m, which includes the two silicate features
     at 9.7 and $18\,\mu$m.
For extinctions ($A_V ({\rm frg}) \gtrsim 5\,$mag) 
the effects of foreground extinction cannot
be ignored, especially in the spectral region around the $9.7\,\mu$m
silicate feature. 

Finally, it is important to note the possible degeneracy between AGN type and
the effects of foreground extinction. Using smooth density  torus
models  Alonso-Herrero et al. (2003) demonstrated that
the reddened near-IR SED
     from a nearly face-on (polar view) torus (underlying
     Seyfert 1) looks similar to the intrinsic
     near-IR SED at a viewing angle closer to a true Seyfert 2
     (equatorial view) given sufficient levels of foreground
     extinction ($A_V {\rm (frg)} \gtrsim 5\,$mag). Given this 
degeneracy in our fitting
     process we choose not
     to leave the foreground extinction as a free parameter 
     (see Section~3.3), but instead we use published values of the extinction as
     estimated from near-IR measurements. Table~4 gives the 
values of the foreground extinction used in this work, only for 
galaxies with evidence of extended dust components and 
with $A_V {\rm (frg)} \gtrsim 5\,$mag. The foreground extinction is
applied to the torus emission, as well as to the AGN emission that is included
for galaxies with a direct view of the BLR.

\subsection{Modeling Technique}

The \textit{CLUMPY} database now
  contains $1.2 \times 10^6$ models, which are calculated for
a fine grid of model parameters.
To fit the data we take a Bayesian approach to dealing with
the inherent degeneracy of the torus model
parameters (see RA09 and RA11).
 In this work we use an updated version of the BayesClumpy code 
developed by Asensio Ramos \& Ramos Almeida (2009). This tool uses a
Bayesian inference to allow as much information as possible to be
extracted from the observations.  
We refer the reader to Asensio Ramos \& Ramos
Almeida (2009) for details on the interpolation methods and algorithms
used by BayesClumpy.  However, it is important to note that in using a
Bayesian approach we do not make use of the original set of models 
of Nenkova et al. (2008a,b), but rather of an interpolated version of
them (see figures 3 and 4 in Asensio Ramos \& Ramos Almeida 2009).
The fineness of the grid of clumpy models makes interpolation an
appropriate methodology for our studies. 

The new version of BayesClumpy allows, in addition to fitting
photometric points, for the possibility of 
fitting spectra. In the case of photometric observations,
BayesClumpy simulates the transmission curve of the corresponding
filter on the model spectra. For the spectroscopic observations, the
full information, flux and associated uncertainty, is used.  When both
photometric SEDs and spectroscopy are fitted together, we use 
the data uncertainties discussed in Section~2 as weights for the fitting
routine. Under the assumption that the torus models
are valid simultaneously for photometric and spectroscopic data and
 that the noise in all
observed points is mutually uncorrelated, the
Bayesian approach is insensitive to the fact that there
are many more spectroscopic points than photometric ones. The reason is
that,
apart from the possible regularization that can be included in the prior
distributions,
we sample the full posterior distribution and obtain marginalized posterior
for each parameter. The marginalization procedure takes into account all
possible values of the parameters producing good fits. This avoids 
the possible
overfitting of the SED which is produced by neglecting the few photometric
points.

%{\rotate

\begin{table*}

\small
\caption{Fitted torus  model parameters from SED + spectroscopy data}
\begin{tabular}{lcccccccccccc}
\hline
\hline
Galaxy & \multicolumn{2}{c}{$\sigma_{\rm torus}$ [deg]} &
\multicolumn{2}{c}{$Y$} &
\multicolumn{2}{c}{$N_0$} &
\multicolumn{2}{c}{$q$} &
\multicolumn{2}{c}{$\tau_{\rm V}$} &
\multicolumn{2}{c}{$i$ [deg]} \\
        & Median & MAP& Median & MAP& Median & MAP& Median & MAP
& Median & MAP& Median & MAP\\
\hline
\multicolumn{13}{c}{Pure Type 1s}\\
\hline
IC~4329A
 & $  24^{+   4}_{  -4}$ &  28
 & $  11^{+   6}_{  -4}$ &   5
 & $  13^{+   2}_{  -3}$ &  15
 & $  0.9^{+  1.2}_{ -0.7}$ &  0.0
 & $ 105^{+  20}_{ -20}$ & 130
 & $  51^{+   8}_{  -8}$ &  54
 \\ 
NGC~3227
 & $  49^{+   9}_{  -6}$ &  44
 & $  17^{+   3}_{  -3}$ &  15
 & $  14^{+   1}_{  -2}$ &  15
 & $  0.2^{+  0.1}_{ -0.1}$ &  0.1
 & $ 146^{+   3}_{  -6}$ & 149
 & $  24^{+  11}_{ -15}$ &  30
  \\ 
NGC~4151
 & $  19^{+   4}_{  -2}$ &  16
 & $  10^{+   1}_{  -1}$ &   9
 & $  15^{+   0}_{-0}$ &  15
 & $  0.2^{+  0.3}_{ -0.1}$ &  0.0
 & $ 120^{+  15}_{ -14}$ & 116
 & $  63^{+   4}_{  -7}$ &  68
  \\ 
NGC~7469
 & $  21^{+   3}_{  -3}$ &  20
 & $  12^{+   2}_{  -1}$ &  11
 & $  15^{+   0}_{  -1}$ &  15
 & $  0.2^{+  0.3}_{ -0.2}$ &  0.0
 & $ 142^{+   5}_{  -9}$ & 148
 & $  58^{+   3}_{  -4}$ &  58
  \\ 
\hline
\multicolumn{13}{c}{Pure Type 2s}\\
\hline
Circinus
 & $  58^{+   7}_{ -11}$ &  45
 & $  15^{+   3}_{  -2}$ &  15
 & $   9^{+   4}_{  -1}$ &  15
 & $  0.7^{+  0.4}_{ -0.4}$ &  0.9
 & $  56^{+   4}_{  -4}$ &  54
 & $  66^{+   7}_{  -4}$ &  61
 \\ 
IC~5063
 & $  60^{+   5}_{  -7}$ &  47
 & $  13^{+   8}_{  -6}$ &   6
 & $  14^{+   1}_{  -1}$ &  15
 & $  2.6^{+  0.2}_{ -1.6}$ &  0.8
 & $ 130^{+   8}_{ -19}$ &  99
 & $  82^{+   5}_{  -9}$ &  84
  \\ 
NGC~1068
 & $  26^{+   6}_{  -4}$ &  21
 & $   6^{+   2}_{  -1}$ &   5
 & $  14^{+   1}_{  -3}$ &  15
 & $  2.2^{+  0.4}_{ -0.3}$ &  2.0
 & $  49^{+   4}_{  -3}$ &  49
 & $  88^{+   2}_{  -3}$ &  89
  \\ 
NGC~7172
 & $  61^{+   6}_{  -8}$ &  68
 & $  16^{+   5}_{  -3}$ &  17
 & $  13^{+   1}_{  -2}$ &  15
 & $  1.1^{+  0.4}_{ -0.5}$ &  1.5
 & $  59^{+   8}_{  -6}$ &  52
 & $  77^{+   8}_{ -14}$ &  85
  \\

\hline
\multicolumn{13}{c}{Type 2s with broad lines detected in near-IR}\\
\hline
MCG~$-$5-23-16
 & $  40^{+   8}_{  -5}$ &  35
 & $  17^{+   7}_{  -5}$ &  14
 & $  12^{+   2}_{  -4}$ &  15
 & $  2.0^{+  0.3}_{ -0.5}$ &  2.1
 & $ 135^{+   8}_{  -9}$ & 133
 & $  57^{+   8}_{  -8}$ &  59
 \\ 
NGC~2110 (+AGN)
 & $  64^{+   4}_{ -10}$ &  70
 & $  17^{+   8}_{ -11}$ &   5
 & $  10^{+   2}_{  -2}$ &  12
 & $  2.7^{+  0.2}_{ -0.2}$ &  2.7
 & $ 147^{+   2}_{  -4}$ & 150
 & $  43^{+   8}_{  -8}$ &  37
  \\ 
NGC~5506
 & $  43^{+   3}_{  -3}$ &  40
 & $  15^{+   2}_{  -2}$ &  15
 & $  14^{+   0}_{  -1}$ &  15
 & $  0.4^{+  0.2}_{ -0.2}$ &  0.3
 & $ 100^{+   6}_{  -6}$ &  99
 & $  34^{+   6}_{  -6}$ &  35
\\
NGC~7582
 & $  48^{+   6}_{  -6}$ &  49
 & $  22^{+   4}_{  -4}$ &  25
 & $  13^{+   1}_{  -3}$ &  15
 & $  0.3^{+  0.2}_{ -0.2}$ &  0.1
 & $  89^{+   9}_{ -11}$ &  97
 & $  12^{+  17}_{  -8}$ &   0
  \\ 
NGC~7674
 & $  28^{+   8}_{  -7}$ &  24
 & $  14^{+   6}_{  -4}$ &  12
 & $  11^{+   2}_{  -3}$ &  15
 & $  1.6^{+  0.5}_{ -0.8}$ &  2.2
 & $ 137^{+   8}_{ -11}$ & 148
 & $  63^{+   9}_{ -10}$ &  69
  \\ 

\hline
\end{tabular}
Notes.--- Torus  model parameters are listed for the median and
$\pm 1\sigma$  values around the median, and the MAP values. 

\end{table*}
%} 

The prior distributions for the model parameters are 
assumed to be truncated uniform distributions in the ranges given 
in Table~3. We note that in the most up-to-date version of the 
  \textit{CLUMPY} models
after correcting for the erroneous AGN scaling factor (Nenkova et
al. 2010), the optical depth of the individual clouds only goes up to
150, instead of $\tau_V=200$ of the older models. The only prior
  information we use in this work is the viewing angle. For those  
galaxies in our sample with H$_2$O maser detections: Circinus (Greenhill et
al. 2003) and NGC~1068 (Greenhill et al. 1996) we
restricted the viewing angles 
$i$ to values in the range 
$60-90\arcdeg$, that is, close to equatorial views through the torus.
We can also use the accretion
disk viewing angles deduced from X-ray observations of the Fe\,K
$\alpha$ line as an additional constraint, if we  
assume that the accretion disk and the torus are coplanar. 
We found estimates for three galaxies in our sample,
for MCG~$-$5-23-16 is $i\sim 53\,\arcdeg$ (Reeves et
al. 2007),  and for NGC~5506 and NGC~2110 is $i\sim 40\,\arcdeg$ (Guainazzi et
al. 2010 and Weaver \& Reynolds 1998). For these three galaxies we
assumed Gaussian distributions for this parameter with a width of 10\,\arcdeg.

In addition to the the six torus  model parameters, there are two 
extra parameters that can be fitted or fixed. The first parameter 
accounts for the vertical displacement
needed to match the fluxes of a given model to an observed  
SED/spectrum. This vertical shift, which we allow to vary freely,  scales with
the AGN bolometric luminosity (see Nenkova et al. 2008b) and will be
discussed in Sections~4.1 and 4.2. The second parameter is the
foreground extinction (see Section~3.2 and Table~4) due to the
host galaxy, which is different 
from that produced by the torus along the LOS (see RA09 and RA11).

For the modelling of the SED and spectroscopy of type 1 Seyferts 
the AGN contribution needs to be added to the torus emission (see
Nenkova et al. 2008a for the assumed shape).  In principle,  
the same should be done for those Seyfert 2s in our sample with broad
lines detected in the 
near-IR, as this means we have a direct view of the BLR. 
However most of these Seyfert 2s in our sample are also heavily affected by
extinction (see Table 4).  Kishimoto 
et al. (2007) demonstrated for  type 1 Seyferts that most of the
unresolved emission at $2.2\,\mu$m is produced by hot dust emission
from the inner walls of the torus, and that there is a very small 
contribution from the big blue bump emission (i.e., AGN
emission) at this wavelength. Then there is the question for AGN with 
broad lines and foreground extinctions $A_V \gtrsim 5\,$mag of whether
we are seeing AGN emission in the range $1-2\,\mu$m or not. We thus
decided for these galaxies to fit the data both including and not
including the AGN emission. This will be discussed in more detail in
Section~4.2.

\section{Results from fits to SED+spectroscopy}

The result of the fitting process are the marginal posterior
distributions for the six free parameters that describe the
\textit{CLUMPY} models plus the vertical shift. These are the
probability distributions 
of each parameter, which are represented as histograms. As explained
in Section~3.1, except for the viewing angle of five galaxies, for the rest of
the parameters we use uniform priors. If the observational data
contain sufficient information for the fit, then the resulting
probability distributions of the fitted torus  model parameters 
will clearly differ from uniform 
distributions. In those cases the probabilities either show trends or
are centered at certain values within the considered intervals. 

For each galaxy 
we translate the fitted torus  model 
parameters into two model spectra. The first one
corresponds to   the maximum-a-posteriori (MAP) values 
 that represent the “best fit” to the data. The
second is the model produced with the median value of the probability
distribution of 
each parameter, which is characteristic of the observed
SED+spectroscopy data. Figures~4 to 6 show for Seyfert 1s, Seyfert 2s,
and Seyfert 2s with broad lines in the near-IR, respectively, 
these fits to the 
SED+spectroscopy. Figure~7 shows the marginal posterior
  distributions of the 
six torus model parameters for IC~4329A. The marginal posterior
distributions for the 
rest of the sample are shown in Appendix~A. In Appendix~B we also show
for Circinus, NGC~1068, NGC~4151, and IC~4329A 
the two-dimensional posterior distributions
for all  combinations of the  torus model parameters. This kind of
two-dimensional distributions can be used to check for possible correlations
and degeneracies between different torus model parameters for a given galaxy. 
Table~5 summarizes the statistics for the fitted torus  model
parameters. In Sections~4.1 to 4.3 we discuss new
constraints on the torus  model
parameters when fitting the SED+spectroscopy
data together, and in Section~4.4 we assess the improvements obtained by adding
mid-IR spectroscopy to the SED data.

\subsection{Fits to pure Seyfert 1s and Seyfert 2s}

In this section we discuss the fits to the {\it pure} Seyfert 1s and
2s\footnote{In this work pure type 1 AGN have 
  broad lines detected in the optical (i.e., include types 1.5, 1.8
  and 1.9), while pure type 2s do not have broad lines detected either
  in the optical or in the near-IR.},
while the fits to those Seyfert 2s with broad lines detected in the near-IR
(including NGC~5506) are
discussed separately in Section~4.2. 
As can be seen from Figures~4 and 5 (in blue), the \textit{CLUMPY}
torus models provide
very good simultaneous fits to the photometric SED and spectroscopy
data of Seyfert 1s and 2s, in particular for those galaxies with low 
host galaxy foreground extinction.  

It is worth noting that all the
Seyfert 1s except IC~4329A show a  slight excess of emission in the
near-IR, above the  median torus+AGN model fits, which might
be attributed to hot dust.
 Mor et al. (2009) included, apart from the torus emission,
 two extra components, hot dust and NLR
  emission, to fit the {\it
  Spitzer}/IRS spectra of PG quasars. These extra components provided the 
additional flux need in the near-IR for their sample. It is not clear
however if such components are needed in our fits, because the
unresolved emission we used for our fits probes typically a few tens of
parsecs, while the Mor et al. (2009) data cover physical sizes on
  scales of a few kpc. Finally, the near-IR photometric
  points of our Seyfert 1s 
are well within the $\pm 1\sigma$ confidence regions of the
  fitted models.

The only galaxy for which we could not get a good fit to both
the SED photometric points and the mid-IR spectroscopy was the Seyfert
2 galaxy NGC~7172. 
This galaxy is highly inclined (see Table~4), has
prominent dust lanes, and thus probably suffers from high extinction in the
host galaxy (see e.g. Roche et al. 2007). Since we could not find
estimates for the host galaxy extinction, we used the foreground
extinction derived from the X-ray column density and the standard
Galactic ratio  
$N_{\rm H}/A_V = 1.9\,10^{21}\,{\rm cm}^{-2}\, {\rm mag}^{-1}$
(Bohlin, Savage, \& Drake 1978).  As can be seen from Figure~5,
while the model provides a  reasonable fit to  
the $9.7\,\mu$m silicate feature, the
near-IR photometric points are well above the model. One possibility is that
there is contamination by extended dust components  in this galaxy, 
and the unresolved flux estimates from  the $K$ and $L$-band 
ground-based data are upper
limits. However, the most likely explanation is that the clumpy torus
emission + cold foreground dust screen are not the appropriate model
for this galaxy. Rather, a spherically symmetric smooth model may be more
appropriate for deeply embedded objects, such as NGC~7172 (see
Levenson et al. 2007).

As can be seen from the marginal posterior distributions in Figure~7 and
Figures~A1 to A7 (blue lines and blue shaded regions), 
the majority of the fitted torus  model
parameters of the {\it pure} Seyfert 1s and Seyfert 2s are well
constrained. In  particular, the width of the 
angular distribution $\sigma_{\rm torus}$, the radial torus thickness
$Y$, and the 
viewing angle $i$, which were also relatively well constrained from
fits to the SED alone (RA09, RA11, and Section~4.4), present narrow
probability distributions. As found by RA11 there is no clear relation
between the derived viewing angle $i$ and the classification of the galaxy
into a type 1 or a type 2. In other words,  not all type 1s are viewed
at relatively low inclination angles and not all type 2 are seen at
directions closer to the equatorial plane of the torus.  However,
  as we shall see in Section~5.2, the relevant quantity for a galaxy
  to be classified as a type 1 or a type 2 is the probability for an
  AGN produced photon to escape unabsorbed or not.

As can be seen from Table~5, our results confirm that 
  \textit{CLUMPY} models of tori with
relatively small radial thicknesses, produce good fits with no need
for very large tori. The fitted values are within the assumed range of
$Y=5-30$,  and the data for most galaxies can be fitted with tori with
radial thickness $Y\sim 10-15$.  In Section~5.1
we will compare in more detail 
our derived  physical sizes of the torus (in parsecs)  
with those derived from the modeling of mid-IR
interferometry.

\setcounter{figure}{7}
\begin{figure}
\vspace{0.5cm}
\includegraphics[width=8.5cm,angle=-90]{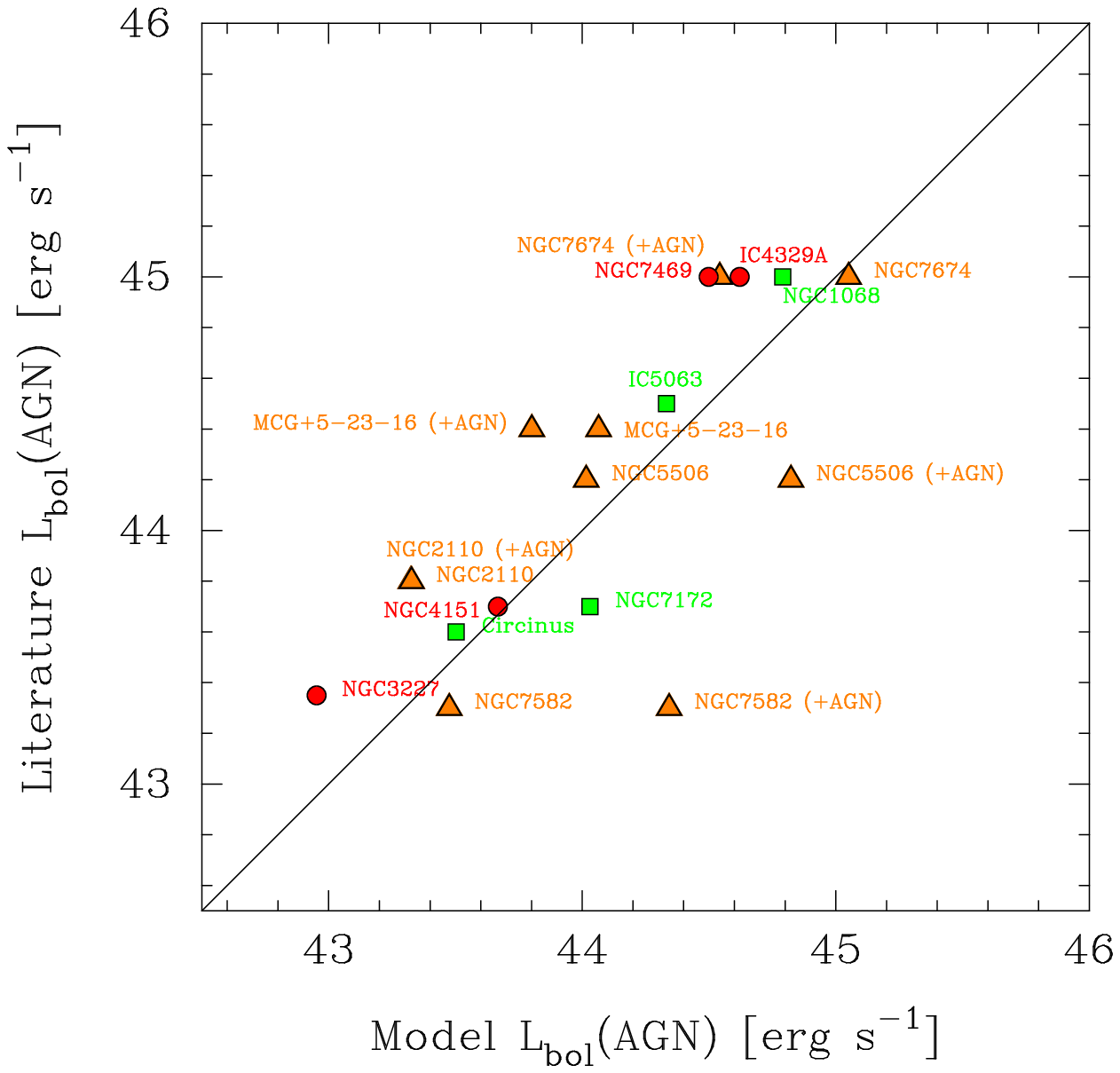}
\caption{Comparison between the AGN bolometric luminosities (median
  values) as derived
  from the SED+spectroscopy fitting and those from the literature. The
  solid line is the 1:1 relation,  not a fit. 
Filled symbols as in Figure~2. 
For those galaxies with broad lines and high extinctions we show the
inferred AGN bolometric luminosities from fits with  and without the 
AGN component. The former are indicated as ``(+AGN)''. Note that the
derived bolometric luminosities for NGC~2110 with and without the AGN
component are the same.}
\vspace{0.5cm}
\end{figure}

The index of the power law distribution $q$ 
controls the placement of clouds between the inner, hotter parts of
the torus and the outer cooler regions (Nenkova et al. 2008b), and
thus plays a role in the shape of the IR SED (see also H\"onig \&
Kishimoto 2010). The effects of changing $q$ on the
  \textit{CLUMPY} models  are easier to see in type
1 objects (low values of $i$). 
For steep radial distributions (clouds concentrated
near the inner radius of the torus) the  near and mid-IR  
SEDs become redder (see figure
9 of Nenkova et al. 2008b), although including the AGN contribution
always makes the model SEDs flatter. We find that the fitted values of
$q$ for our Seyfert 1s are relatively small. However the reverse for
Seyfert 2s  (seen at high inclinations) 
is not necessarily true. We also note that the sensitivity of the SED
  to $q$ for very 
small values of $Y$ is highly reduced. This is because for such 
small tori the SED does not change much whether the
clouds are distributed along the whole extent of the torus ($q=0$)
or highly concentrated in its inner part (large $q$). This lack
of sensitivity could result in posterior distributions that depend
on the quality of the interpolation technique used by BayesClumpy.
Consequently, whenever the inferred $Y$ is very small (e.g.,
NGC~1068),  the inferred value of $q$ should be treated with caution.

For a larger sample and using fits
of the SEDs alone RA11 found statistically significant differences
of $q$ between type 1s and type 2s. RA11 found that the SEDs 
of type 1s were fitted
with small values of $q$, while those of type 2s required larger 
 $q$. H\"onig et al. (2010), on the other hand, found the 
$q$ parameter ($\left | a \right |$, in their notation) 
to be in the range $0-1.5$ for both type 1s and 2s.
The other torus model
parameters, $N_0$ and $\tau_V$, derived from the SED+spectroscopy data
 will be
discussed in the context of the fits of the spectral region around the
$9.7\,\mu$m silicate
feature (Section~4.3). For a detailed discussion of the fits to
  individual sources and comparison with clumpy torus models fits in
  the literature, we refer the reader to Appendix~B.

As explained in Section~3.3, the shift applied to scale the 
  \textit{CLUMPY} models
to the data is directly related to the bolometric luminosity of the
AGN. In this work we chose to leave the shift as a free parameter, but other
works (e.g., Mor et al. 2009) used it as a constraint for the fits. 
Figure~8 shows a comparison between the fitted AGN bolometric
luminosities  ($L_{\rm bol}^{\rm model}({\rm AGN})$)
and AGN bolometric luminosities from the literature. The latter were
computed with different methods, including applying bolometric
corrections and modelling of the SEDs (see Table~1 and
Section~2.1). The typical uncertainties of the AGN bolometric
luminosities from scaling the \textit{CLUMPY} 
models are $0.1-0.2$dex. It is clear from this 
figure that the agreement between the bolometric luminosities 
is good for the majority of {\it pure} Seyfert 1s
and Seyfert 2s, with most of our 
estimates within $\sim 0.4$ dex of the 1:1
  relation.  
The only exception is the Seyfert 1 
NGC~7469. The fitted AGN bolometric luminosity
for this galaxy is  below two 
independent literature estimates (e.g., Woo \&
Urry; Vasudevan et
al. 2010). There is strong nuclear ($\sim 0.2\arcsec \sim 65\,$pc)
star formation in this 
galaxy, but it only contributes a small fraction of the $K$-band
luminosity within 0.2\arcsec \, (see e.g., Davies et al. 2007 and
references therein). However, 
the ground-based mid-IR nuclear spectrum of this galaxy 
(H\"onig et al. 2010) show faint $11.3\,\mu$m PAH emission,
and the PAH emission becomes very prominent in the circumnuclear
regions (Roche et al. 1991). It is then likely that
the mid-IR nuclear  fluxes and spectroscopy  of NGC~7469 
contain a contribution from star
formation.  A similar situation may be the case for NGC~3227.

\subsection{Fits to Seyfert 2s with near-IR broad lines}
Four galaxies in our sample are classified as Seyfert 2s, but there
are also reports in the literature of detections of  
broad lines  in the near-IR (see Table~1 for references).
Additionally, NGC~5506 has different spectral
classifications in the literature and there is some controversy about
whether it 
has broad components of the near-IR
emission lines (Blanco et al. 1990; Ruiz et al. 1994: Goodrich et
al. 1994; Veilleux et al. 1997). Nagar et al. (2002) clearly detected
BLR emission in the 
near-IR and classified NGC~5506 as a narrow line Seyfert 1
(NLSy1). Given these facts we discuss NGC~5506 in this section.

As the near-IR data suggest, for this kind of galaxies
we may have a direct view of the BLR, and thus in
principle we should include the AGN component 
when modelling the data. However, some of these galaxies also tend to 
suffer from relatively high values
of foreground extinction (Table~4). For these Seyferts we 
performed the fits with and without the AGN components. In this
section we use other observational properties of these galaxies 
to determine whether the AGN component should be included or not for
fitting their SED+spectroscopy data.

\begin{table*}
\caption{AGN and torus model properties derived from the fits.}
\begin{center}
\begin{tabular}{lccccc}
\hline
\hline
Galaxy & $L_{\rm bol}^{\rm model}$(AGN)  &
$r_{\rm torus}$  &
$P_{\rm esc}$  &
\multicolumn{2}{c}{Covering factors}\\
 &  [erg s$^{-1}$] & [pc] & [\%] & $f(i)$ & $f_2$\\
\hline
\multicolumn{6}{c}{Pure Type 1s}\\
\hline
IC~4329A   
 &  44.6 &   2.7& $41.8^{+43.8}_{-27.3}$ & $0.80^{+  0.22}_{-0.22}$
 &$  0.24^{+  0.11}_{-0.08}$\\ 
NGC~3227   
 &  43.0 &   0.7& $11.6^{+ 8.8}_{-4.9}$ &$  1.36^{+  0.06}_{ -0.10}$
 &$  0.86^{+  0.07}_{-0.12}$ \\
NGC~4151   
 &  43.7 &   0.9& $15.7^{+13.1}_{-8.2}$ &$  0.46^{+  0.22}_{-0.10}$
 &$  0.16^{+  0.08}_{-0.04}$ \\
NGC~7469   
 &  44.5 &   2.6& $27.8^{+17.3}_{-11.5}$ &$  0.61^{+  0.14}_{-0.11}$
 &$  0.20^{+  0.06}_{-0.05}$\\
\hline
\multicolumn{6}{c}{Pure Type 2s}\\
\hline
Circinus  
 &  43.5 &   1.0& $ 0.07^{+ 0.06}_{-0.06}$ &$  0.85^{+  0.05}_{ -0.09}$ &$  0.89^{+  0.04}_{-0.10}$\\
IC~5063    
 &  44.3 &   2.4& $ 0.0002^{+ 0.0005}_{-0.0001}$ &$  0.85^{+  0.04}_{-0.10}$ &$  0.96^{+  0.02}_{-0.07}$\\
NGC~1068   
 &  44.8 &   2.0& $ 0.0001^{+ 0.0010}_{-0.0001}$ &$  0.32^{+  0.11}_{ -0.07}$ &$  0.30^{+  0.11}_{-0.08}$\\
NGC~7172   
 &  44.0 &   2.0& $ 0.001^{+ 0.01}_{-0.001}$ &$  0.85^{+  0.10}_{-0.13}$ &$  0.93^{+  0.05}_{-0.11}$\\
\hline
\multicolumn{6}{c}{Type 2s with near-IR broad lines}\\
\hline
MCG$-$5-23-16     
 &  44.1 &   2.3& $ 0.4^{+ 1.3}_{-0.3}$ &$  0.80^{+  0.19}_{-0.19}$
 &$  0.63^{+  0.11}_{-0.14}$\\ 
NGC~2110 (+AGN)
 &  43.3 &   1.0& $ 0.4^{+ 0.8}_{-0.3}$ &$  1.00^{+  0.06}_{-0.06}$
 &$  0.94^{+  0.04}_{-0.11}$ \\
NGC~5506   
 &  44.0 &   1.9& $ 8.6^{+ 5.1}_{-3.7}$ &$  1.30^{+  0.08}_{-0.10}$
 &$  0.72^{+  0.07}_{-0.07}$ \\
NGC~7582   
 &  43.5 &   1.5& $36.8^{+15.1}_{-15.8}$ &$  1.62^{+  0.11}_{-0.20}$
 &$  0.83^{+  0.06}_{-0.14}$\\
NGC~7674   
 &  45.1 &   6.1& $ 1.8^{+ 4.3}_{-1.5}$ &$  0.61^{+  0.25}_{-0.24}$
 &$  0.36^{+  0.15}_{-0.14}$ \\
\hline
\end{tabular}

Notes.--- $L_{\rm bol}^{\rm model}$(AGN) and $r_{\rm torus}$ were 
derived from the median values of the marginal posterior
distributions. 
\end{center}

\end{table*}
\smallskip

In Section~4.1 we showed that the model fits for {\it pure} 
Seyfert 1s and 2s provided
good constraints to the AGN bolometric luminosity. We can then use the 
comparison between the fitted AGN luminosity and those taken from
the literature to determine if we need to add the AGN
component. Figure~8 
suggests that for NGC~5506, NGC~7582, NGC~7674, and
MCG~$-$5-23-16 we do not need to add the AGN component. 
For NGC~2110 we obtained comparable bolometric luminosities (and other
torus  model parameters) with and
without the AGN component. This is probably due to the lack of
photometric points in the near-IR for this galaxy. 
Finally we note that when we  included the AGN component to fit the
SED + spectroscopy data of NGC~5506 we could not
  fit the data with the viewing angles inferred from X-ray data
  (Section~3.2). This is 
 in agreement with the luminosity comparison.

The fits of the Seyfert 2s with broad lines detected in the near-IR are shown
in Figure~6, while the marginal posterior distributions are
  displayed in  Appendix~A
(Figures~A8 to A12). 
For the three galaxies with moderate silicate features
(MCG~$-$5-23-16, NGC~2110, and NGC~7674, see Table~4) we are able to reproduce
reasonably well the photometric SED points and the mid-IR spectroscopy,
and the torus  model parameters are mostly well constrained. As 
for the pure Seyfert 2 NGC~7172, the two galaxies with relatively deep
silicate features in absorption (NGC~5506 and NGC~7582), it is 
possible to do a 
simultaneous fit of the photometry and mid-IR spectroscopy, although
it is not a good fit. 
Imanishi (2000) measured an unusual ratio between the depths of the
$3.4\,\mu$m carbonaceous dust absorption and the $9.7\,\mu$m silicate
dust in NGC~5506. He suggested that the 
obscuration towards the nucleus of NGC~5506 could be ascribed mostly
to dust in this host galaxy ($>$100-pc scale) and not to the obscuring
torus. This may also reflect differences in grain populations (see
Roche et al. 2007). The X-ray emission of NGC~7582 indicates the presence
of at least two absorbers of columns densities of $\sim 10^{24}\,{\rm cm}^{-2}$ and
$4-5\times10^{22}\,{\rm cm}^{-2}$, with the latter probably associated with
large scale obscuration (Bianchi et al. 2009).

\subsection{Fits to the silicate feature}

The insets of the upper panels of Figures~4 to 6 show in detail the
fits in the spectral region around 
the $9.7\,\mu$m silicate feature. For NGC~1068 we also show in
  Figure~5 the fit to the $Q$-band spectrum. It is clear that the
interpolated version of the Nenkova
et al. (2008a,b) clumpy torus models produce excellent fits to the
$N$-band silicate
feature, and  the $\sim 20\,\mu$m spectrum of NGC~1068. 
In particular for galaxies with low host galaxy 
extinctions and moderate silicate strengths ($S_{9.7} >-1$), 
the fits to the silicate feature spectral region 
are also compatible with
those to the SED photometric points. This suggests that the data mostly
correspond to the torus emission. These galaxies are all the pure Seyfert 1s
(IC~4329A, NGC~3227, NGC~4151 and NGC~7469), and NGC~1068, IC~5063,
MCG~$-5$-23-16, and NGC~7674.

We now discuss the fits of those galaxies with deep silicate
features. For the Circinus galaxy,
which has a relatively deep silicate feature ($S_{9.7}=-1.8/-2.4$,
Roche et al. 2007 and Table~4),  \textit{CLUMPY} 
torus models with the parameters given
in Table~5 plus  a foreground absorbing screen with $A_V({\rm
  frg})=9\,$mag reproduce reasonably well  the silicate feature
and the photometric SED, 
except for the $18.3\,\mu$m flux. For the other galaxies
 (NGC~5506, NGC~7172 and
NGC~7582), although the fits to the feature are qualitatively good,
the models do not reproduce at the same time all the photometric
points. One possibility for  these three galaxies is that there is an
important contamination from extended dust structures, while in
Circinus, the closest galaxy in our sample ($d=4\,$Mpc) this
contamination is minimized. 

Nenkova et al. (2008b) emphasized that the clumpy torus distributions
produce more elaborate patterns of the $9.7\,\mu$m and the $18\,\mu$m
silicate features (see also H\"onig \& Kishimoto 2010), while for
smooth density models the silicate feature is always in emission for
face-on views and in absorption for edge-on views. Besides, the 
  \textit{CLUMPY} torus models 
never produce very deep silicate features, in contrast with
smooth density models (e.g., Pier \& Krolik 1993; Efstathiou \&
Rowan-Robinson 1995). When fitted together with the
photometric SEDs we  expect the silicate
feature to be mostly sensitive to the optical depth of the clouds
$\tau_V$, and the average number of clouds along radial equatorial rays $N_0$. 
As an illustration of the complicated behavior of the silicate
feature in terms of the torus model parameters, we can see that 
galaxies of the same type and with similar apparent depths of the
$9.7\,\mu$m silicate feature  
(e.g., type 1s IC~4329A and NGC~7469, and type 2s  MCG$-5$-23-16 and
IC~5063, see H\"onig et al. 2010) 
have different fitted values of $N_0$ and $\tau_V$ (see Table~5, and
Appendix~A). 

A general result for our sample of Seyfert
galaxies is that the average number of
clouds along radial equatorial rays is  never very low,  and is in the
range $N_0=8-15$. This is in good agreement with
the  results of Nenkova et al. (2008b). However,  H\"onig et al. (2010)
inferred fewer clouds along equatorial rays
from their fits to the mid-IR spectroscopy of 
Seyfert galaxies, but we note these authors fixed the value of
$\sigma_{\rm torus}$, among other parameters (see Appendix B for
  a more detailed discussion). We also note here, that the values of the
optical depth of the clouds $\tau_V$ in the corrected version of
the models (see Nenkova et al. 2010) only go up to 150, whereas the
older version were up to 200. In some cases (e.g., NGC~3227, 
NGC~7469) it appears as if to 
compensate for limited values of $\tau_V$, 
the fits are achieved with more clouds along radial equatorial rays,
close to the maximum value allowed by the models of $N_0=15$. 

While for most Seyferts in our sample  \textit{CLUMPY}
torus models with a typical value of
$N_0=12$ produced good fits to the data, the {\it
  Spitzer}/IRS spectra of PG quasars were well fitted with torus models
containing a mean value of $N_0=5$ (Mor et al. 2009). This is well understood
because most PG quasars show the 
$9.7\,\mu$m silicate feature in emission (Shi et al. 2006; Hao et
al. 2007), and the Nenkova et al. 
clumpy torus models with $\sigma_{\rm torus} \gtrsim 30\arcdeg$
and $N_0 \gtrsim 10$ almost always produce the feature in absorption, for all
viewing angles (see also discussions by Nenkova et al. 2008b and
Nikutta et al. 2009).

The optical depths of our sample of Seyferts show a broad distribution,
$\tau_V \sim 50-150$, 
with no obvious dependence on 
other torus model parameters or AGN
type (see also Mor et al. 2009). This result 
may be understood  because the
  \textit{CLUMPY} model near and mid-IR SEDs  
and the strength of silicate silicate feature  vary only slightly
  with varying  $i$ for a given set of $N_0$ and $\sigma_{\rm torus}$, at
$\tau_V>100$ (see
Nenkova et al. 2008b). The IR data of the PG quasars  studied by Mor et
  al. (2009) were fitted with $\tau_V \lesssim 100$ (Mor et al. 2009).

\setcounter{figure}{8}
\begin{figure}
\vspace{0.5cm}
\includegraphics[width=8.5cm,angle=-90]{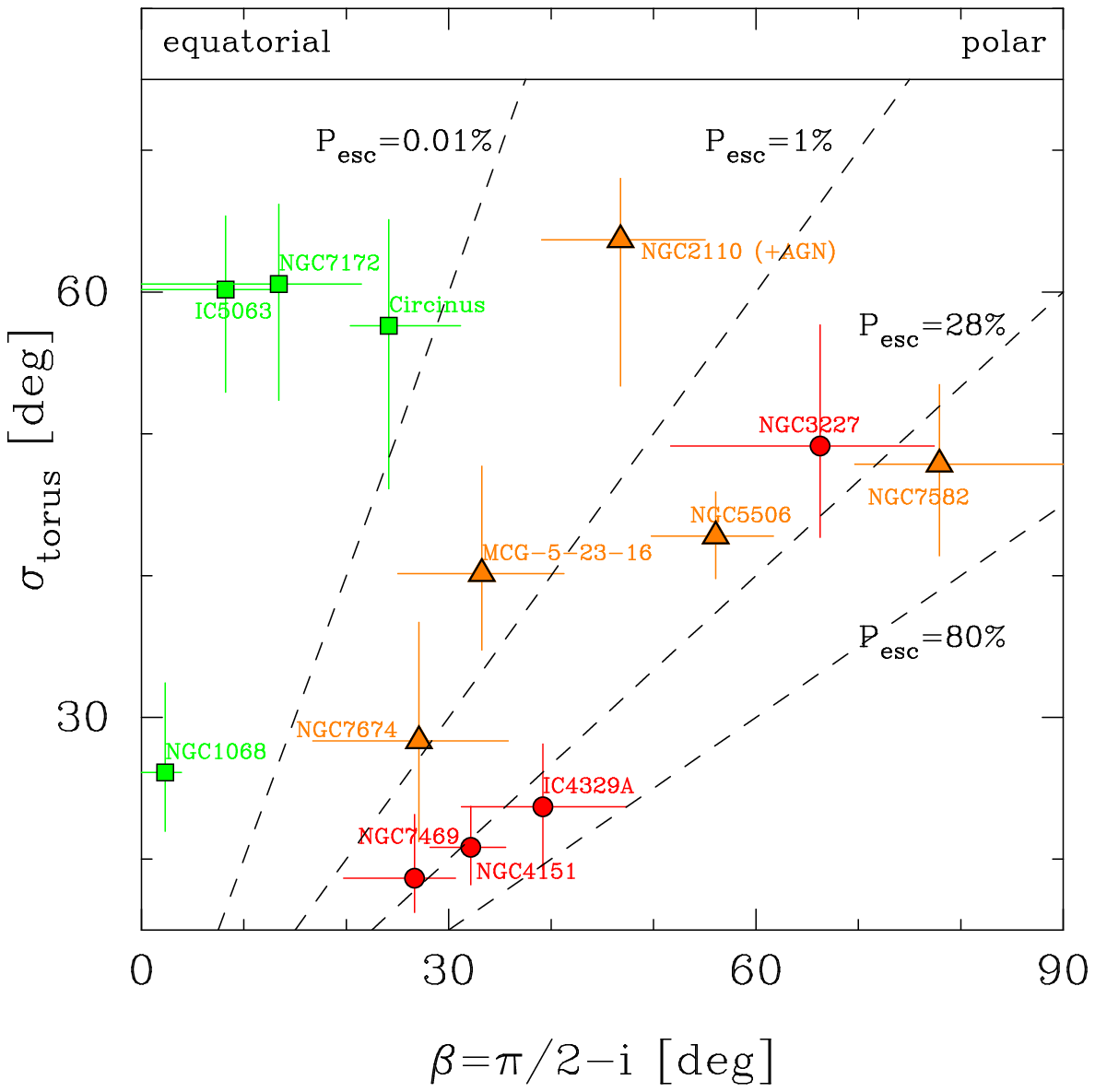}
\caption{Width of the torus $\sigma_{\rm torus}$ vs. the complementary
  to the viewing
  angle $\beta=90-i$. We plot the median and the $\pm
  1\sigma$ values of the probability distributions of the fitted
  values. The dashed lines are lines of constant 
AGN escape
  probabilities $P_{\rm esc}$ 
(Equation~3)  for an average number of clouds
  along an radial equatorial rays of $N_0=12$, which is representative of
  our sample (see Section~4.3). Filled symbols as in Figure~2.}
\vspace{0.5cm}
\end{figure}

\subsection{Comparison with fits of SEDs alone}
In Figures~4-6 we also compare the fits done with SED+spectroscopy
(blue lines and blue shaded regions) and SEDs alone (orange lines and
orange shaded regions), except in cases of sparse SEDs. 
 Some of our galaxies are in common with the works of Ramos
  Almeida et al. (2009, 2011). We have redone the SED alone fits  for those
  galaxies to use consistent errors with our work. 
In general we find that the main difference is that fits to SEDs alone
tend to infer  
broader probability distributions of these torus  model
parameters (see  Appendix~A).
In particular, for galaxies with low foreground extinctions 
the fits with SEDs alone when compared to the fits with SED+spectroscopy
produce compatible values (within the $\pm 1\sigma$ values of the
probability distributions) of
$\sigma_{\rm torus}$, $Y$,
and $i$. 

Most disagreements in the fitted parameters tend to occur for $N_0$,
$q$, and $\tau_V$.  Some of these differences are explicable. For instance, 
 $N_0$ and $q$ may trade off to yield an approximately constant number of clouds
in the inner region.  
Also, the optical depth of the individual clouds
$\tau_V$ and the average number of clouds along an equatorial ray
$N_0$ have a strong influence on the depth of the
$9.7\,\mu$m silicate feature and the shape of the $\sim 8-13\,\mu$m emission. 
Note also that the SED alone fits of some {\it pure} type 1
Seyferts always tend to produce the $9.7\,\mu$m  silicate feature
 in emission (see Figure~4), while the observations show an
almost featureless spectrum (IC~4329A, NGC~3227) or 
the feature slightly in emission (NGC~4151). For the Seyfert 2s with low host
galaxy foreground extinction there is a better agreement for these two
parameters 
between fits with and without mid-IR spectroscopy.

The other important fact to note is that including mid-IR spectroscopy
helps assess whether there is an important contribution from extended dust
structures causing absorption (and possibly emission too). This is
clearly the case for Circinus, NGC~5506, NGC~7172 and NGC~7582, which
have relatively deep silicate features. Indeed,
the fits to the SEDs alone  of Seyfert 2s of 
RA09, which were done without foreground extinction,  
only predicted  moderately deep silicate features for
these galaxies. However, we reiterate that it is not always possible
to produced good simultaneous fits to the SEDs and spectroscopy of Seyferts with deep
silicate features even when the host galaxy foreground extinction is  
included. It is possible that for the Seyfert galaxies
  with the deepest silicate features a clumpy medium in a torus-like
  configuration  may 
  not be appropriate to explain the observations 
(see e.g., Levenson et al. 2007).

\section{Properties of the torus and AGN}

\subsection{Torus size and  Angular width}

In this section we discuss the two torus model parameters that 
can be  compared with 
observations, the torus size  and the angular width. 

For each AGN, using the  $Y$ parameter and the bolometric luminosity of the 
system derived from the fits, and Equation~1 we computed the
physical radius $r_{\rm torus}$  of the torus. We used 
the median value of the fitted torus
radial thickness $Y$ and AGN bolometric luminosity. As can be seen from Table~6 the torus radii derived 
from our fits to the SED+spectroscopy are between $r_{\rm torus} 
\sim 1$ and 6\,pc. Our
fitted  physical sizes  are consistent with the mid-IR interferometric
results that the torus is relatively
compact (Tristram et al. 2007, 2009; Burstscher et al. 2009; Raban et
al. 2009).

Before we compare the torus radii with
the  $12\,\mu$m sizes inferred from mid-IR interferometric
observations, it is necessary to discuss some caveats. First, the 
modelling of the $12\,\mu$m interferometric observations requires
structures on different physical 
scales, as demonstrated for Circinus and NGC~1068 
(see Tristram et al. 2007, 2009; Raban et al. 2009, and Table~1). 
 However, the FWHM sizes of the cooler, more extended component
  could still be compared with model predictions.
Second, Nenkova et al. (2008b) showed that for tori with    
$Y=10-30$ and various viewing angles, at $12\,\mu$m $\sim 60-75\%$ of
the total flux is enclosed 
within angular radii of $\theta \sim 2-5 \theta_{\rm d}$. Here
$\theta_{\rm d}$ is the angular size at the inner radius of the
torus. This is because the $12\,\mu$m emission traces warm dust, with  
$T \gtrsim 200\,$K,  and is rather insensitive to cooler material further
from the nuclei. Finally the {\it compact} dusty torus  is expected to 
blend into the galaxy disk  (e.g., Shi et al. 2006; Packham et
al. 2007; Nenkova et al. 2008b), and thus deriving the ¨true
  size¨ of the torus might not be straightforward.

With these considerations in mind, 
we can compare  the
sizes derived  from the modelling of the $12\,\mu$m interferometric
data may  not be appropriate.   However, 
values of $Y$ larger than the $12\,\mu$m
sizes from interferometric observations are a necessary, although not
sufficient, condition for consistency between our modelling and the
interferometric results.
There are estimates of the $12\,\mu$m physical
size of the torus through mid-IR interferometry 
for six galaxies in our sample (see Table~1 for
 sizes and corresponding references). 
We find a relatively good agreement, within a factor of two, 
between the fitted torus sizes\footnote{For comparison
with the $12\,\mu$m sizes (FWHM) from mid-IR interferometric observations,
the $r_{\rm torus}$ values given in Table~6 
need to be multiplied by two.} and the sizes inferred
from mid-IR interferometry. The galaxy with the largest discrepancy is
NGC~7469.  Tristram et al. (2009) inferred $s_{12\mu{\rm 
m}  }=10.5\,$pc, whereas our fitted torus size is about 5\,pc. We
note however, that the interferometric observations of NGC~7469 were
affected by large uncertainties in the determination of the total flux
that may have compromised the derived physical size of the torus of
this galaxy.

The  torus model width of the angular distribution of the
  clouds, $\sigma_{\rm torus}$, is related to  
the opening angle of the ionization cones, $\Theta_{\rm cone}$. The
latter can be measured 
from emission line images, usually [O\,{\sc
  iii}]$\lambda$5007 and/or H$\alpha$. 
Table~1 gives the measured opening angles (measured from the
  polar direction) for our sample and corresponding references. It is 
important to note that this 
comparison is not straightforward because the relation between the true
opening angle of the cone and the measured opening 
angle from line emission depends on the gas distribution. For thin gas
disks, Mulchaey, Wilson, \& Tsvetanov (1996) simulations showed that cone-shaped
emission can be produced for most viewing angles and that the observed
angles are  less than the true opening angle. We can try to make this
comparison for galaxies with the observed widest ionization cones,
for which we can set more meaningful upper limits to the widths of the torus.
For the two galaxies with $\Theta_{\rm cone} \sim 100\arcdeg$
(NGC~5506 and NGC~7582, Table~1) the fitted $\sigma_{\rm torus}$ are compatible with the approximate
limits set by the observations ($\sigma_{\rm obs}\le 40-50\arcdeg$). Note
however, the difficulties encountered for fitting the observations of
these galaxies (Section~4.2).

\setcounter{figure}{9}
\begin{figure*}

\vspace{0.5cm}
\includegraphics[width=8.cm,angle=-90]{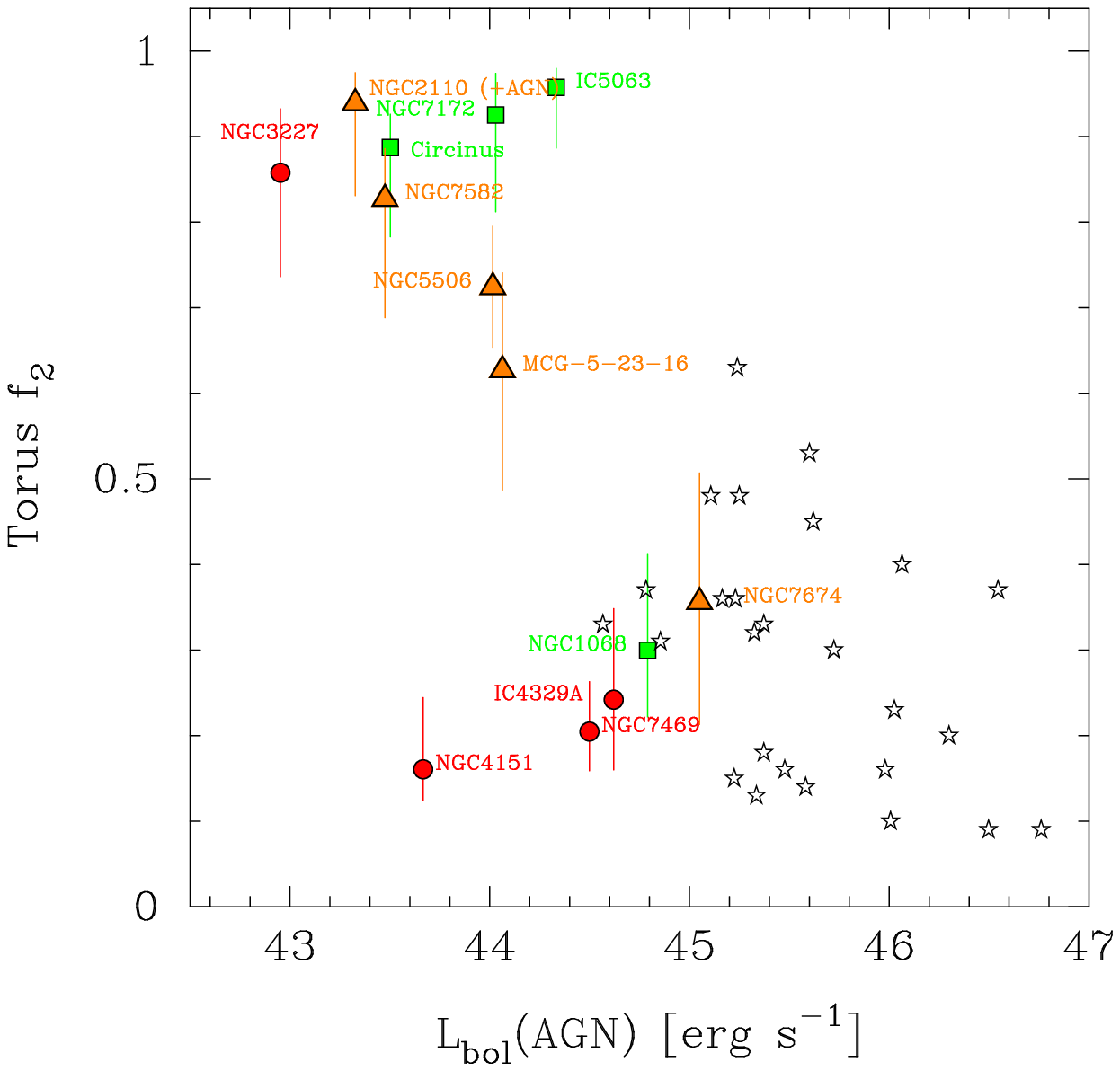}
\hspace{0.1cm}
\includegraphics[width=8.cm,angle=-90]{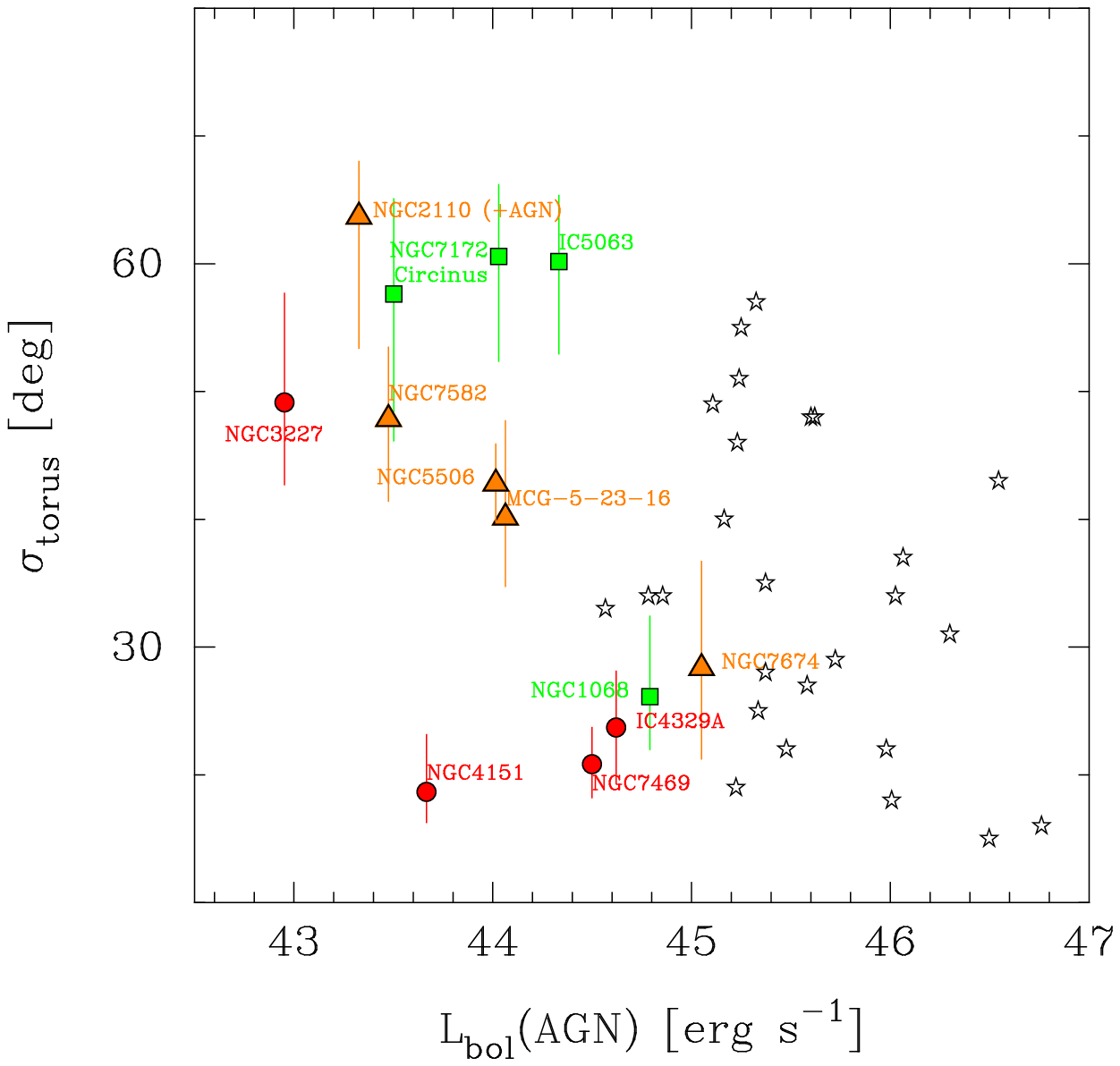}
\caption{AGN bolometric luminosities 
vs. the torus geometrical covering factor $f_2$ (left panel) and 
AGN bolometric luminosities vs. the angular width 
of the cloud distribution of the torus $\sigma_{\rm torus}$ (right
panel). The quantities  for our sample of Seyferts were derived 
  from the SED+spectroscopy fits 
(filled symbols, as in Figure~2). The plotted uncertainties for $f_2$ 
and $\sigma_{\rm torus}$ are the $\pm 1\sigma$ values of the posterior
distributions of the fitted parameters. Also plotted are PG
  quasars (star-like symbols), 
for which Mor et al. (2009) derived
the torus $f_2$ and $\sigma_{\rm torus}$   using
  the Nenkova et al. (2008b) models. The bolometric luminosities of
  the PG quasars are derived  from the
  optical luminosities (see Mor et al. 2009 for details).} 
\vspace{0.5cm}
\end{figure*}

\subsection{AGN Escape Probability}
One of most salient properties of the clumpy torus models is that
there is always a finite probability for an obscured view of the AGN,
irrespective of the viewing angle (Nenkova et al. 2008a,b). 
While in 
the simplest version of the unified model type 1s would be observed
in directions closer to the pole of the torus (low values
of $i$) than type 2s, in a clumpy medium this is not necessarily so. 
Indeed, it is possible to have a type 1 viewed at a relatively high
$i$ provided that the width of the angular size of the torus is small (for
a given $N_0$), as is the case for NGC~4151 and
IC~4329A.  Note that two-dimensional posterior
  distributions of these two galaxies show that 
the fitted values of $i$ and
  $\sigma_{\rm torus}$ are inversely correlated (Figures~B3 and B4 
in the Appendix.)
Seyfert 2s can also be viewed from any angle, as long as
there is at least one
cloud along the line of sight.  The probability of encountering a
cloud is higher for increasing values of 
$\sigma_{\rm torus}$, $N_0$, and $i$. The fact that not all type 2s 
 in our sample require very broad
angular distributions (large $\sigma_{\rm torus}$, e.g., NGC~1068 in Figure~9), 
suggests that $N_0$ and $i$ must be high  in those cases. 
In Figure~9 the error
bars reflect the $\pm 1\sigma$ uncertainty around the median 
of the probability distributions  of 
the fitted parameters.

A more relevant quantity for an AGN to be classified as a type 1 or a
type 2 is the probability for an AGN-produced photon to escape
unabsorbed (Equation~3). Since the escape probability is a highly
non-linear function of $N_0$, $i$, and $\sigma_{\rm torus}$, 
we took full advantage of our Bayesian approach. We thus generated the 
posterior distributions for $P_{\rm esc}$ 
from the posterior distributions of the relevant fitted parameters 
given in Table~5. As can be seen from the
values in Table~6, all {\it pure} type 1s and type 2s have
respectively, relatively 
high ($\sim 12-44\%$) and low ($\le 0.1\%$)
AGN escape probabilities, as expected. The type 2 Seyferts with
broad-lines detected in the near-IR show a range of escape
probabilities. We note that these probabilities do not
include the effects of foreground extinction. 

\subsection{Covering Factors and a Receding Torus?}
Observational (e.g., Simpson 2005) and theoretical (e.g., H\"onig \&
Beckert 2007) arguments provide evidence of the so-called receding torus
(Lawrence 1991). In this scenario the  higher fraction of type 1 AGN
at high AGN luminosities is explained in terms of a decreasing covering
factor of the torus. Furthermore,  H\"onig \& Beckert (2007) and 
Nenkova et al. (2008b) argued that the
decreasing fraction of type 2s at high luminosities not only arises
from decreasing $\sigma_{\rm torus}$, but also from decreasing $N_0$, or both.

In this work we calculated two different covering factors using
  our Bayesian approach. The first
one is the geometrical covering factor $f_2$, which does not depend on
the viewing angle, and is derived by  
 integrating the AGN escape probability (Equation~3) over all angles
as put forward by Mor et al. (2009). If we define $\beta =\pi/2 - i$,
then:
\begin{equation}
f_2= 1 - \int_0^{\pi/2} P_{\rm esc}(\beta)\cos(\beta)d\beta.
\end{equation}

A related quantity is the  the ratio between
the torus luminosity integrated over the entire wavelength
  range covered by the torus models ($\sim 0.2-700\,\mu$m) 
and the inferred AGN bolometric luminosity:

\begin{equation}
f(i)=\frac{L_{\rm
  torus}(i)}{L_{\rm bol}^{\rm model}({\rm AGN})}. 
\end{equation}

This ratio is the {\it apparent}
covering factor (see Mor et al. 2009), because it depends
on the viewing angle $i$ (as well as on $\sigma_{\rm torus}$ and $N_0$), and
can be 
understood as a reprocessing efficiency. 
As we did for $P–{\rm esc}$ we generated the
posterior distributions for $f_2$ and $f(i)$ 
from the posterior distributions of the relevant fitted parameters 
given in Table~5. The median values and the $\pm 1\sigma$ values 
of the distributions of the geometrical covering
factor and 
reprocessing efficiencies
of our sample of Seyferts are given in Table~6. 

The galaxies in our
sample show a  range of reprocessing efficiencies, 
although in general there is no clear dependency with any torus
 model parameters. 
In contrast, the  torus geometrical covering factor $f_2$
appears to depend on the AGN bolometric luminosity, as can be seen
from Figure~10 (left panel). We also included in this comparison
high luminosity AGN. These are PG quasars, 
for which Mor et al. (2009) computed the torus
$f_2$ from $N_0$ and  $\sigma_{\rm
  torus}$ as inferred from fits to {\it Spitzer}/IRS
spectroscopy.  We note that these authors also  used
the Nenkova et al. (2008a,b) torus models, but added extra components
to fit their data (see Section~4.1). Mor et al. (2009) calculated the PG
quasars  AGN bolometric luminosities using the observed optical
luminosities (see Mor et al. 2009 for more details). 
For our sample of Seyfert galaxies the uncertainties in $f_2$ take
into account the $\pm 1\sigma$ values around the median 
of the probability distribution  of
$\sigma_{\rm torus}$. When combining the PG quasars of Mor et al. (2009) and
our sample of Seyferts,  there is  a tendency 
for the  torus
covering factor $f_2$ to decrease from $f_2 \sim 0.9-1$  at  low  AGN  
luminosities ($10^{43-44}\,{\rm erg\,s}^{-1}$) to $f_2 \sim 0.1-0.3$
at high luminosities ($\gtrsim 10^{45}\,{\rm erg\,s}^{-1}$).  
Such tendency for the torus covering factor to decrease with the bolometric
luminosity of the AGN was already noted for PG quasars by Mor et
al. (2009) when including an additional hot black-body component to fit
their data. 

A similar trend  may also be  present when comparing 
$\sigma_{\rm torus}$ vs. $L_{\rm bol}^{\rm model}({\rm AGN})$, 
with high luminosity AGN having narrower cloud angular distributions 
(Figure~10, right panel) than low luminosity ones.  
We also found in Section~4 that on average 
Seyfert galaxies (with bolometric luminosities in the range
$10^{43-45}\,{\rm erg\,s}^{-1}$) are fitted with more clouds along a
radial equatorial ray than PG quasars (bolometric luminosities  
$\gtrsim 10^{45}\,{\rm erg\,s}^{-1}$). Given these trends, it is not
surprising that the  torus geometrical covering factor $f_2$, which depends
on both  $N_0$ and $\sigma_{\rm torus}$, tends
to be lower in the most luminous objects in our sample than in the
less luminous ones. 

The decreased torus covering factor at high AGN 
luminosities  may provide some support for a receding torus
at high luminosities. We note however, that in these comparisons there
are  few type 2 objects at
high AGN luminosities, so we cannot rule out a possible dependency
of $\sigma_{\rm torus}$ with the 
AGN type. That is, type 2s might have wider tori than type 1s as suggested
by the results of RA11. Finally a note of caution. A large fraction of
the low luminosity AGN in our sample with wide angular cloud
distributions are in highly inclined galaxies. Therefore, we cannot
exclude the possibility that 
 contamination from extended dust features not directly related to the
 dusty torus has 
resulted in large fitted values of  $\sigma_{\rm torus}$. 
Finally, as we explained in Section~2.1, our
  sample is not complete, and may not be representative, and thus this
result needs to be explored for a complete sample.

\section{Summary and Conclusions}

This is the third paper in a series performing detailed fits to the 
unresolved IR emission of AGN using  the clumpy torus models of
  Nenkova et al. (2008b). 
In the first two papers of the series
(RA09 and RA11) we fitted only the photometric SEDs, while in this
paper we also included high angular resolution ($0.3-0.4\arcsec$)
ground-based mid-IR spectroscopy. The sample in this work is composed 
of 13 nearby (at a median distance of 31\,Mpc) Seyfert
galaxies with 
bolometric luminosities $L_{\rm bol}({\rm AGN}) \sim
10^{43}-10^{45}\,{\rm erg\, s}^{-1}$. 
The sample contains {\it pure} type 1 and type 2 Seyferts,
as well as Seyfert 2s with broad-lines detected in the near-IR.  
The galaxies also span a range in
the observed strength of the $9.7\,\mu$m silicate 
feature, going from  mild emission to deep absorption, and in  
X-ray column densities, from Compton thin to Compton thick objects.  

We compiled near and mid-IR  imaging and spectroscopy data
taken with high-angular resolution ($\sim 0.3-0.8\arcsec$). We also
presented new ground-based mid-IR spectroscopic observations in the $N$-band
of NGC~4151 and in the $Q$-band of NGC~1068.
We have used an interpolated version
of the clumpy torus models of Nenkova et al. (2008a,b and 2010) and a Bayesian
approach to fit together the photometric SEDs and mid-IR
spectroscopy. The fits provided the probability distributions for the
six torus  model parameters that describe the 
  \textit{CLUMPY} models (see Table~3), plus 
the AGN bolometric luminosity. We included the effects of
foreground extinction in the host galaxy for galaxies with evidence
of extended dust structures and $A_V({\rm
  frg}) \gtrsim 5\,$mag. When compared to fits done using SEDs alone, 
the fits to the
SED+spectroscopy data result in smaller uncertainties (narrower probability
distributions) for the angular distribution of the clouds $\sigma_{\rm
  torus}$, torus radial thickness $Y$, and viewing angle $i$.

The interpolated version of the Nenkova et al. (2008a,b; 2010) clumpy
torus models
provided good fits to the photometric SEDs and mid-IR
spectroscopy of our sample, and 
in particular for those Seyfert galaxies with low or
moderate foreground extinctions ($A_V({\rm frg})\lesssim 5-10\,$mag). 
For three AGN with relatively
deep $9.7\,\mu$m silicate features and  hosted in inclined galaxies with
prominent dust lanes
(NGC~5506, NGC~7172, and 
NGC~7582), we were not able to fit
simultaneously the photometric SEDs and mid-IR spectroscopy. These
three galaxies show evidence of extended dust structures on tens of parsec
scales, not directly related to the dusty torus. In contrast, we were
able to fit reasonably well the data of Circinus. One possibility is
that because Circinus is  the closest galaxy in our sample (distance of
4\,Mpc), the contamination by extended dusty structures is reduced.

\textit{CLUMPY} models with small tori (torus radial
thickness in the  range $Y\sim 10-15$) 
provided adequate fits to the data. Combining the modelled 
AGN bolometric luminosities and the values of $Y$, we inferred that
the physical radii of the tori of Seyfert galaxies are between $\sim
1\,$pc and 6\,pc. For the six 
galaxies in our sample with   $12\,\mu$m sizes derived in the literature 
from  the modelling of mid-IR interferometric observations, we found a
reasonably good agreement. 

The  Nenkova et al. (2008b) models were also able to produce good
fits to mid-IR high angular 
resolution spectroscopy, in terms of the shape of the silicate feature
and continuum slope and not only the apparent strength of the feature.
When combined with the 
photometric SEDs, the mid-IR spectroscopy allowed us to constrain the number of
clouds along radial equatorial rays $N_0$ and the optical depth of the
individual clouds $\tau_V$. We find that the tori of Seyfert galaxies
have typically $N_0\sim 8-15$, and the optical depth of the clouds
are in the range $\tau_V=50-150$. By comparison the tori of PG
quasars ($L_{\rm bol}({\rm AGN})>10^{45}\,{\rm erg\,s}^{-1}$) 
appear to contain fewer clouds along radial equatorial rays, typically
$N_0 \sim 5$,  with optical depths of 
$\tau_V\lesssim 100$ (Mor et al. 2009).

From the scaling of the models to the data we derived the 
AGN bolometric luminosities, which 
were in good agreement with those derived with other methods in the literature. 
As also found by RA11, the viewing angle $i$ is not the only parameter
controlling the classification of a galaxy
into a type 1 or a type 2. For instance, type 2s tend to be viewed
at directions closer to the equatorial plane of the torus, but small
values of $i$ are permitted 
as long as there is one cloud along the LOS. This is because in clumpy
media, there is always a finite probability for an AGN photon to
escape absorption, and this probability is 
a function of $i$, $\sigma_{\rm torus}$, and $N_0$. From our fits, 
we found that this
probability is relatively high in type 1s ($P_{\rm esc} \sim 12-42\%$),
while it is found to be low in {\it pure } type 2s ($P_{\rm esc} \le 0.1\%$). 

We finally discussed the
possibility of a receding torus. We compared the results for our
sample of relatively low luminosity AGN with those of the PG
quasars analyzed by Mor et al. (2009). Both samples combined together
span a range of $L_{\rm bol}({\rm AGN}) \sim 10^{43-47}\,{\rm erg
  \,s}^{-1}$. We found that high
luminosity AGN ($L_{\rm bol}({\rm AGN}) \gtrsim 10^{45}\,{\rm erg
  \,s}^{-1}$) tend to have lower torus covering factors ($f_2 \sim
0.1-0.3)$ because they have
narrower tori (smaller $\sigma_{\rm
  torus}$) and contain fewer clouds along radial equatorial rays
$N_0$. In contrast lower luminosity ones (at $L_{\rm bol}({\rm AGN}) \sim
10^{43-44}\,{\rm erg 
  \,s}^{-1}$) tend to have high covering factors ($f_2 \sim 0.9-1$). 
This  might explain the decreased observed fraction
of type 2 AGN at high luminosities. We note  however, that this result
may be subject to some caveats. In particular most of the AGN in our 
sample at the low end of $L_{\rm bol}({\rm AGN})$ are hosted in 
highly inclined galaxies, and it is possible that contamination by
extended dust components not related to the dusty torus 
results in larger fitted values of
$\sigma_{\rm torus}$. Also,  as discussed in Section~2.1, our
  sample is not complete, and in particular there are very few type 2s
at high AGN luminosities. Therefore  our results need to be explored
further,  and to this end we have been allocated time on the Gran
Telescopio Canarias with the mid-IR instrument 
CanariCam (Telesco et al. 2003) for a large mid-IR survey of AGN
(Levenson et al. 2008).

\smallskip

\acknowledgments
%The authors would like to thank 

We thank an anonymous referee for scientific suggestions that
  helped improve the paper.
We are very grateful to Sebastian H\"onig and Konrad Tristram for
making their ground-based mid-IR spectroscopy available to us and for
enlightening discussions. A.A.-H. thanks the Astrophysics Department
at Oxford University, where most of this research was conducted, for
their warm hospitality.

A.A.-H. acknowledges support from the Spanish Plan Nacional del Espacio
under grant ESP2007-65475-C02-01 and Plan Nacional de Astronom\'{\i}a
y Astrof\'{\i}sica under grant AYA2009-05705-E. 
C.R.A. and J.M.R.E. 
acknowledge the Spanish Ministry of Science and Innovation (MICINN)
through Consolider-Ingenio 2010 Program grant CSD2006-00070: First
Science with the 
GTC
(http://www.iac.es/consolider-ingenio-gtc/).
C.R.A. acknowledges financial support from STFC PDRA (ST/G001758/1).
A.A.R. acknowledges financial support from 
AYA2010-18029 (Solar Magnetism and Astrophysical
Spectropolarimetry) and Consolider-Ingenio 2010 CSD2009-00038.
C. P. acknowledges the National Science Foundation under gran number 
0904421.

This work is based on observations obtained at the Gemini Observatory,
which is operated by the Association of 
Universities for Research in Astronomy, Inc., under a cooperative
agreement with the NSF on behalf of the Gemini 
partnership: the National Science Foundation (United States), the
Science and Technology Facilities Council (United 
Kingdom), the National Research Council (Canada), CONICYT (Chile), the
Australian Research Council (Australia), 
Minist\'erio da Ci\^encia e Tecnologia (Brazil), and Ministerio de
Ciencia, Tecnolog\'{\i}a e Innovaci\'on Productiva (Argentina).

This research has made use of the NASA/IPAC Extragalactic Database
(NED) which is operated by the Jet Propulsion Laboratory, California
Institute of Technology, under contract with the National Aeronautics
and Space Administration.  

\appendix
\section{Marginal Posterior Distributions}
Here we present the marginal 
posterior distributions of the fits to the
SED+spectroscopy and SED alone for all the galaxies in the sample
except for IC~4329A, which is shown in Figure~7.

\begin{figure*}[h]
\hspace{2cm}
\includegraphics[width=10.cm,angle=-90]{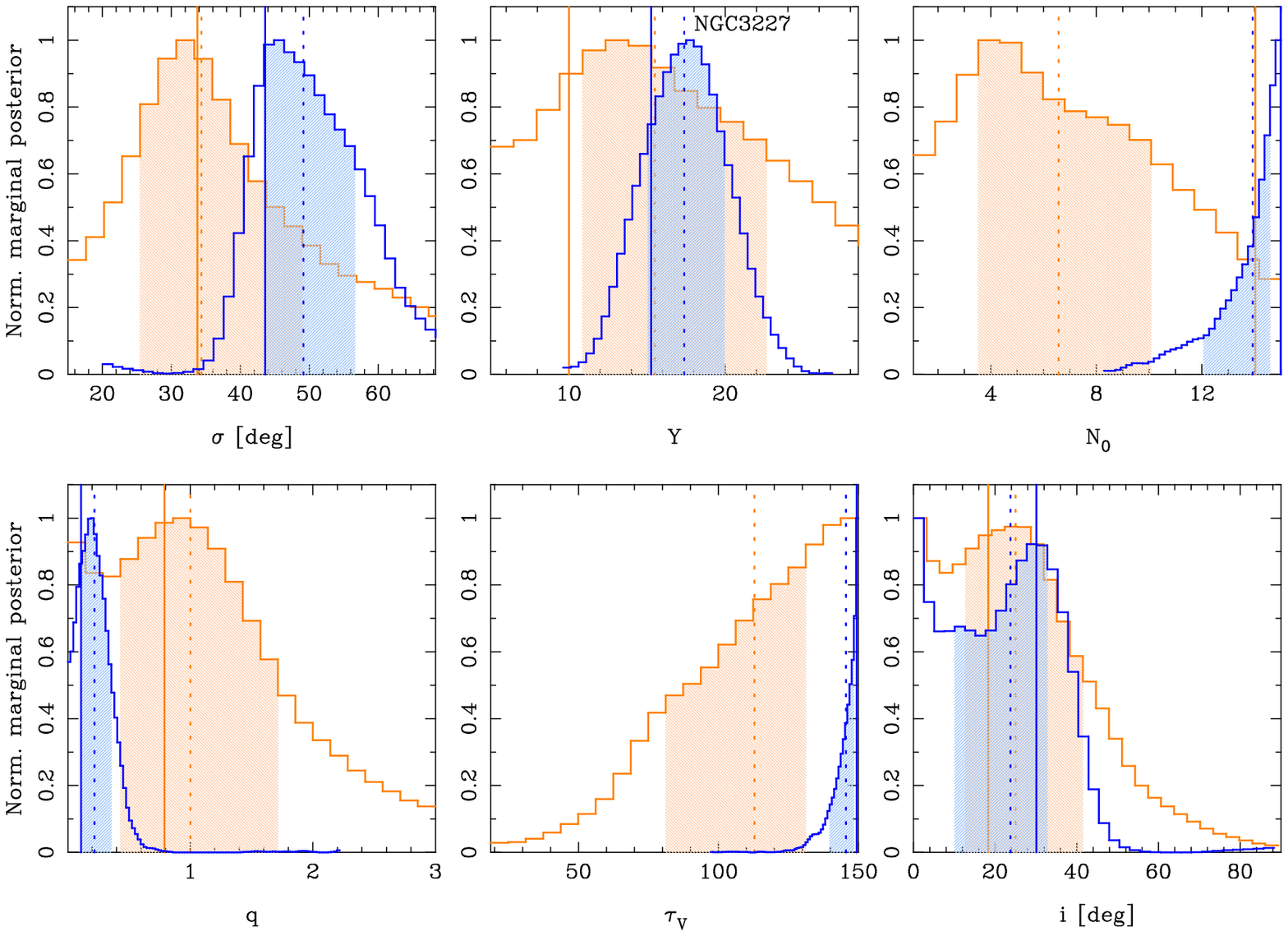}

{Fig. A1.--- Marginal posterior distributions of the free parameters that
  describe the \textit{CLUMPY}
models resulting from fitting the data for
  NGC~3227.  The
  SED alone fits are shown in orange and the SED+spectroscopy fits are
  shown in blue. The
  solid and dashed lines indicate the MAP and median values of the
  distributions, and the shaded areas are the $\pm 1\sigma$ values.
We did not
  use a foreground extinction. }
\end{figure*}

\begin{figure*}
\hspace{2cm}
\includegraphics[width=10.cm,angle=-90]{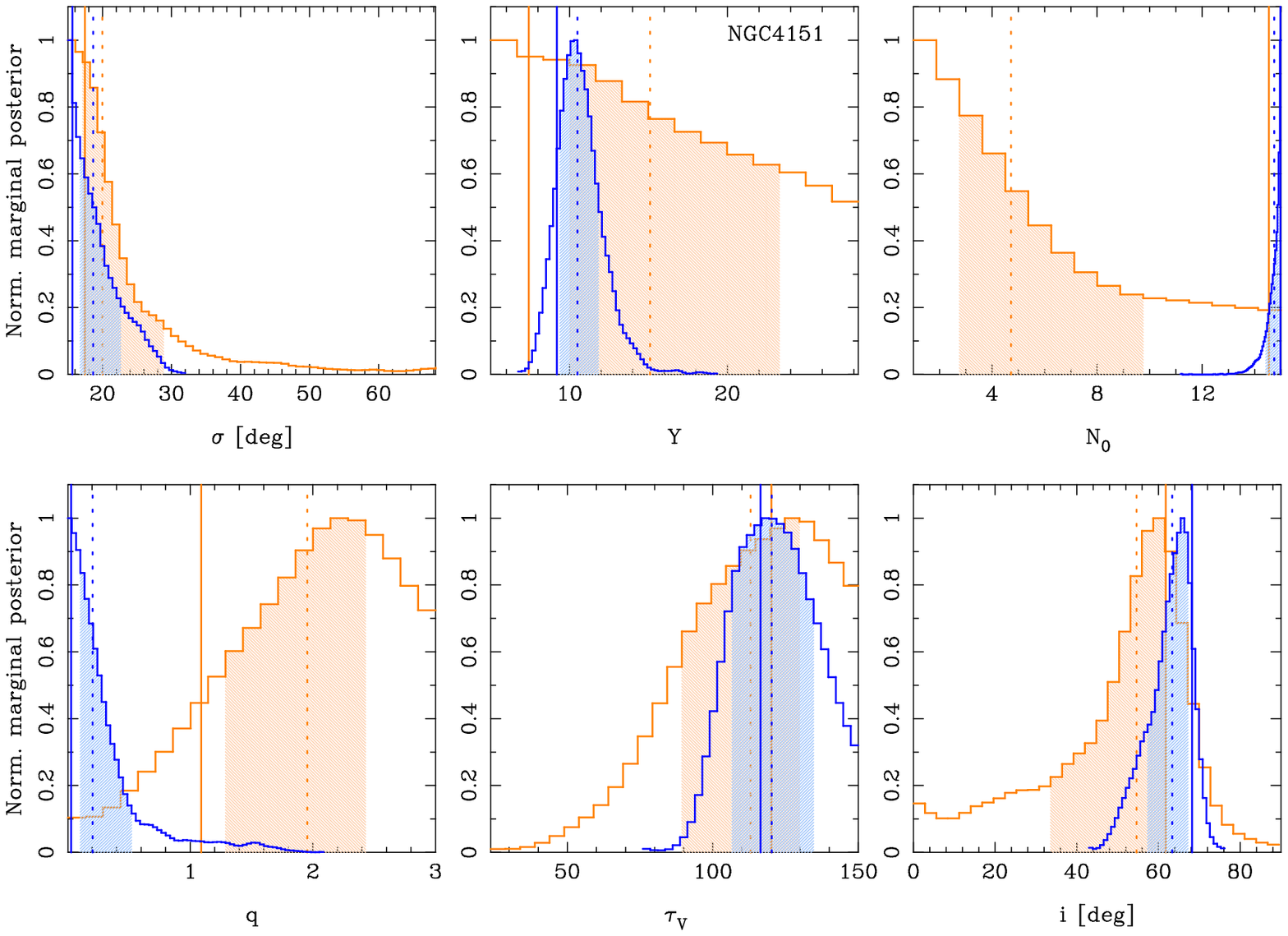}

{Fig. A2.--- Same as Figure~A1 but for NGC~4151. }
\end{figure*}

\begin{figure*}
\hspace{2cm}
\includegraphics[width=10.cm,angle=-90]{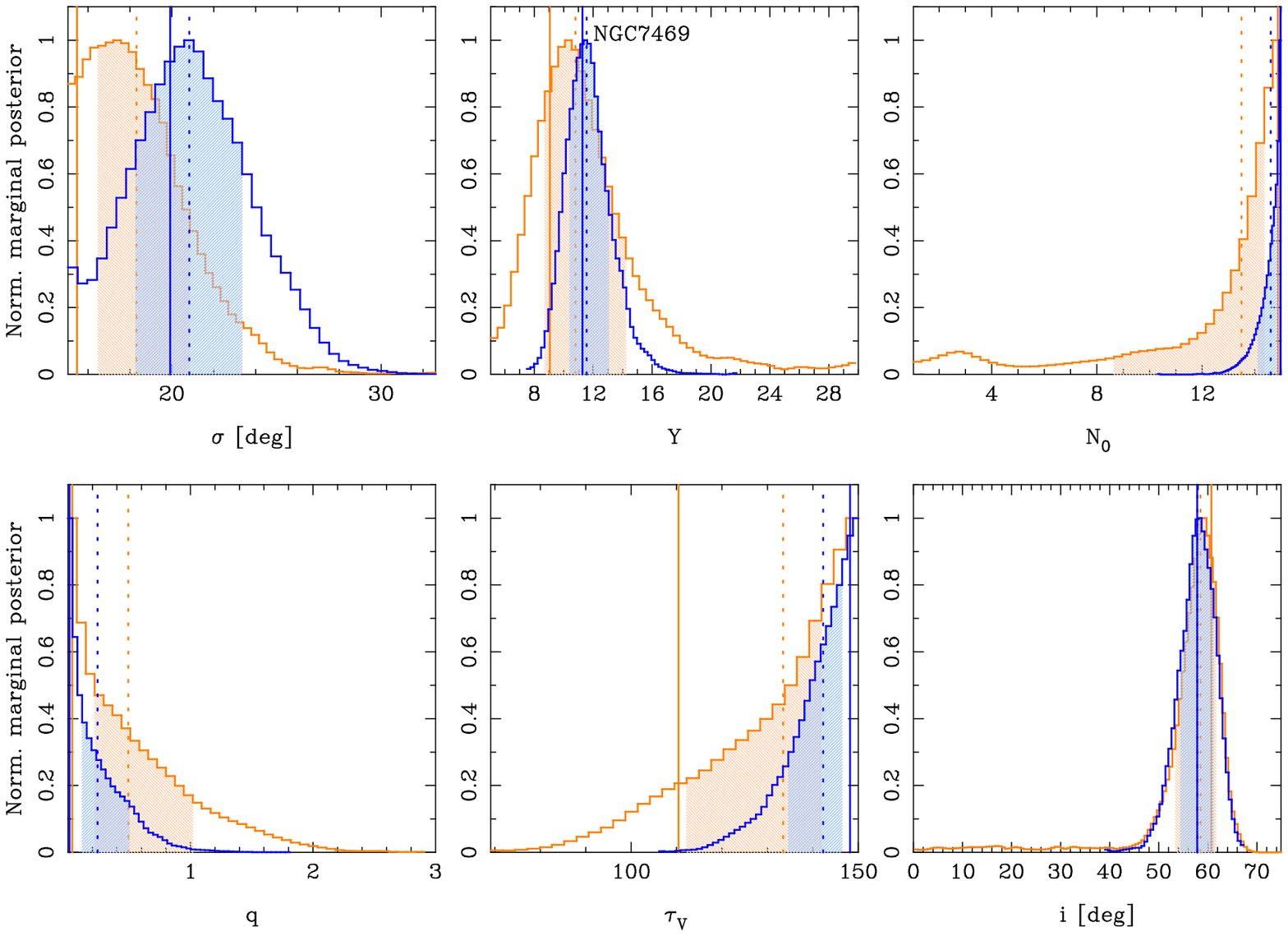}

{Fig. A3.---  Same as Figure~A1 but for NGC~7469. }
\end{figure*}

\begin{figure*}
\hspace{2cm}
\includegraphics[width=10.cm,angle=-90]{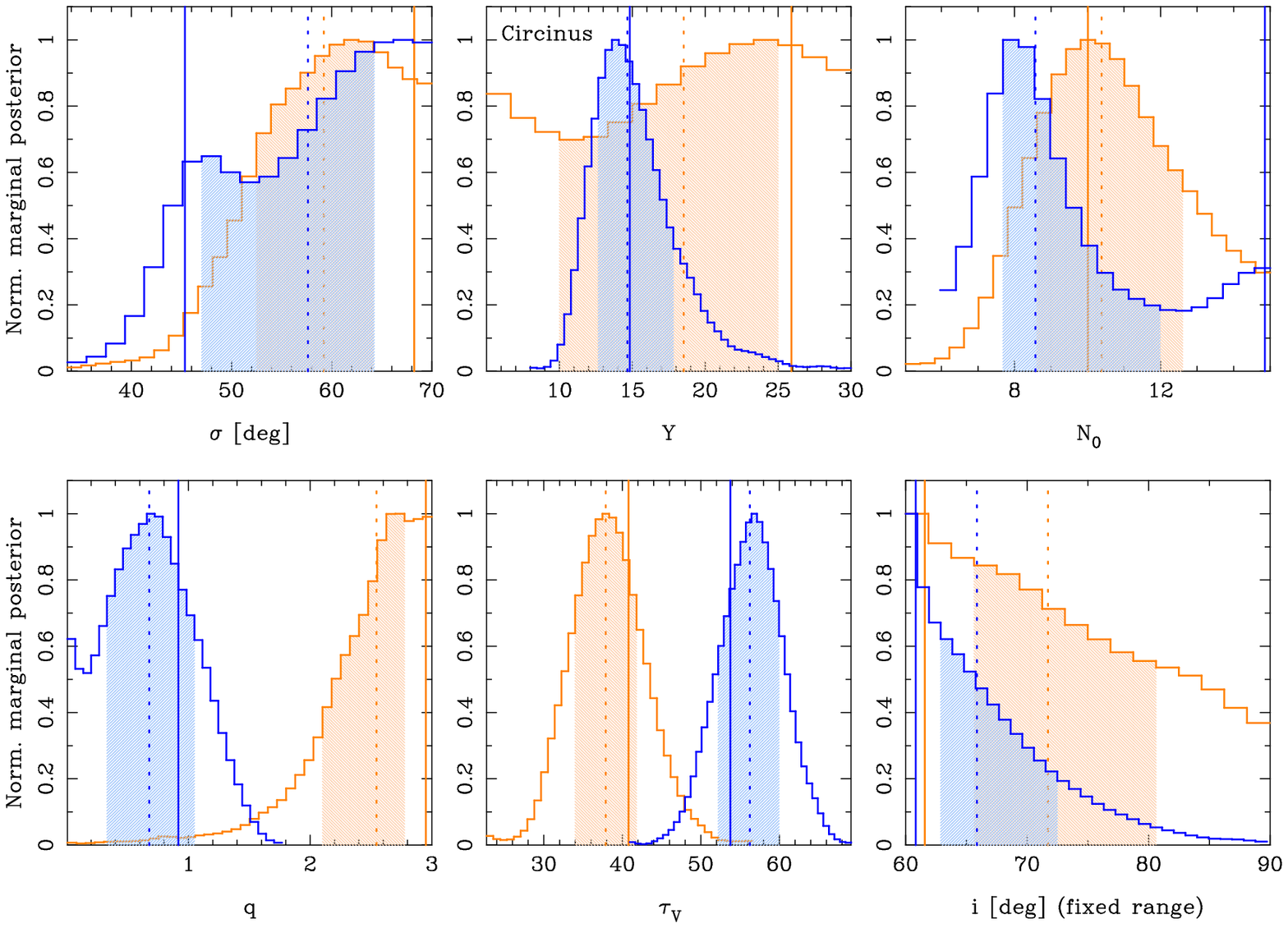}

{Fig. A4.---  Same as Figure~A1 but for Circinus. 
We fixed the 
foreground extinction to $A_V({\rm frg})=9\,$mag (Maiolino et
al. 2000). We restricted the viewing angle to the range
$i=60-90\arcdeg$. 
}
\end{figure*}

\begin{figure*}
\hspace{2cm}
\includegraphics[width=10.cm,angle=-90]{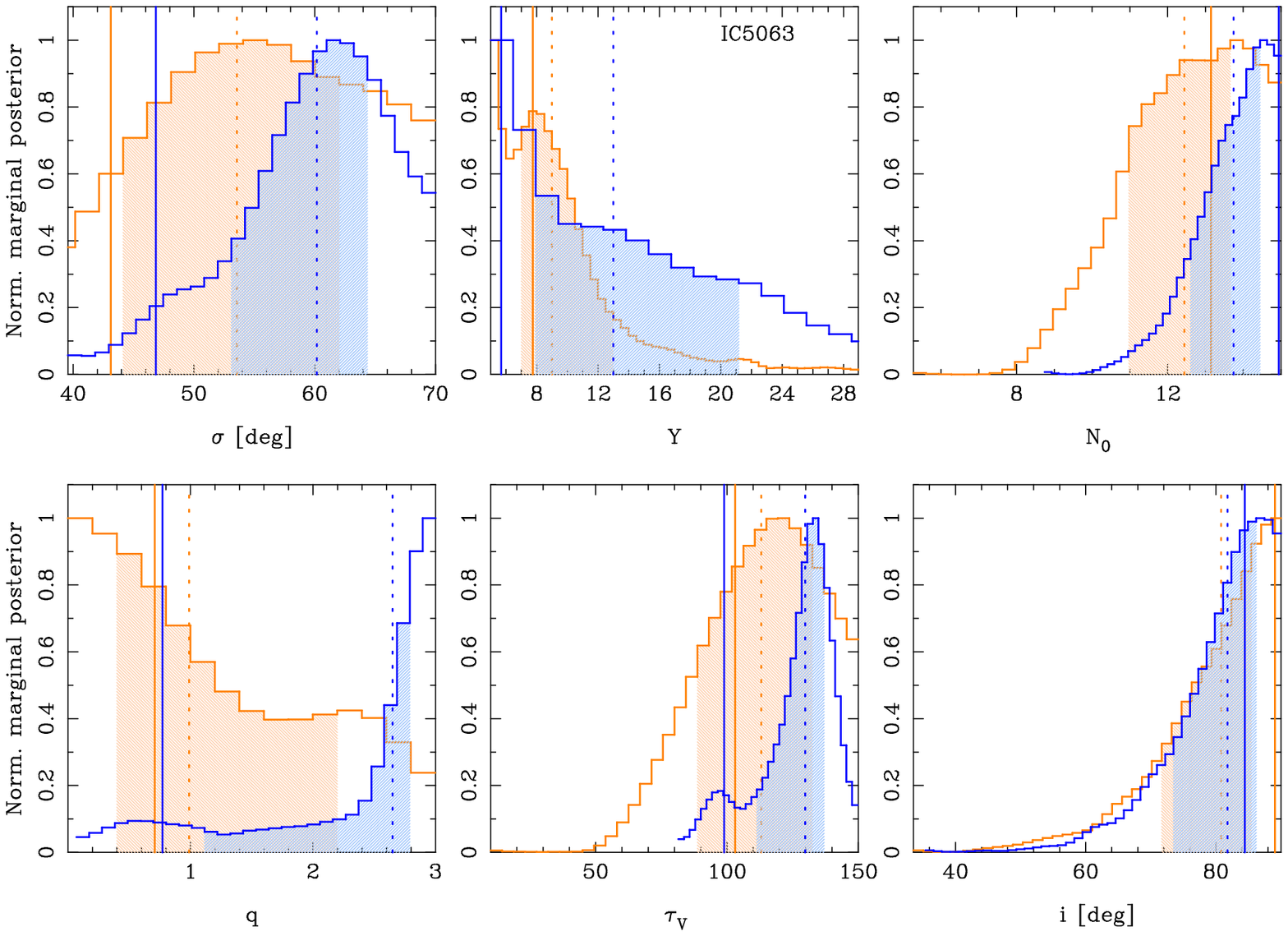}

{Fig. A5.--- Same as Figure~A1 but for IC~5063. We did not restrict 
any of the torus  model parameters.  
The foreground extinction was fixed to A$_V{\rm (frg)}=7\,$mag based
on results by Heisler \& de Robertis (1999).  }
\end{figure*}

\begin{figure*}
\hspace{2cm}
\includegraphics[width=10.cm,angle=-90]{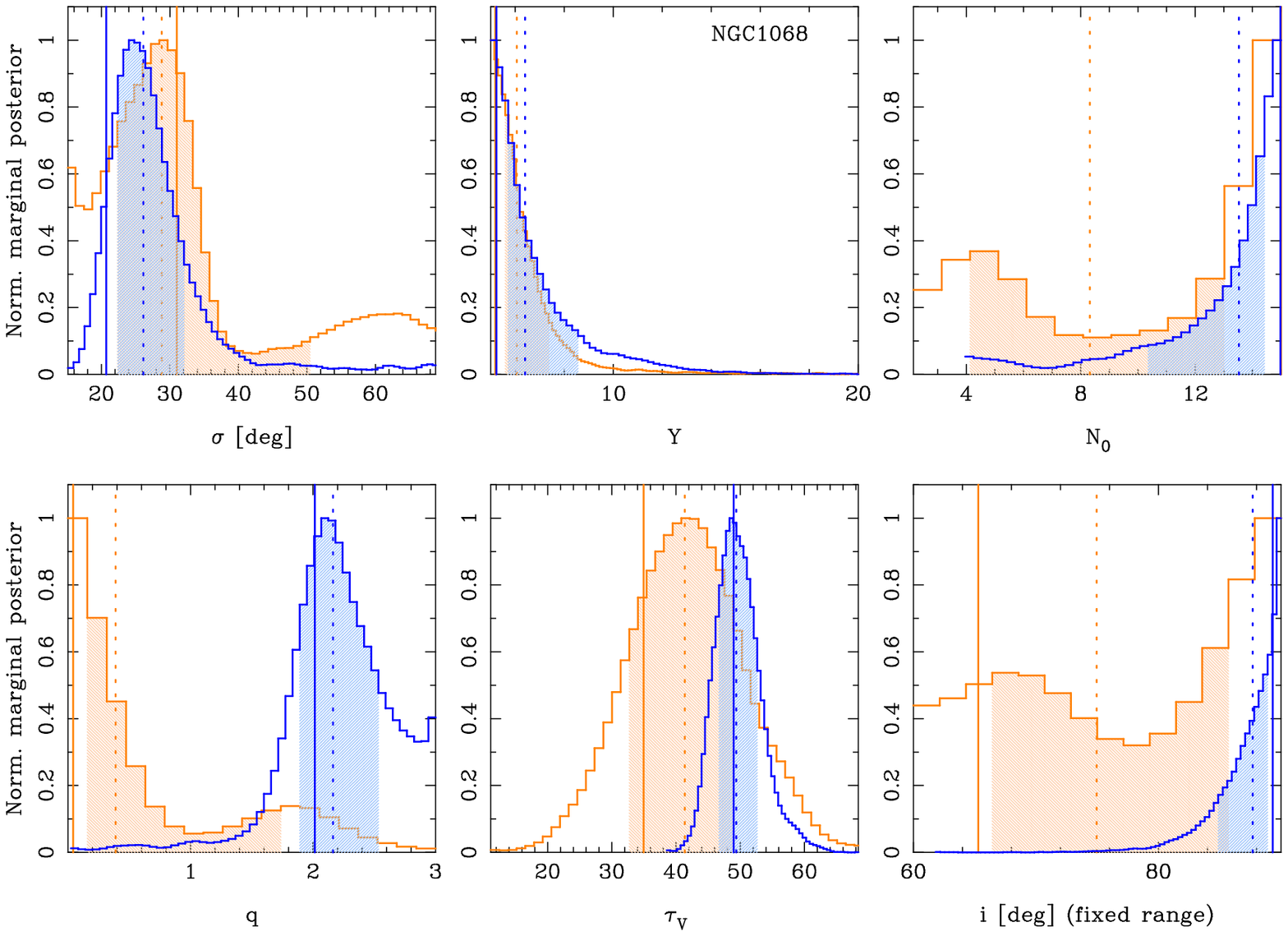}

{Fig. A6.--- Same as Figure~A1 but for NGC~1068. We restricted the viewing
  angle to the range $i=60-90\arcdeg$ (see Section~3.3). }
\end{figure*}

\begin{figure*}
\hspace{2cm}
\includegraphics[width=10.cm,angle=-90]{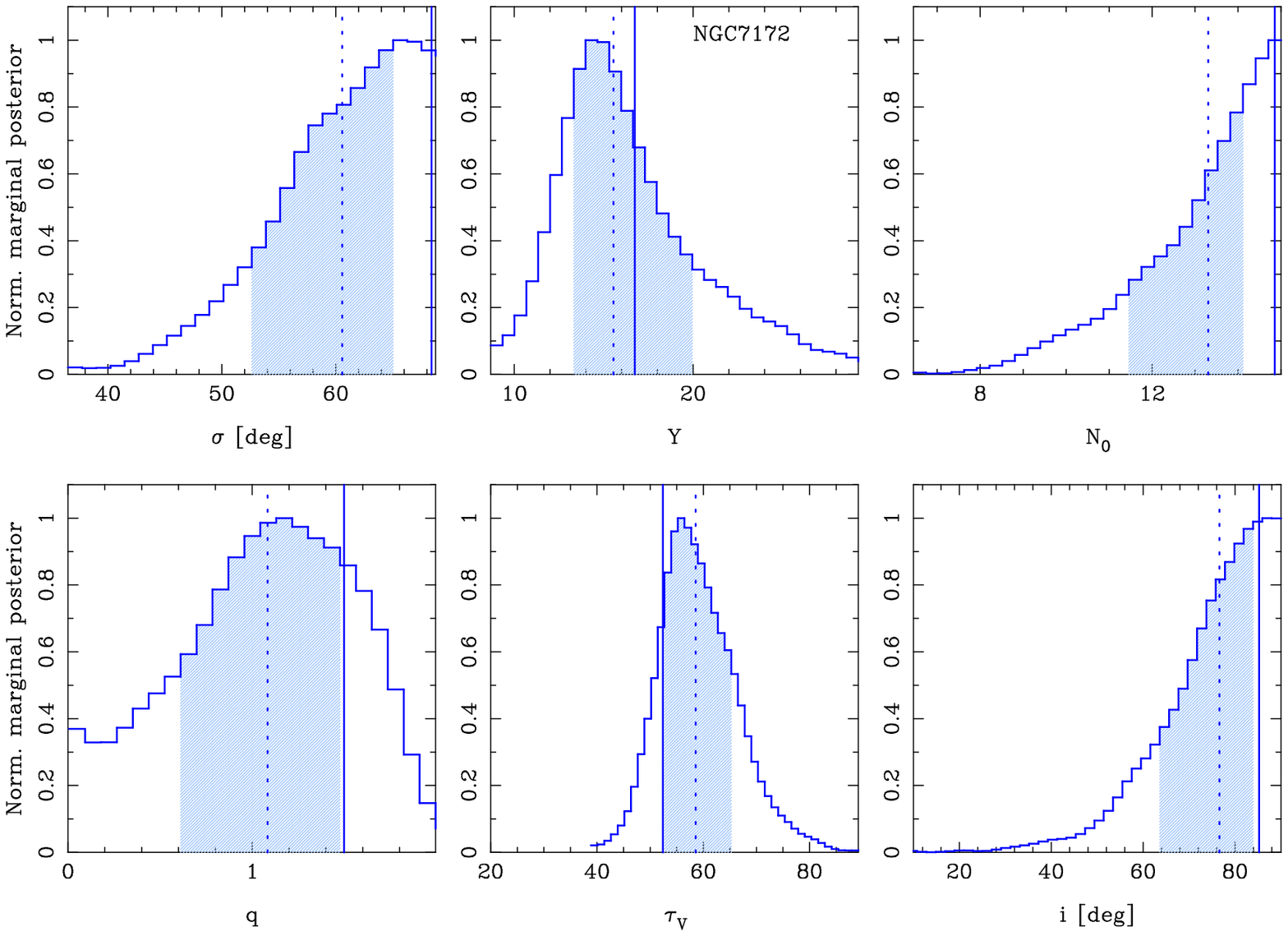}

{Fig. A7.--- Same as Figure~A1 but for NGC~7172. Only the SED+spectroscopy
  fit is shown due to the limited number of 
photometric points in the near-IR. 
We did not restrict any of the torus  model 
  parameters. The foreground extinction was fixed to 
$A_V({\rm frg})=40\,$mag.}
\end{figure*}

\clearpage

\begin{figure*}
\hspace{2cm}
\includegraphics[width=10.cm,angle=-90]{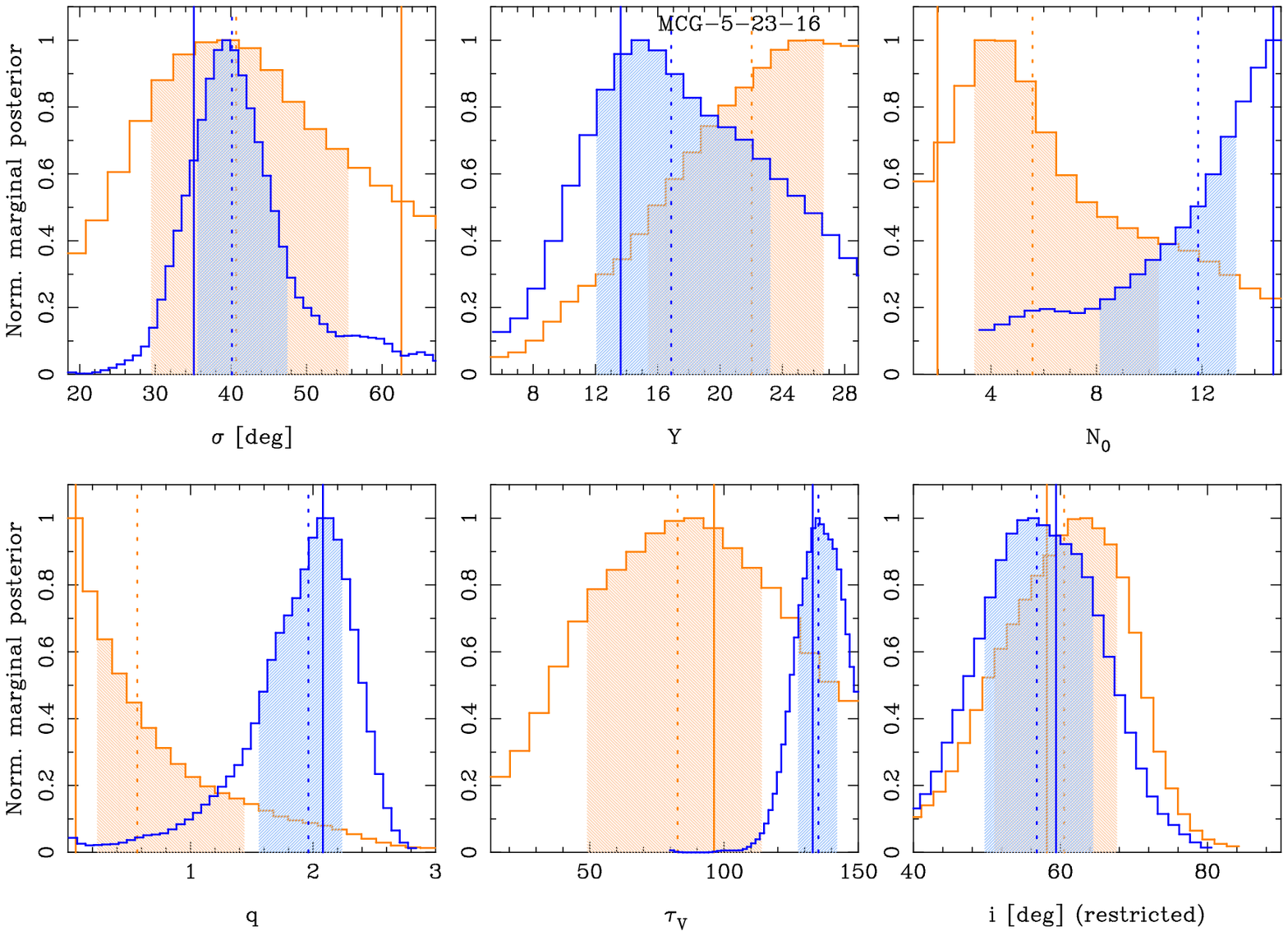}

{Fig. A8.--- Same as Figure~A1 but for MCG~$-$5-23-16. 
The viewing angle was restricted to a Gaussian distribution
  centered at $i=55\arcdeg$ \, with a 10\arcdeg \, width (see Section~3.3).  
The foreground extinction was fixed to A$_V{\rm (frg)}=7\,$mag based
on results by Veilleux et al. (1997).  }
\end{figure*}

\begin{figure*}
\hspace{2cm}
\includegraphics[width=10.cm,angle=-90]{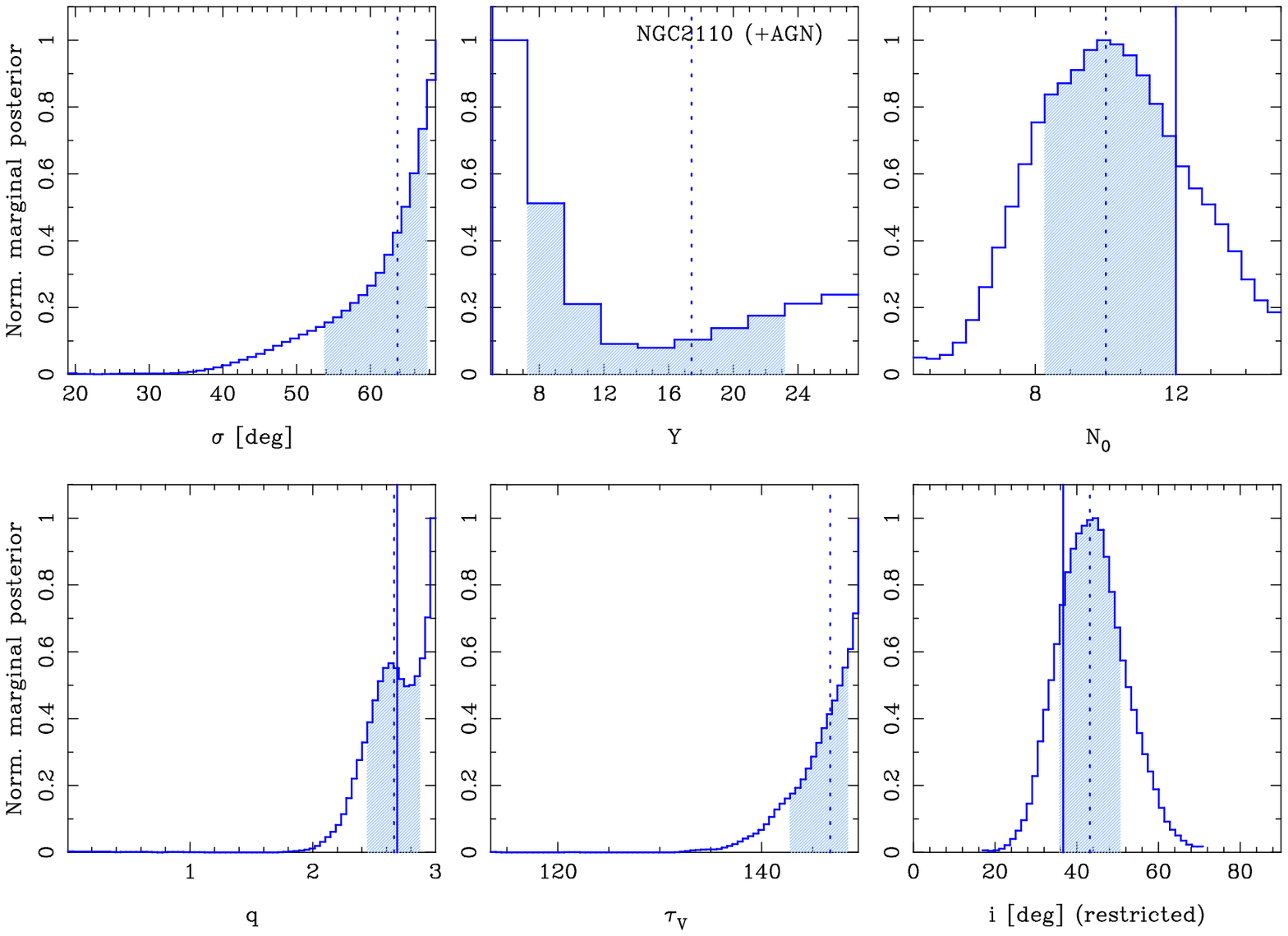}

{Fig. A9.--- Same as Figure~A1 but for NGC~2110. The viewing angle was
  restricted to a Gaussian distribution 
  centered at $i=40\arcdeg$ \, with a 10\arcdeg \, width (see
  Section~3.3).  
Only the SED+spectroscopy fit is shown due to the limited number of
photometric points in the near-IR.
The foreground extinction was fixed to 
$A_V({\rm frg})=5\,$mag (Storchi-Bergman et al. 1999).
}

\end{figure*}

\begin{figure*}
\hspace{2cm}
\includegraphics[width=10.cm,angle=-90]{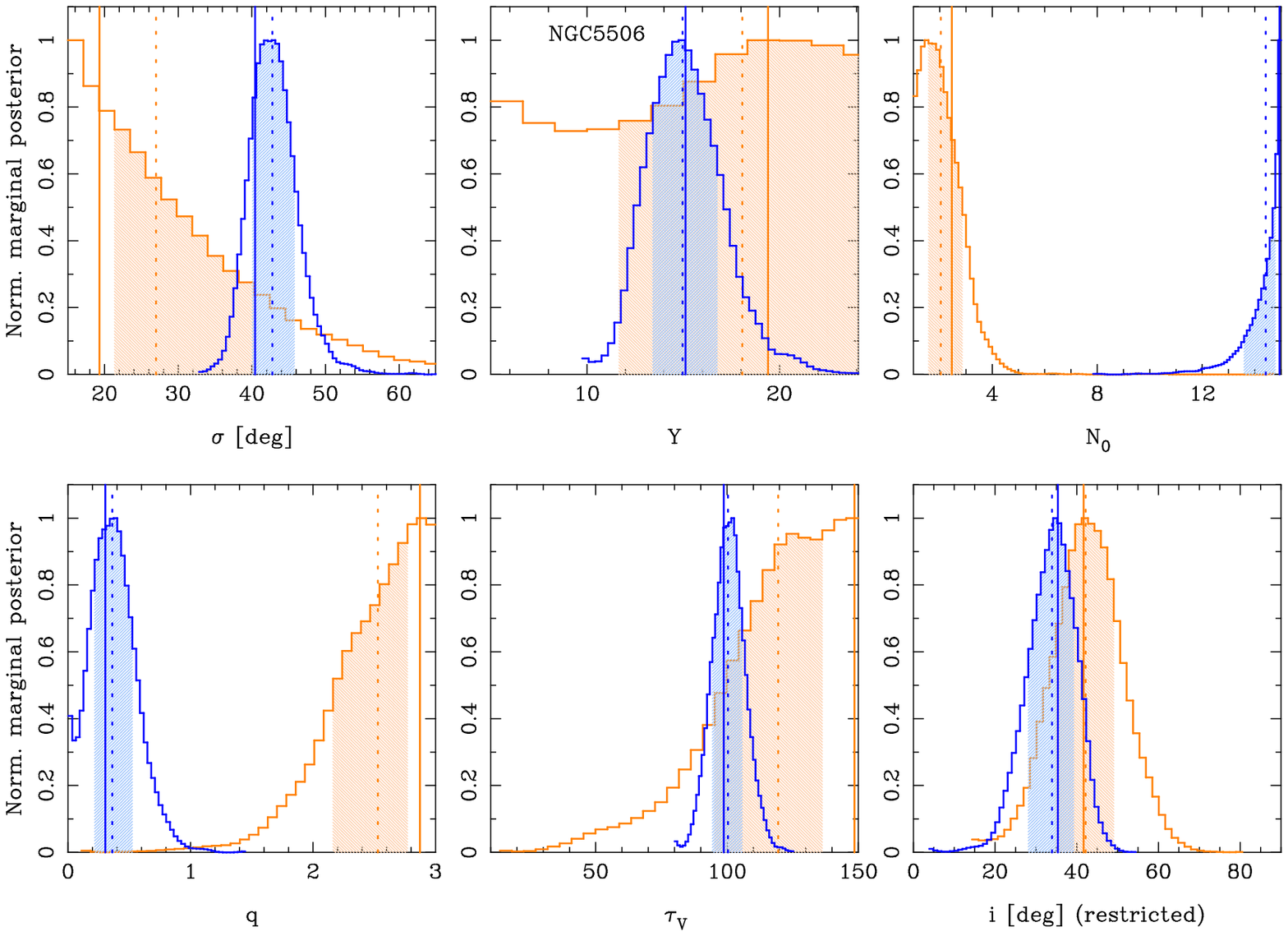}

{Fig. A10.--- Same as Figure~A1 but for NGC~5506. 
The viewing angle was restricted to a
  Gaussian distribution 
  centered at $i=40\arcdeg$ \, with a 10\arcdeg \, width (see
  Section~3.3). The foreground extinction was fixed to  
$A_V({\rm frg})=11\,$mag (Goodrich et al. 1994).}
\end{figure*}

\begin{figure*}
\hspace{2cm}
\includegraphics[width=10.cm,angle=-90]{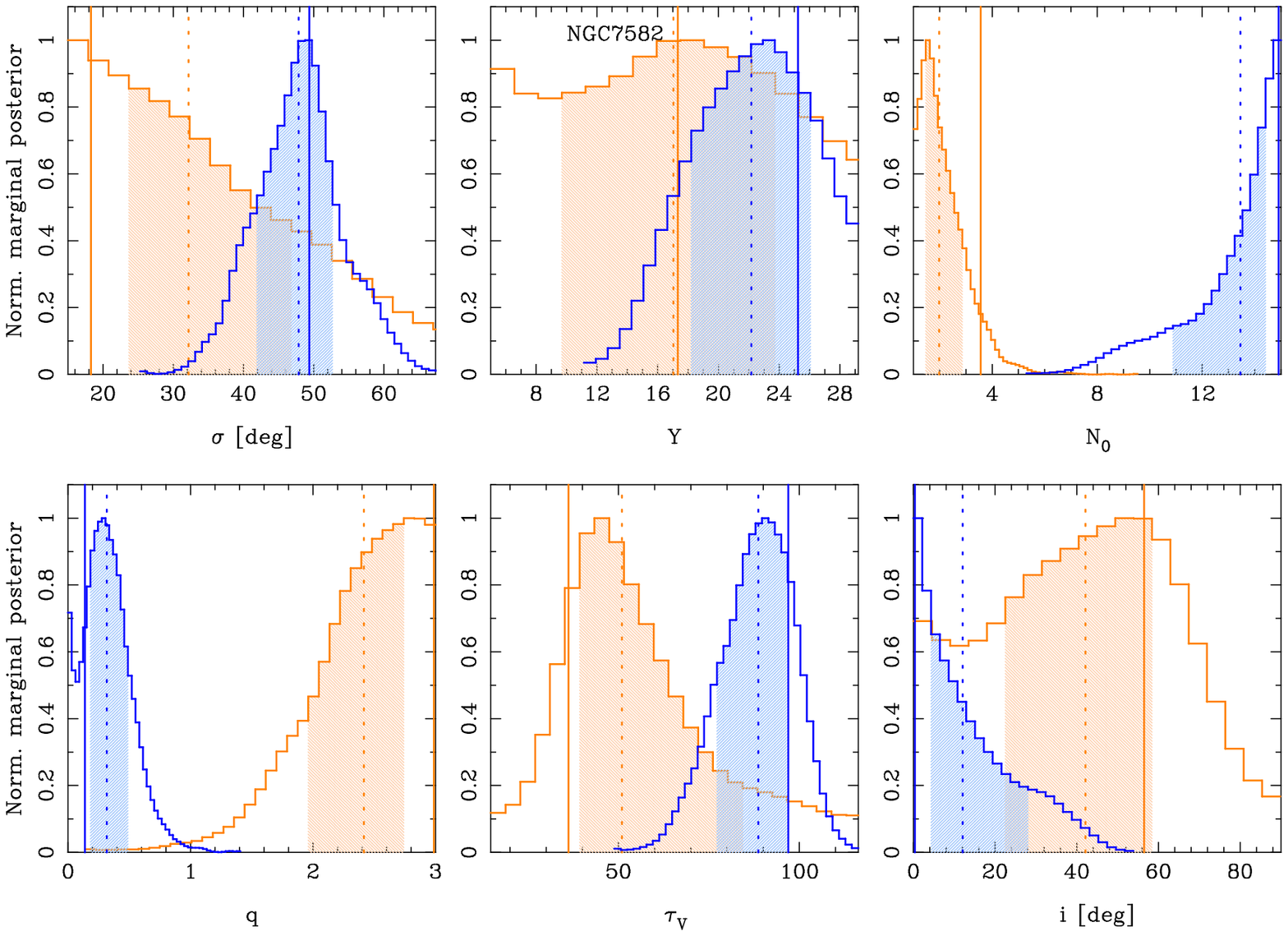}

{Fig. A11.--- Same as Figure~A1 but for NGC~7582. We did not restrict any
  of the torus  model  
  parameters. The foreground extinction was fixed to 
$A_V({\rm frg})=13\,$mag (Winge et al. 2000).}
\end{figure*}

\begin{figure*}
\hspace{2cm}
\includegraphics[width=10.cm,angle=-90]{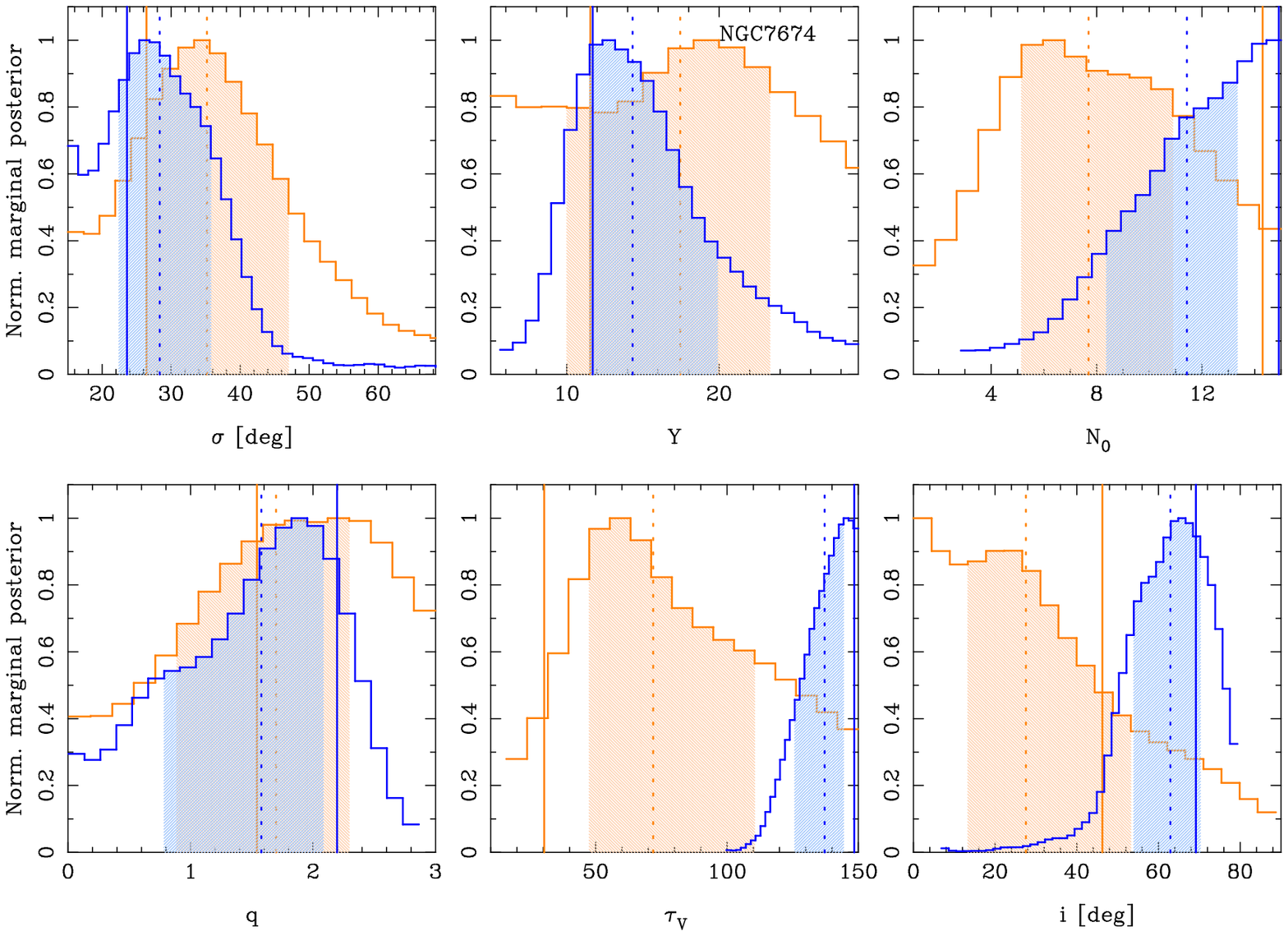}

{Fig. A12.--- Same as Figure~A1 but for NGC~7674. Model fits are for 
torus emission alone. We did not restrict any of the torus
model   parameters. The foreground extinction was fixed to 
$A_V({\rm frg})=5\,$mag.
 }
\end{figure*}

\clearpage

\appendix

\section{Notes on individual sources}
In this section we discuss in detail the fits to individual sources
and compare them with similar fits done in the literature with clumpy
dusty torus models. We also show in Figures~B1 to B4 
the two-dimensional marginal distributions of four selected sources:
Circinus, NGC~1068, NGC~4151, and IC~4239A. In these figures the contours
indicate the $68\%$ and $95\%$
confidence regions. These two-dimensional distributions can be used to
assess whether or not certain torus model parameters are correlated. 

{\bf Circinus.} Circinus is a Seyfert 2 galaxy located at a distance
of 4\,Mpc. The 
water maser observations of this galaxy (Greenhill et al. 2003) 
trace a warped edge-on disk, thus indicating that the torus is
seen at a high inclination. We then used  the viewing angle as a prior
for our fits and restricted it to the range $i=60-90\deg$. The fit to
the photometric SED and mid-IR spectroscopy is shown in Figure~5,
while the marginal posterior distributions are in Figure~A4. As can be
seen from the latter figure all the torus model parameters are nicely
constrained and our estimated value of the AGN bolometric luminosity 
$\log L_{\rm bol}^{\rm model}({\rm AGN})=43.5\,{\rm erg \, s}^{-1}$ 
is in excellent agreement with that of Moorwood et al. (1996). 
Figure~B1 shows that the fitted torus model
parameters of Circinus are not degenerate at the 68\% confidence level

Schartmann et al. (2008) fitted the photometric SED and mid-IR
spectroscopy of Circinus using their clumpy torus models, although due
to long computational times they did not explore the entire parameter
space of their models. They found an appropriate fit for the
SED+spectroscopy as well as for the mid-IR interferometric
visibilities with  $i=70\deg$,
$\theta_{\rm open} = 65\deg$ (this is equivalent to our $\sigma_{\rm
torus}$ parameter) and $\alpha = -0.5$ ($\left|\alpha\right|$ is 
equivalent to our $q$
parameter). These are in excellent agreement with our fits using the
interpolated version of the
Nenkova et al. (2008b) models of $i=66^{+7}_{-4}\deg$, 
$\sigma_{\rm torus} = 58^{+7}_{-11}\deg$,
and $q = 0.7^{+0.4}_{-0.4}$ (see Table~5). The largest disagreement is
in the torus size since their modelling favors an outer radius of
30\,pc. Our fitted AGN bolometric luminosity sets the dust sublimation radius
(see Equation~1) 
of Circinus at $0.07\,$pc, whereas the fitted radial extent of the
torus $Y$ translates into an outer radius of $r_{\rm torus}=2\,$pc
(see Table~5).

{\bf NGC~1068. } NGC~1068 is a Seyfert 2 galaxy at a distance of
15\,Mpc. As with 
Circinus, the water maser observations (Greenhill et al. 1996) 
suggest that the torus is viewed at a high inclination. We used
$i=60-90\deg$ as a prior for our fits. The fit to
the photometric SED and mid-IR spectroscopy is shown in Figure~5. We
note that the Nenkova et al. (2008a, b) 
models provide an excellent fit to the photometric SED
as well as to the $N$-band and
the $Q$-band spectroscopy of this galaxy. 
The  marginal posterior distributions of NGC~1068 are shown in
Figure~A6, while the the two-dimensional marginal
distributions are shown in Figure~B2. These figures indicate that 
all the fitted torus model parameters for NGC~1068 are nicely
constrained. 
From our fits we derived an AGN bolometric luminosity of 
$\log L_{\rm bol}^{\rm
  model}({\rm AGN})=44.8\,{\rm erg \, s}^{-1}$, which is in good
agreement with the estimates  
of Bock et al. (2000) and Vasudevan et al. (2010). 

H\"onig, Prieto, \&  Beckert (2008) modelled the IR SED and VLTI/MIDI data 
using the H\"onig et al. (2006) models. Their modeled torus 
was found to have a line-of-sight inclination of $90\deg$ with the
clouds are distributed according to a power law of $r^{-3/2}$,  a 
half opening angle of the torus of $\sim 40\deg$. The  average
number of clouds, which obscure the AGN along the line of sight
towards the observer, was  $N_0=10$. Their fitted torus model 
 parameters are in a relatively good agreement with our fitted values
 of $i=88^{+2}_{-3}\deg$,  
$\sigma_{\rm torus} = 26^{+6}_{-4}\deg$,
$q = 2.2^{+0.4}_{-0.3}$, and $N_0=14^{+1}_{-3}$ (see Table~5). H\"onig
et al. (2008) also estimated the diameter of the dust torus to be
$\sim1.4-2\,$pc, depending on wavelength and orientation, whereas our
fitted diameter is 4\,pc.

\begin{figure*}[h]
\includegraphics[width=17.cm]{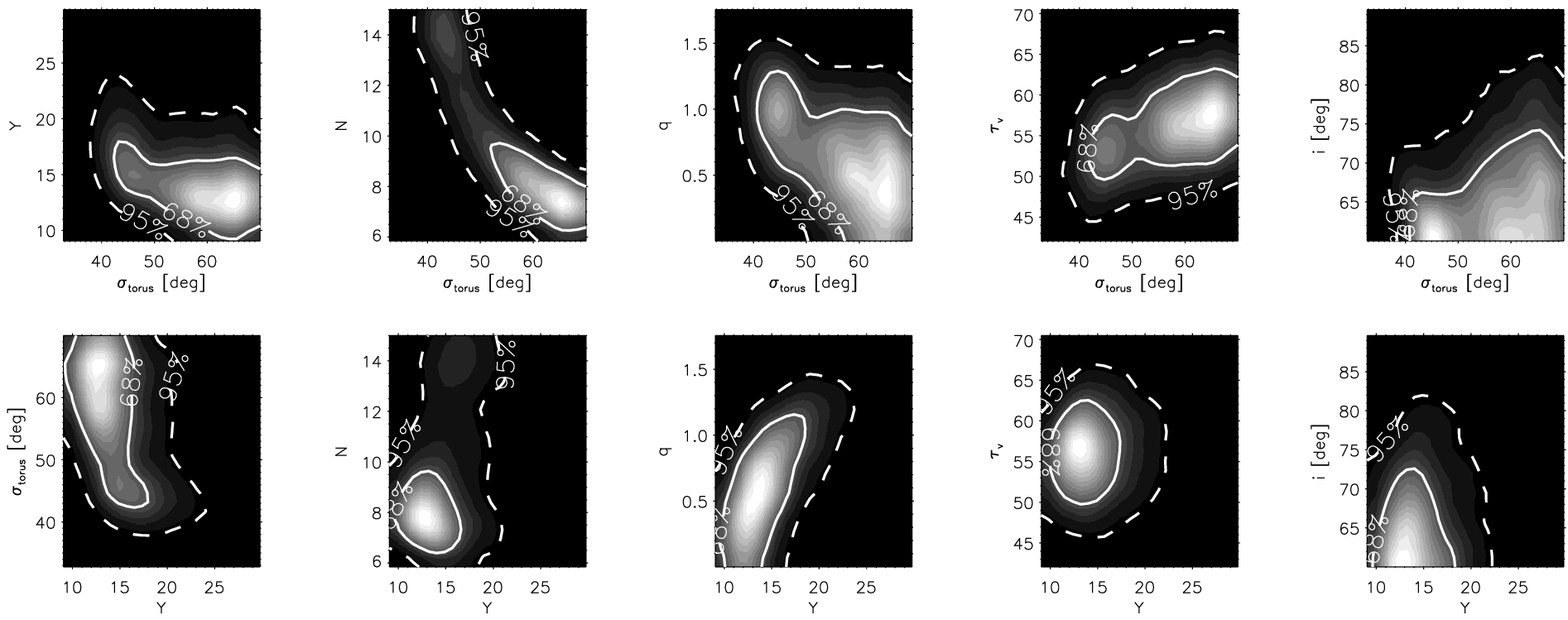}
\includegraphics[width=17.cm]{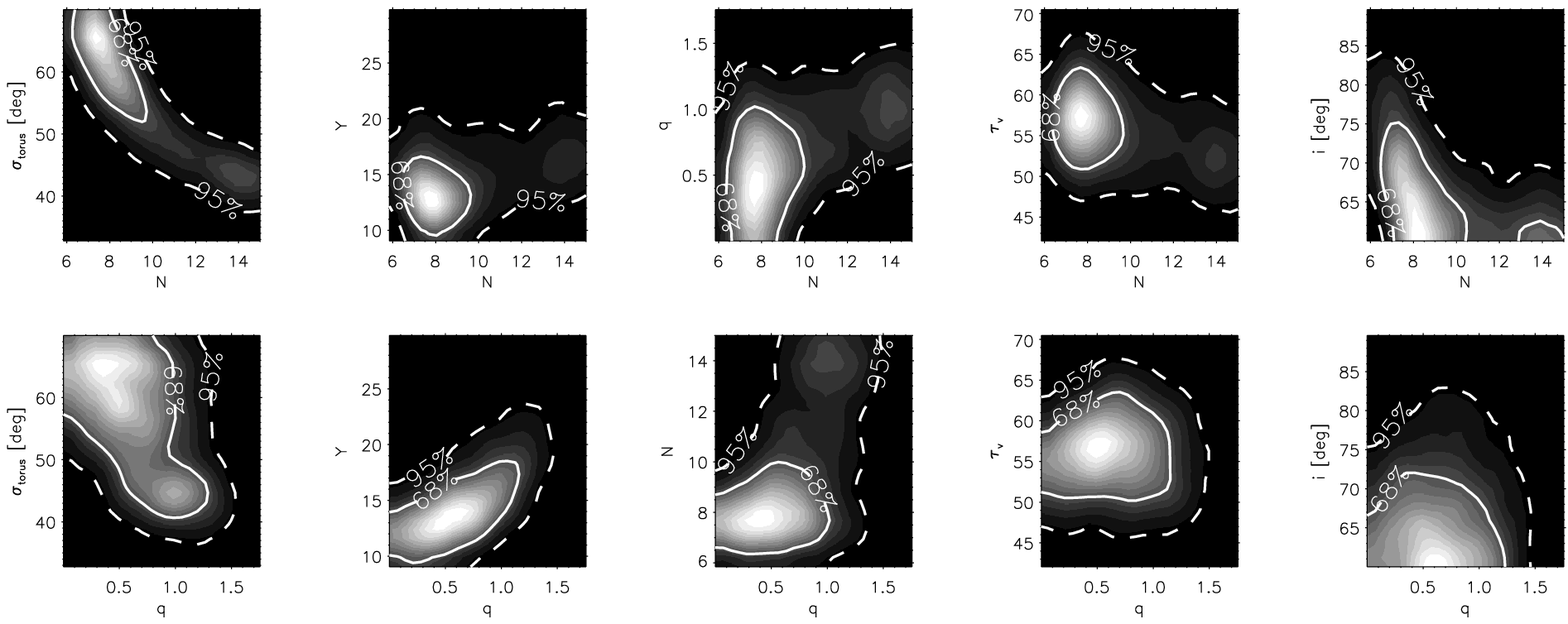}
\includegraphics[width=17.cm]{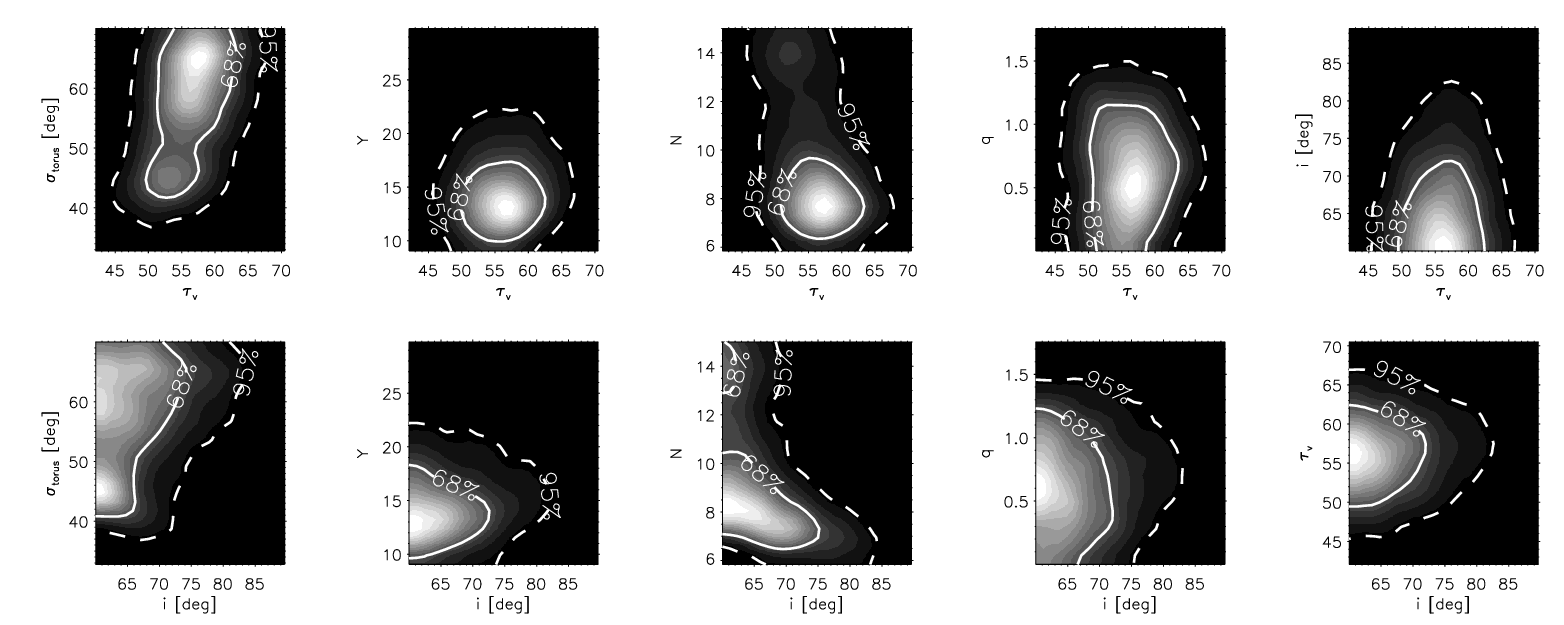}

{Fig. B1. --- Circinus. Two-dimensional marginal posterior distributions
  (joint distributions) for all combinations of the six torus model
  parameters for the SED+spectroscopy fit. 
The contours indicate the regions at $68\%$
 (solid lines) and $95\%$ (dashed lines) confidence levels. 
See Asensio Ramos \& Ramos
 Almeida (2009) for more details. }
\end{figure*}

\begin{figure*}[h]
\includegraphics[width=17.cm]{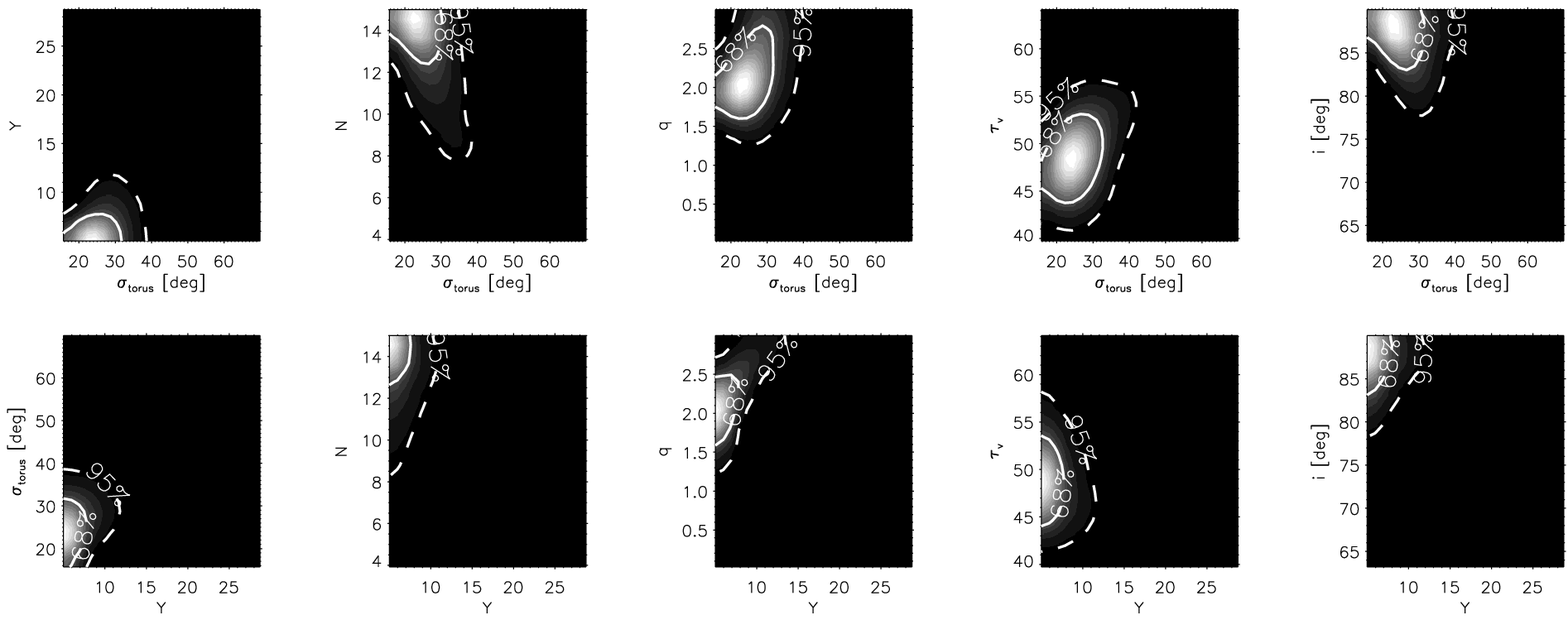}
\includegraphics[width=17.cm]{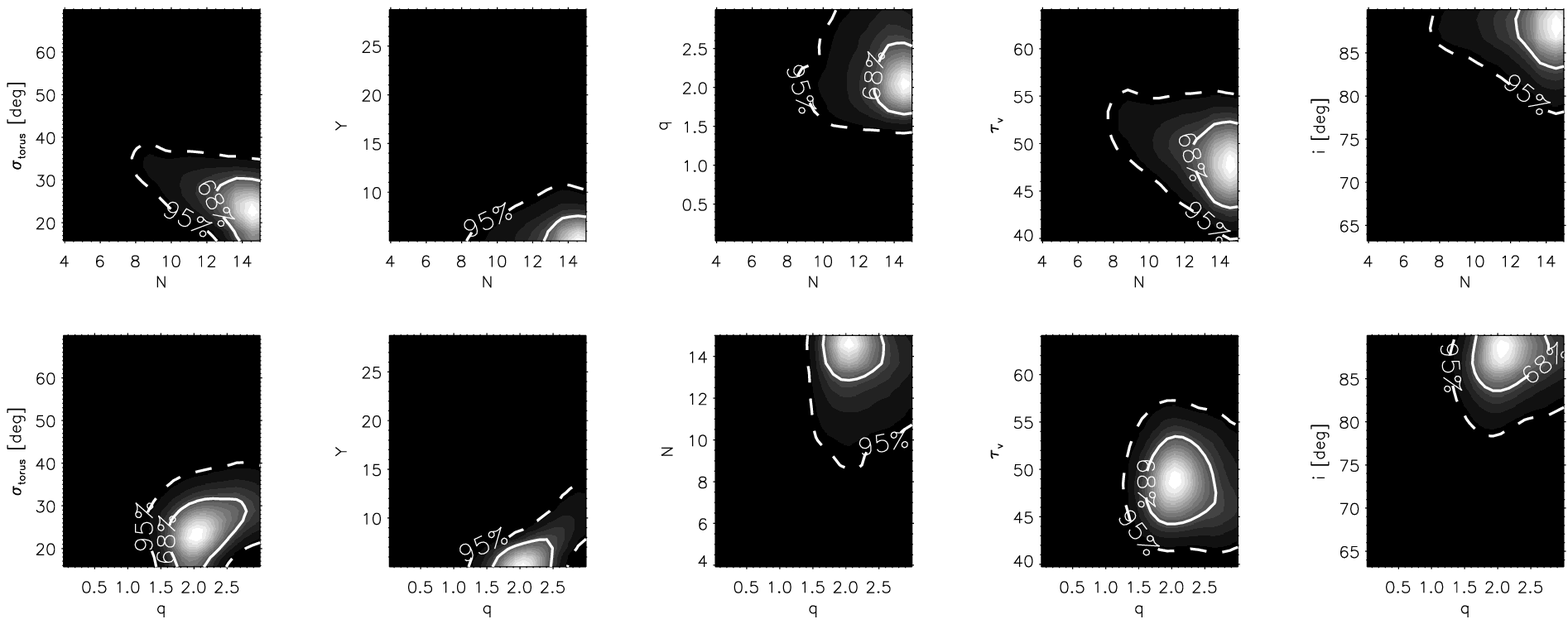}
\includegraphics[width=17.cm]{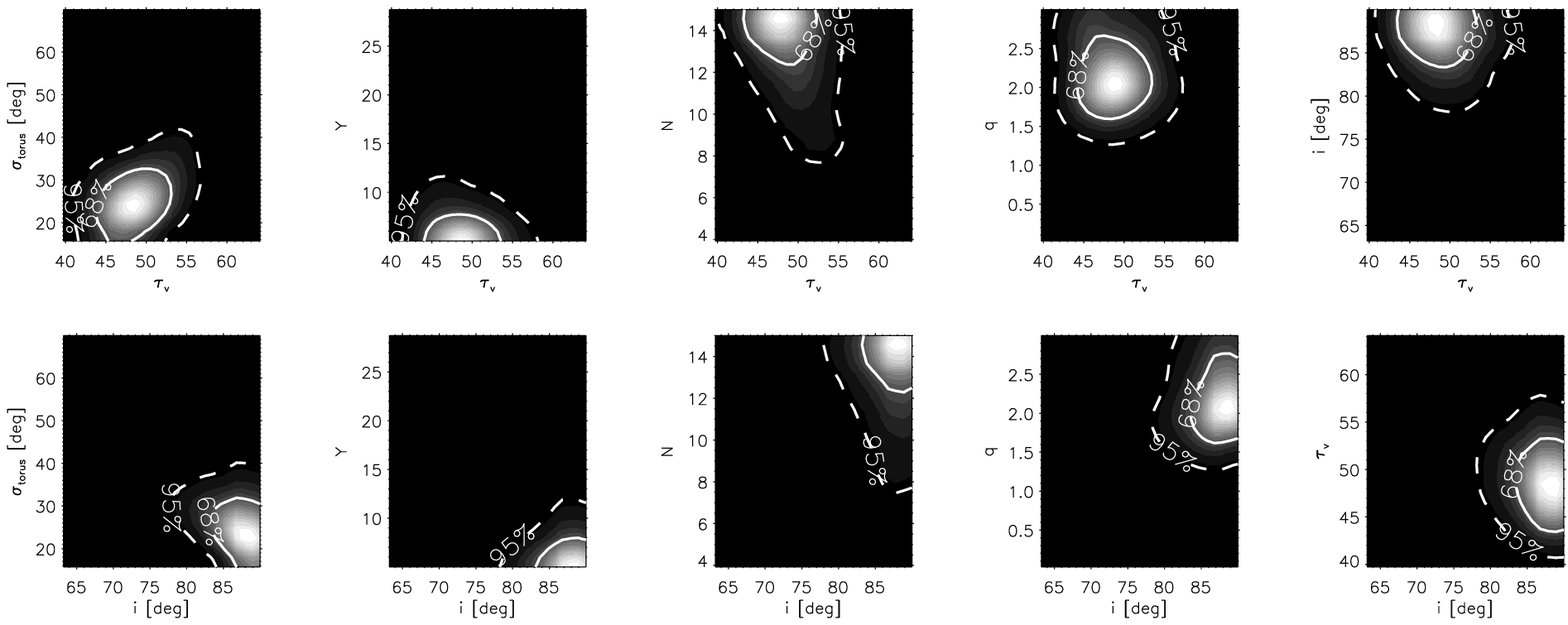}

{Fig. B2.--- As Figure~B1 but for NGC~1068. }
\end{figure*}

\begin{figure*}[h]
\includegraphics[width=17.cm]{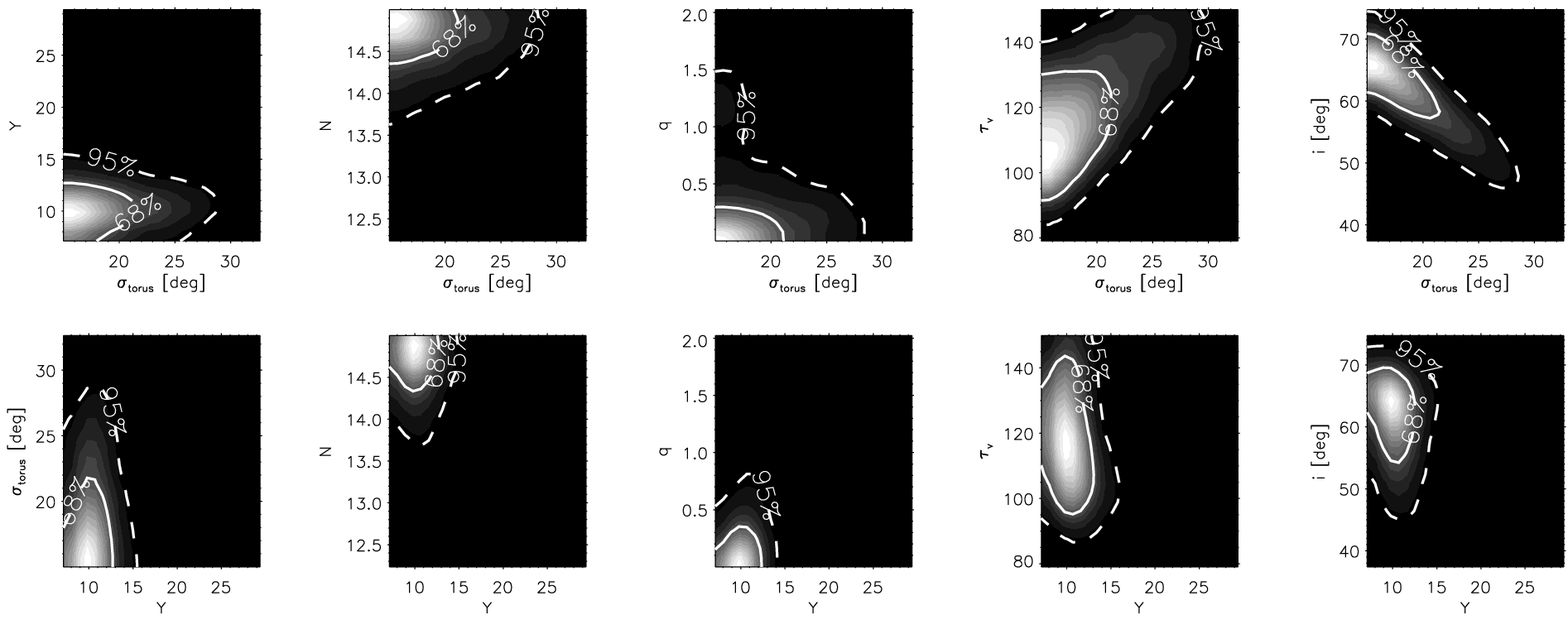}
\includegraphics[width=17.cm]{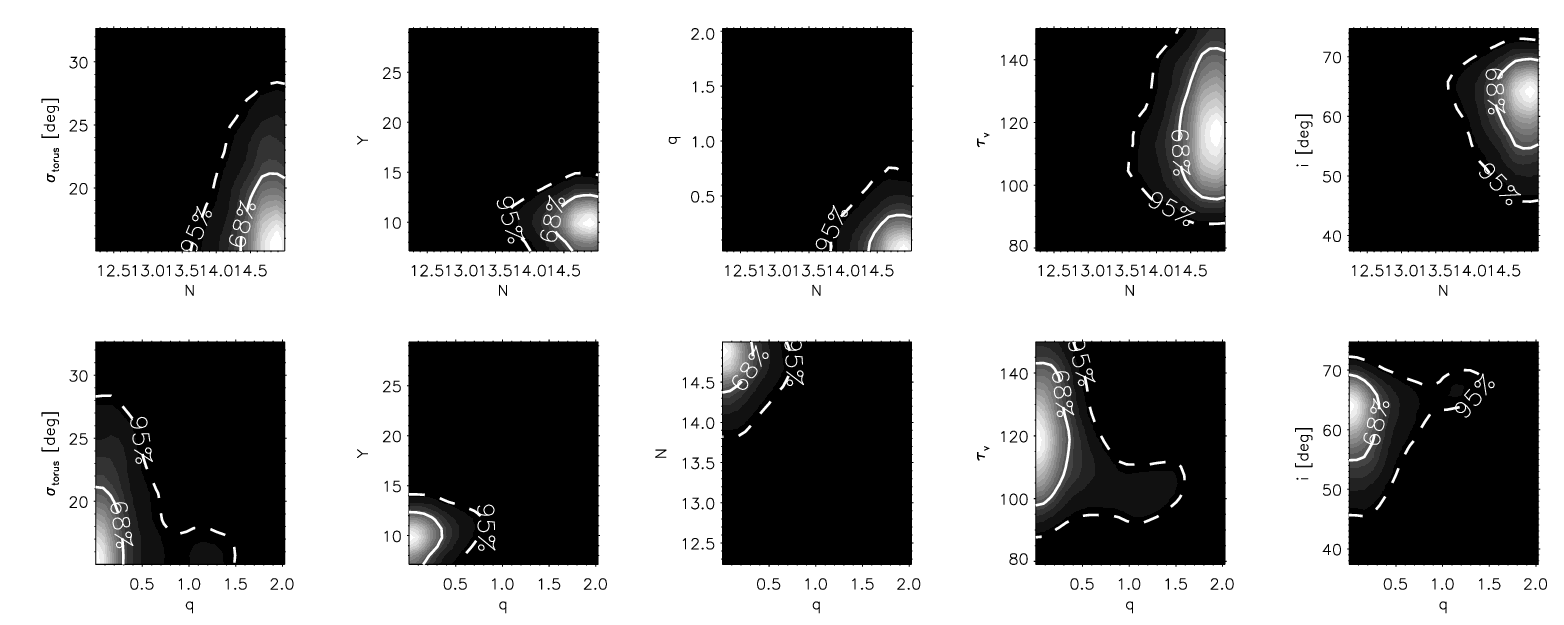}
\includegraphics[width=17.cm]{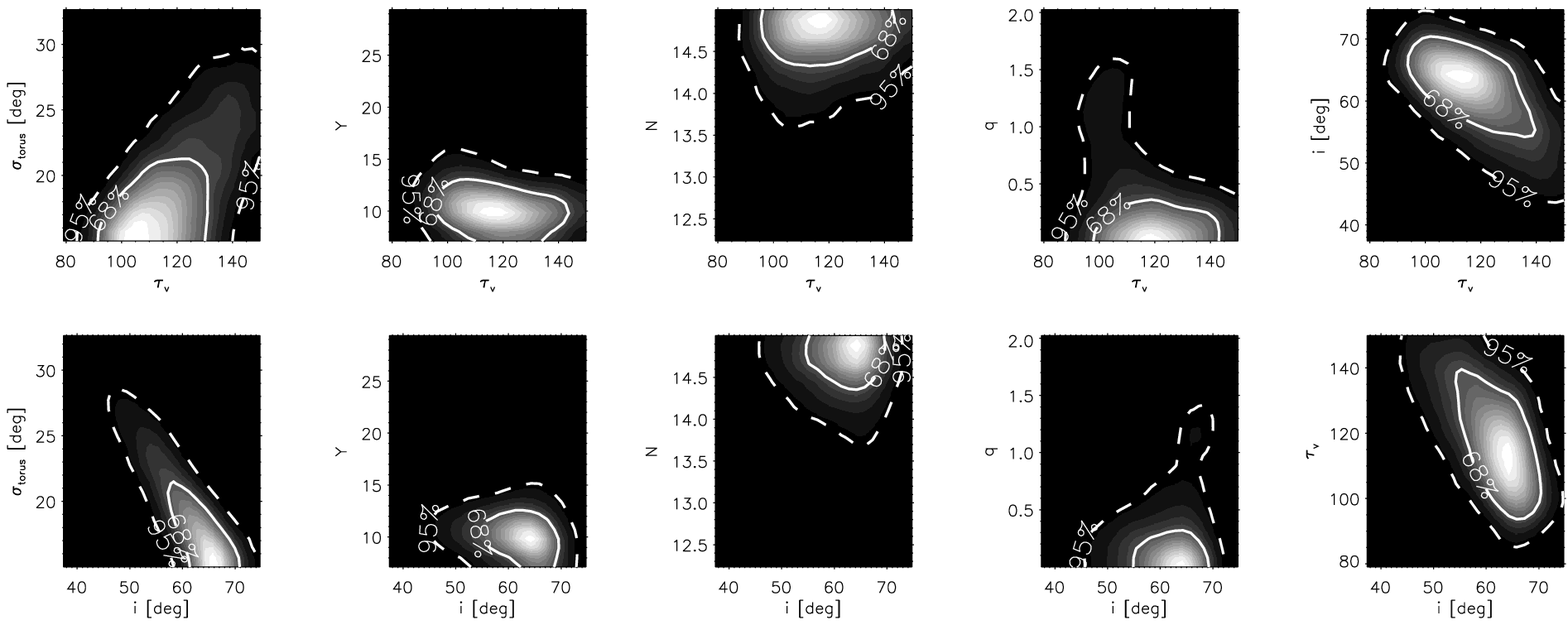}

{Fig. B3.--- As Figure~B1 but for NGC~4151. }
\end{figure*}

\begin{figure*}[h]
\includegraphics[width=17.cm]{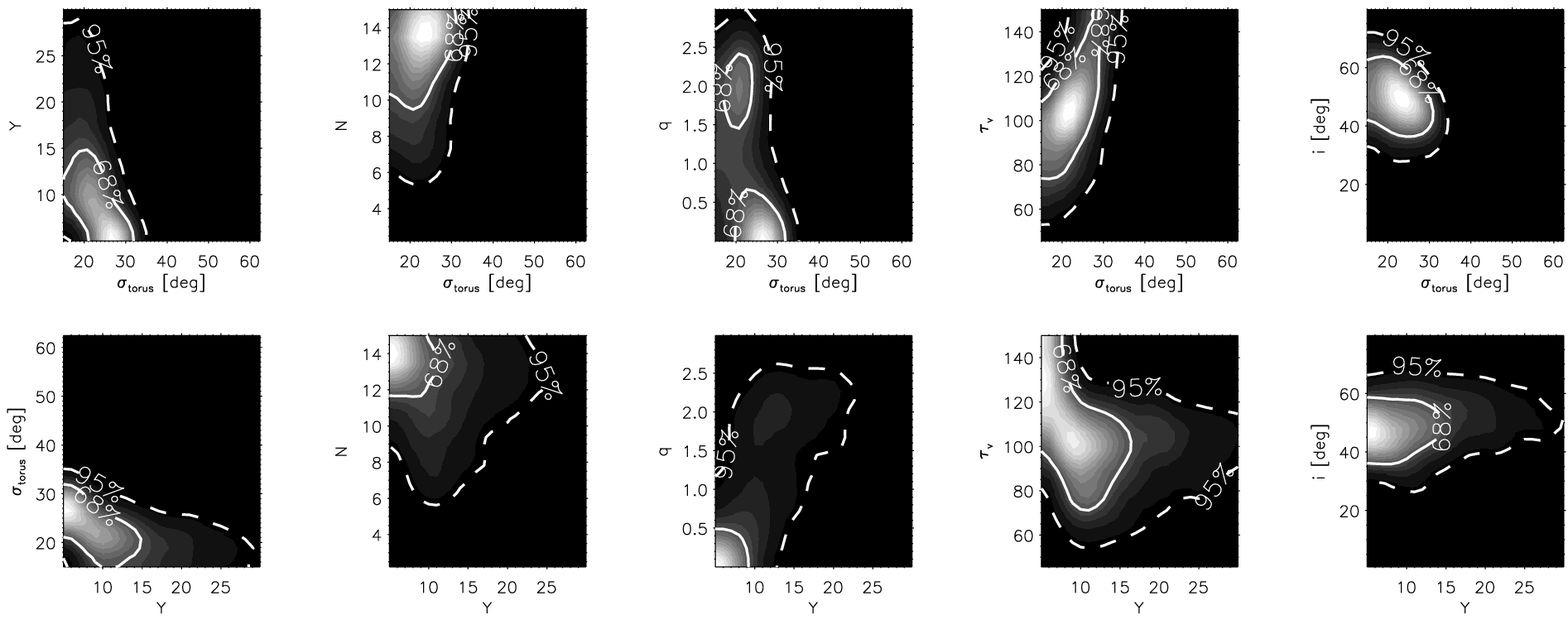}
\includegraphics[width=17.cm]{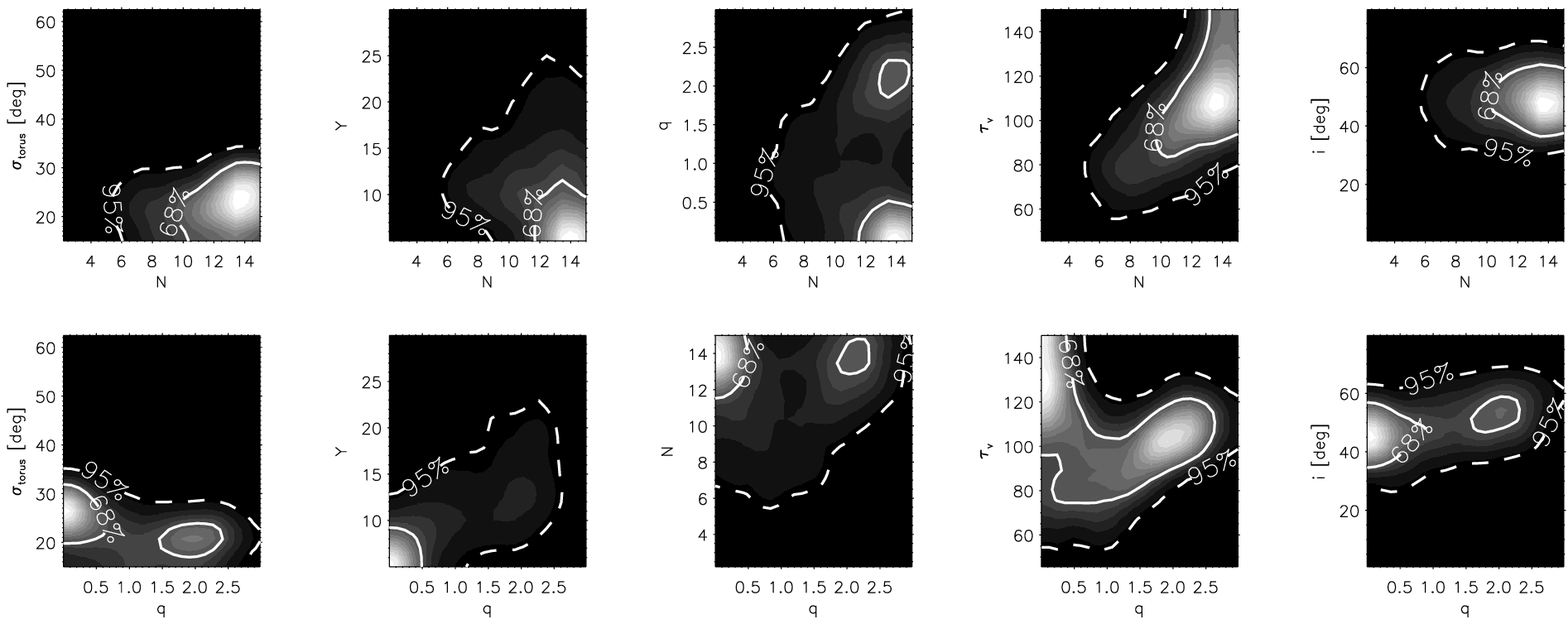}
\includegraphics[width=17.cm]{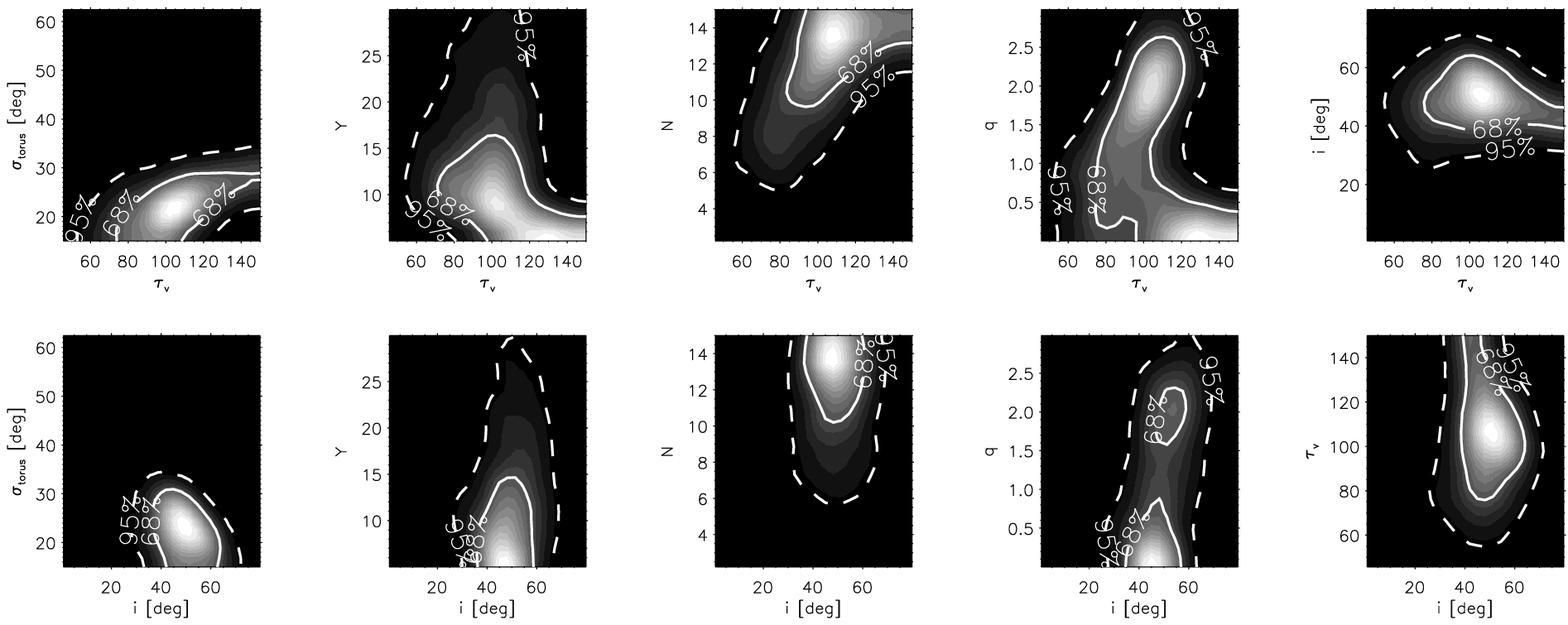}

{Fig. B4.--- As Figure~B1 but for IC~4329A. }
\end{figure*}

{\bf Type 1s: IC~4329A, NGC~3227, and NGC~7469.}
These three Seyfert 1s are in common with the work of H\"onig et
al. (2010). They modelled the mid-IR spectroscopy of a sample of
AGN using their newest clumpy torus models (H\"onig \& Kishimoto
2010). H\"onig et al. (2010) showed that the mid-IR spectroscopy
provides sensitive 
constraints to the number of clouds along the equatorial direction and
the index of power law ($a$ in their notation) 
of the cloud distribution. They fixed other
torus parameters such as the opening angle ($45\deg$), the radial
extent to $R_{\rm out} =150$,  $\tau_{\rm cloud} = 50$ (see H\"onig
\& Kishimoto 2010 for a detailed discussion).  We
note that their model parameters were also chosen to  provide 
good estimates of the VLTI/MIDI observed 
visibilities (see H\"onig et al. 2010 for more details). For the three Seyfert 1
galaxies in common with our study they find indexes of the power law
distribution of $\left | a \right |\sim 0-1 $ and $N_0=7.5-10$, for
their assumed typical viewing angles of $i=30\deg$.   These
compare well with our fitted values of $q \sim 0.2-0.9$, although we
tend to find 
slightly more clouds along the equatorial line of sight (see Table~5).  

{\bf Type 2s: IC~5063, MCG~$-$5-23-16, NGC~2110, and NGC~7674.}
These four type 2 galaxies in our sample are in common with the work of 
H\"onig et al. (2010).  They modelled the
mid-IR spectroscopy of these galaxies with the same parameters as the
type 1s in the previous section but with typical viewing angles of
$i=75\deg$. Their modelling  finds indexes of the power law
distribution of the clouds $\left | a \right |\sim 0-1.5 $ and
$N_0\sim 5$. More clouds along the equatorial line of sight are
allowed in their modelling 
if the viewing angles for Seyfert 2s are smaller ($i \sim
60\deg$). Our fitted values of $q$, except for NGC~2110, are 
consistent with their highest values of $\left |a \right |$, whereas
we find $N_0>10$ (median values, see Table~5).


\clearpage

\begin{thebibliography}{}

\bibitem[Alloin et al. (2000)]{alloin}Alloin, D., Pantin, E., Lagage,
  P. O., \& Granato, G. L. 2000, A\&A, 363, 926

\bibitem[Alonso-Herrero et al.(1996)]{AAH96} Alonso-Herrero, A., Ward,
  M. J., \& Kotilainen, J. K. 1996, \mnras, 278, 902 

\bibitem[Alonso-Herrero et al. (1997)]{AAH97}Alonso-Herrero, A., Ward,
  M. J., \& Kotilainen,  J. K. 1997, MNRAS, 288, 977

\bibitem[Alonso-Herrero et al. (1998)]{AAH98}Alonso-Herrero, A.,
  Simpson, C., Ward, M. J., \& Wilson, A. S. 1998, ApJ, 495, 196

\bibitem[Alonso-Herrero et al. (2001)]{AAH01}Alonso-Herrero, A.,
  Quillen, A. C., Simpson, C., 
  Efstathiou, A., \& Ward, M. J. 2001, AJ, 121, 136

\bibitem[Alonso-Herrero et al. (2003)]{AAH03}Alonso-Herrero, A.,
  Quillen, A. C., Rieke, G. H., 
  Ivanov, V. D., \& Efstathiou, A. 2003, AJ, 126, 81

\bibitem[Alonso-Herrero et al. (2006)]{AAH06}Alonso-Herrero, A.,
  Colina, L., Packham, C., D\'{\i}az-Santos, T., Rieke, G. H.,
  Radomski, J. T., \& Telesco, C. M. 2006, ApJ, 652, L83

\bibitem[Antonucci \& Miller (1985) ]{antonuccimiller}Antonucci, R.,
  \& Miller, J. 1985, ApJ, 297, 621 

\bibitem[Antonucci(1993)]{antonucci}Antonucci, R. 1993, ARA\&A, 31, 473 


\bibitem[Asensio Ramos \& Ramos Almeida(2009)]{BC}Asensio Ramos, A.,
  \& Ramos Almeida, C. 2009, ApJ, 696, 
  2075  

\bibitem[Barvainis (1987)]{Barvainis}Barvainis, R. 1987, ApJ, 320, 537

\bibitem[Beckmann et al. (2005)]{Beckmann2005}Beckmann, V., et
  al. 2005, ApJ, 634, 939

\bibitem[Bianchi et al. (2009)]{Bianchi}Bianchi, S., Piconcelli, E.,
  Chiaberge, M., Jim\'enez Bail\'on, E., Matt, G., \& Fiore, F. 2009,
  ApJ, 695, 781

\bibitem[Blanco et al. (1990)]{Blanco}Blanco, P. R., Ward, M. J., \&
  Wright, G. S. 1990, MNRAS, 242, 4

\bibitem[Bock et al. (2000)]{bock}Bock, J. J. et al. 2000, AJ, 120,
  2904 

\bibitem[Bohlin et al. (1978)]{Bohlin}Bohlin R. C., Savage B. D., \& 
Drake J. F., 1978, ApJ, 224, 132

%\bibitem[Braatz et al. (1994)]{Braatz}Braatz, J. A., Wilson, A. S., \&
%  Henkel, C. 1994, ApJ, 437, L99

\bibitem[Burtscher et al. (2009)]{burtscher}Burtscher, L. et al. 2009,
  ApJ, 705, L53 

\bibitem[Chiar \& Tielens (2006)]{chiar}Chiar, J. E., \& Tielens,
  A. G. G. M. 2006, ApJ, 637, 774

\bibitem[Colina et al. (1991)]{Colina}Colina, L., Sparks, W. B., \&
  Macchetto, F. 1991, ApJ, 370, 102

\bibitem[Colling et al. (2009)]{Colling}Colling, M. D., Roche, P. F., \& Mason,
  R. E. 2009, MNRAS, 394, 2043

\bibitem[Davies et al. (2006)]{davies06}Davies, R. I. et al. 2006, ApJ,
  646, 754 

\bibitem[Davies et al. (2007)]{davies}Davies, R. I. et al. 2007, ApJ,
  671, 1388 

\bibitem[de Vaucouleurs et al. (1991)]{rc3}de Vaucouleurs, G., de
  Vaucouleurs, A., Corwin, H. G. Jr.,  Buta, R. J., 
  Paturel, G., \& Fouque, P. 1991, in Third Reference Catalogue of
  Bright Galaxies, Volume 1-3, XII, 2069 pp. 7  
  figs.. Springer-Verlag Berlin Heidelberg New York 

\bibitem[Deo et al. (2009)]{Deo}Deo, R. P., Richards, G. T., Crenshaw,
  D. M., \& Kraemer, S. B. 2009, ApJ, 705, 14

\bibitem[Efstathiou \& Rowan-Robinson(1995)]{efstathiou}Efstathiou, A.,
\& Rowan-Robinson, M. 1995, MNRAS, 273, 649 

\bibitem[Elitzur (2008)]{Elitzur08}Elitzur, M. 2008, NewAR, 52, 274

\bibitem[Elvis et al. (1994)]{Elvis94}Elvis, M., et al. 1994, ApJS,
  95, 1

\bibitem[Gallimore \& Matthews (2003)]{gallimore}Gallimore, J. F., \&
  Matthews, L. D. 2003, ASPC, 290, 501

\bibitem[Glass (2004)]{glass}Glass, I. S. 2004, MNRAS, 350, 1049

\bibitem[Glasse et al. (1997)]{glasse}Glasse, A. C., Atad-Ettedgui,
  E. I., \& Harris, J. W. 1997, in Proc. SPIE 2871, Optical Telescopes
  of Today and Tomorrow, ed.A. L.Ardeberg (Bellingham, WA: SPIE), 1197–1203

\bibitem[Goodrich et al. (1994)]{goodrich94}Goodrich, R. W., Veilleux,
  S., \& Hill, G. J. 1994, ApJ, 422, 521

\bibitem[Granato \& Danese (1994)]{granato}Granato, G. L., \& Danese,
  L. 1994, MNRAS, 268, 235 

\bibitem[Greenhill et al. (1996)]{greenhill}Greenhill, L. J., Gwinn,
  C. R., Antonucci, R., \& Barvainis, R. 1996, ApJ, 472, L21

\bibitem[Greenhill et al. (2003)]{greenhill2003}Greenhill, L. J., et
  al. 2003, ApJ, 590, 162

\bibitem[Guainazzi et al. (1994)]{guainazzi}Guainazzi, M. et al. 1999,
  ApJ, 436, L35

\bibitem[Guainazzi et al. (2005)]{guainazzi05}Guainazzi, M., Matt, G.,
  \& Perola, G. C. 2005, A\&A, 444, 119

\bibitem[Guainazzi et al. (2010)]{guainazzi10}Guainazzi, M., et
  al. 2010, MNRAS, 406, 2013

\bibitem[Hao et al. (2007)]{Hao}Hao, L., Weedman, D. W., Spoon,
  H. W. W., Marshall, J. A., Levenson, N. A., Elitzur, M., \& Houck,
  J. R. 2007, ApJ, 655, L77

\bibitem[Heisler \& de Robertis (1999)]{heisler}Heisler, C. A., \& de
  Robertis, M. M. 1999, AJ, 118, 2023

\bibitem[Hicks et al. (2009)]{Hicks}Hicks, E. K. S. et al. 2009, ApJ, 696, 448

\bibitem[H\"onig \& Kishimoto (2010)]{HoenigKishimoto}H\"onig, S. F.,
  \& Kishimoto, M. 2010, A\&A, 523, 27

\bibitem[H\"onig et al.(2010)]{Hoenig10}H\"onig, S. F., Kishimoto, M.,
  Gandhi, P., Smette, A., Asmus, D., Duschl, W., Polletta, M., \&
  Weigelt, G. 2010, A\&A, 
  515, 23

\bibitem[H\"onig et al. (2008)]{Hoenig08}H\"onig, S. F., Prieto, 
M. A. \& Beckert, T. 2008, A\&A, 485, 33

\bibitem[H\"onig \& Beckert (2007)]{Hoenig07}H\"onig, S. F., \& Beckert,
  T. 2007, MNRAS, 380, 1172

\bibitem[H\"onig et al.(2006)]{Hoenig06}H\"onig, S. F., Beckert, T.,
  Ohnaka, K., \& Weigelt,   G. 2006, A\&A, 452, 459

\bibitem[Horst et al. (2008)]{horst}Horst, H., Gandhi, P., Smette, A.,
  \&  Duschl, W. J. 2008, A\&A, 479, 389

\bibitem[Imanishi (2000)]{imanishi2000}Imanishi, M. 2000, MNRAS, 313,
  165 

\bibitem[Jaffe et al. (2004)]{jaffe}Jaffe, W. et al. 2004, Nature, 429. 47

\bibitem[Jarrett et al. (2003)]{jarrett}Jarrett, T. H. et al. 2003,
  AJ, 125, 525

\bibitem[Koshida et al. (2009)]{Koshida}Koshida, S. et al. 2009, ApJ,
  700, L109

\bibitem[Kishimoto et al. (2007)]{Kishimoto}Kishimoto, M., H\"onig,
  S. F., Beckert, T., \& Weigelt, G. 2007, A\&A, 476, 713

\bibitem[Kishimoto et al. (2009)]{Kishimoto09}Kishimoto, M., H\"onig,
  S. F., Antonucci, R., Kotani, T., Barvainis, R., Tristram, K. R. W.,
 \& Weigelt, G. 2009, A\&A, 507, L57

\bibitem[Kotilainen et al. (1992)]{koti92}Kotilainen, J. K., Ward,
  M. J., Boisson, C., Depoy, D. L., Smith, M. G., \& Bryant,
  L. R. 1992, MNRAS, 256, 125

\bibitem[Kulkarni et al. (1998)]{kulkarni}Kulkarni, V. et al. 1998,
  ApJ, 492, L121 

\bibitem[Lawrence et al. (1985)]{lawrence}Lawrence, A., Ward, M.,
  Elvis, M., Fabbiano, G., 
  Willner, S. P., Carleton, N. P., \& Longmore, A. 1985, ApJ, 291, 117

\bibitem[Lawrence (1991)]{lawrence91}Lawrence, A. 1991, MNRAS, 252, 586

\bibitem[Leinert et al. (2003)]{Leinert}Leinert, C. et al. 2003, ApSS,
  286, 73

\bibitem[Levenson et al. (2007)]{Levenson07}Levenson, N. A.,
  Sirocky, M. M., Hao, L., Spoon, H. W. W., Marshall, J. A., Elitzur,
  M., \& Houck, J. R. 2007, ApJ, 654, L45 

\bibitem[Levenson et al. (2008)]{Levenson08}Levenson, N. A. et al. 
2008, SPIE, 7014, 83

\bibitem[Lira et al. (2011)]{Lira}Lira, P., et al. 2011, ApJ, in 
  preparation 

\bibitem[Maiolino et al. (1994)]{maio94}Maiolino, R., Stanga, R.,
  Salvati, M., \& Rodr\'{\i}guez Espinosa, J. M. 1994, A\&A, 290, 40 

\bibitem[Maiolino \& Rieke (1995)]{maio95}Maiolino, R., \& Rieke,
  G. H. 1995, ApJ, 454, 95

\bibitem[Maiolino et al. (2000)]{maio2000}Maiolino, R.,
  Alonso-Herrero, A., Anders, S., Quillen, A., Rieke, M. J., Rieke,
  G. H., \& Tacconi-Garman, L. E. 2000, ApJ, 531, 219

\bibitem[Malaguti et al. (1998)]{malaguti}Malaguti, G. et al. 1998,
  A\&A, 331, 519 

\bibitem[Malkan et al. (1998)]{Malkan}Malkan, M. A., Gorjian, V., \&
  Tam, R. 1998, ApJS, 117, 25

\bibitem[Marco \& Alloin (2000)]{marco}Marco, O., \& Alloin, D. 2000,
  A\&A, 353, 465

\bibitem[Martini et al. (2001)]{martini01}Martini, P., 
Pogge, R. W., Ravindranath, S., \& An, J. H. 2001, ApJ, 562, 139

\bibitem[Martini et al. (2003)]{martini03}Martini, P., Regan, M. W.,
  Mulchaey, J. S., \& Pogge, R. W. 2003, ApJ, 589, 774

\bibitem[Mason et al. (1996)]{mason06}Mason, R. E., Geballe, T. R.,
  Packham, C., Levenson, 
  N., Elitzur, M. et al. 2006, ApJ, 640, 612

\bibitem[Mason et al. (2007)]{mason07}Mason, R. E., Levenson, N. A.,
  Packham, C., Elitzur, M., Radomski, J., Petric, A. O., \& Wright,
  G. S. 2007, ApJ, 659, 241

\bibitem[Mason et al. (2009)]{mason09}Mason, R. E., Levenson, N. A.,
  Shi, Y., Packham, C., 
  Gorjian, V., Cleary, K., Rhee, J., \& Werner, M. 2009, ApJ, 693,
  L136

\bibitem[Matt et al. (1999)]{matt99}Matt, G. et al. 1999, A\&A, 341, L39

 \bibitem[Matt et al. (2000)]{matt}Matt, G. et al. 2000, MNRAS, 318,
  173 

\bibitem[Moorwood et al. (1996)]{moorwood}Moorwood, A. F. M. et
  al. 1996, A\&A, 315, L109  

\bibitem[Mor et al. (2009)]{mor}Mor, R., Netzer, H., \& Elitzur,
  M. 2009, ApJ, 705, 298 


\bibitem[Mulchaey et al. (1994)]{mulchaey}Mulchaey, J. S. et al. 1994,
  ApJ, 436, 586 

\bibitem[Mulchaey et al. (1996)]{mulchaey96}Mulchaey, J. S., Wilson,
  A. S., \& Tsvetanov, Z. 1996, ApJ, 467, 197

\bibitem[Mundell et al. (1995)]{mundell}Mundell, C. G. et al. 1995,
  MNRAS, 275, 67

\bibitem[Nagar et al. (1999)]{nagar99}Nagar, N. M., Wilson, A. S.,
  Mulchaey, J. S., \& Gallimore, J. E. 1999, ApJS, 120, 209

\bibitem[Nagar et al. (2002)]{nagar}Nagar, N. M., Oliva, E., Marconi,
  A. \& Maiolino,   R. 2002, A\&A, 391, L21

\bibitem[Nenkova et al. (2002)]{nenkova02}Nenkova, M., Ivezic, Z., \&
  Elitzur, M. 2002, ApJ,   570, L9 

\bibitem[Nenkova et al. (2008a)]{nenkova08a}Nenkova, M., Sirocky,
  M. M., Ivezi\'c, Z.,   \& Elitzur, M. 2008a, ApJ, 685, 145

\bibitem[Nenkova et al. (2008b)]{nenkova08b}Nenkova, M., Sirocky,
  M. M., Nikkuta, R., Ivezi\'c, Z.,   
\& Elitzur, M. 2008b, ApJ, 685, 160

\bibitem[Nenkova et al. (2010)]{nenkova10}Nenkova, M., Sirocky,
  M. M., Nikkuta, R., Ivezi\'c, Z.,   
\& Elitzur, M. 2010, ApJ, 723, 1827

\bibitem[Nikutta et al. (2009)]{Nikutta}Nikutta, R., Elitzur, M., \&
  Lacy, M. 2009, ApJ, 707, 1550 

\bibitem[Oliva et al. (1994)]{Oliva94}Oliva, E., Salvati, M.,
  Moorwood, A. F. M., \& Marconi, A. 1994, A\&A, 288, 457

\bibitem[Ossenkopf et al. (1992)]{ossenkopf}Ossenkopf, V., Henning,
  T., \& Mathis, J. S. 1992, A\&A, 261, 567 

\bibitem[Osterbrock \& Martel (1993)]{osterbrock}Osterbrock, D. E. \&
  Martel, A. 1993, ApJ, 414, 552 

\bibitem[Packham et al. (1997)]{packham97}Packham, C., Young, S.,
  Hough, J. H., Axon, D. J., \& Bailey, J. A. 1997, MNRAS, 288, 375

\bibitem[Packham et al. (2005)]{packham}Packham, C. et al. 2005, ApJ,
  618, L17

\bibitem[Packham et al. (2007)]{packham07}Packham, C. et al. 2007,
  ApJ, 661, L29

\bibitem[Perola et al. (2002)]{perola}Perola, G. C. et al. 2002, A\&A, 389, 802 

\bibitem[Pier \& Krolik (1993)]{pierkrolik}Pier, E. A., \& Krolik,
  J. H. 1993, ApJ, 418, 673 

\bibitem[Pogge (1989)]{pogge}Pogge, R. W. 1989, ApJ, 345, 730

\bibitem[Polletta et al. (2008)]{polletta}Polletta, M., Weedman, D., H\"onig,
  S., Lonsdale, C. J., Smith, H. E., \& Houck, J. 2008, ApJ, 675, 960

\bibitem[Pott et al. (2010)]{pott}Pott, J.-U., Malkan, M. A., 
Elitzur, M., Ghez, A. M., Herbst, T. M., Sch\"odel, R., \& Woillez,
J. 2010 ApJ, 715, 736

\bibitem[Prieto et al. (2010)]{prieto2010}Prieto, M. A., Reunanen, J.,
  Tristram, K. R. W.,  
Neumayer, N., Fernandez-Ontiveros, J. A., Orienti, M., \&
Meisenheimer, K. 2010, MNRAS, 402, 724

\bibitem[Quillen et al. (1999)]{quillen99}Quillen, A. C.,
  Alonso-Herrero, A., Rieke, M. J., 
  Rieke, G. H., Ruiz, M., \& Kulkarni, V. 1999, ApJ, 527, 696

\bibitem[Quillen et al. (2000)]{quillen00}Quillen, A. C., Shaked, S.,
  Alonso-Herrero, A., McDonald, C., Lee, A., Rieke, M. J., \& Rieke,
  G. H. 2000, ApJ, 532, L17

\bibitem[Quillen et al. (2001)]{quillen01}Quillen, A. C., McDonald,
  C., Alonso-Herrero, A., 
  Lee, A., Shaked, S., Rieke, M. J., \& Rieke, G. H. 2001, ApJ, 547, 129 

\bibitem[Raban et al. (2009)]{Raban}Raban, D. et al. 2009, MNRAS, 394, 1325

\bibitem[Radomski et al. (2003)]{radomski}Radomski, J. T. et al. 2003,
  ApJ, 587, 117

\bibitem[Ramos Almeida et al. (2009a)]{RamosAlmeida}Ramos Almeida, C.,
 P\'erez-Garc\'{\i}a, A. M., \& Acosta-Pulido, J. A. 2009, ApJ, 694, 1379

\bibitem[Ramos Almeida et al. (2009b)]{RA09}Ramos Almeida, C.,
  Levenson, N. A., Rodr\'{\i}guez 
  Espinosa, J. M., Alonso-Herrero, A., Asensio Ramos, A., Radomski,
  J., T., Packham, C., Fisher, R. S., \& Telesco, C. M. 2009, ApJ,
  702, 1127 (RA09)

\bibitem[Ramos Almeida et al. (2011)]{RA11}Ramos Almeida, C.,
  Levenson, N. A., Alonso-Herrero, A., 
  Asensio Ramos, A., Rodr\'{\i}guez Espinosa, J. M., P\'erez-Garc\'{\i}a,
  A. M., Packham, C., Mason, R., Radomski, J. T., \& D\'{\i}az-Santos,
  T. 2011, ApJ, 731, 92 (RA11)

\bibitem[Reeves et al. (2007)]{reeves}Reeves, J. N. et al. 2007, PASP,
  59, 301

\bibitem[Regan \& Mulchaey (1999)]{regan}Regan, M. W., \& Mulchaey,
  J. S. 1999, AJ, 117, 2676

\bibitem[Reunanen et al. (2003)]{reunanen03}Reunanen, J., et al. 2003,
  MNRAS, 343, 192 

\bibitem[Reunanen et al. (2010)]{reunanen}Reunanen, J., Prieto, M. A.,
  \& Siebenmorgen, 
  R. 2010, MNRAS, 402, 879

\bibitem[Riffel et al. (2006)]{riffel}Riffel, R.,
  Rodr\'{\i}guez-Ardila, A., \& Pastoriza, M. G. 2006, A\&A, 457, 61

\bibitem[Risaliti et al. (1999)]{risaliti}Risaliti, G., Maiolino, R.,
  \& Salvati, M. 1999, ApJ, 522, 157

\bibitem[Roche et al. (1991)]{roche91}Roche, P. F., Aitken, D. K.,
  Smith, C. H., \& Ward, M. J. 1991, MNRAS, 248, 606

\bibitem[Roche et al. (2006)]{roche06}Roche, P. F., Packham,
  C. Telesco, C. M., Radomski, 
  J. T., Alonso-Herrero, A., Aitken, D. K., Colina, L., \& Perlman,
  E. 2006, MNRAS, 367, 1689

\bibitem[Roche et al. (2007)]{roche07}Roche, P. F., Packham,
  C. Aitken, D. K., \& Mason, 
  R. E. 2007, MNRAS, 375, 99

\bibitem[Ruiz et al. (1994)]{ruiz}Ruiz, M., Rieke, G. H., \& Schmidt,
  G. D. 1994, ApJ,   423, 608

\bibitem[Schartmann et al. (2008)]{schartmann}Schartmann, M.,
  Meisenheimer, K., Camenzind, M.,  
Wolf, S., Tristram, K. R. W., \& Henning, T. 2008, A\&A, 482, 67

\bibitem[Schmitt et al. (2003)]{schmitt}Schmitt, H.R. et al. 2003,
  ApJS, 148, 327  

\bibitem[Shi et al. (2006)]{Shi2006}Shi, Y., et al. 2006, ApJ, 653,
  127 

\bibitem[Siebenmorgen et al. (2004)]{siebenmorgen}Siebenmorgen, R.,
  Kr\"ugel, E., \& Spoon, H. W. W. 2004, A\&A, 414, 123

\bibitem[Simpson (2005)]{Simpson}Simpson, C. 2005, MNRAS, 360, 565

\bibitem[Sirocky et al. (2008)]{sirocky}Sirocky, M. M., Levenson, N.,
  A., Elitzur, M., Spoon, H. W. W., \& Armus, L. 2008, ApJ, 678, 729

\bibitem[Storchi-Bergman et al. (1999)]{storchi}Storchi-Bergman, T.,
  Winge, C., Ward, M. J., \& Wilson, A. S. 1999, MNRAS, 304, 25

\bibitem[Suganuma et al. (2006)]{Suganuma}Suganuma, M. et al. 2006,
  ApJ, 639, 46

\bibitem[Telesco et al. (1998)]{telesco98}Telesco, C. M., Pina, R. K., 
Hanna, K. T., Julian, J. A., Hon, D. B., \& Kisko, T.
M., 1998, Proc. SPIE, 3354, 534

\bibitem[Telesco et al. (2003)]{telesco03}Telesco, C. M. et al. 2003,
  SPIE, 4841, 913 

\bibitem[Thompson et al. (2009)]{thompson}Thompson, G. D., Levenson,
  N. A., Uddin, S. A., \& Sirocky, M. M. 2009, ApJ, 697, 182

\bibitem[Tomono et al. (2001)]{tomono01}Tomono, D., Doi, Y.,Usuda, T.,
  \& Nishimura, 
  T. 2001, ApJ, 557, 637

\bibitem[Tran et al. (1992)]{Tran}Tran, H. D., Miller, J. S., \& Kay,
  L. E. 1992, ApJ,   397, 452

\bibitem[Tristram et al. (2007)]{tristram07}Tristram, K. R. W. et
  al. 2007, A\&A, 474, 837 

\bibitem[Tristram et al. (2009)]{tristram09}Tristram, K. R. W. et
  al. 2009, A\&A, 502, 67 

\bibitem[Vasudevan et al. (2010)]{vasudevan}Vasudevan, R. V., Fabian,
  A. C., Gandhi, P., 
  Winter, L. M., \& Mushotzky, R. F. 2010, MNRAS, 402, 1081

\bibitem[Veilleux et al. (1997)]{veilleux97}Veilleux, S., Goodrich,
  R. W., \& Hill,   G. J. 1997, ApJ, 477, 631 

\bibitem[Veron-Cetty \& Veron (2006)]{veron}V\'eron-Cetty, M.-P., \&
  V\'eron, P. 2006, A\&A, 455, 773  

\bibitem[Videla et al. (2011)]{Videlaa}Videla, L., et al. 2011, ApJ, in
  preparation 

\bibitem[Ward et al. (1987)]{ward87}Ward, M., Elvis, M., Fabbiano, G.,
  Carleton, N. P.,   Willner, S. P. \& Lawrence, A. 1987, ApJ, 315, 74  

\bibitem[Weaver \& Reynolds (1998)]{weaver}Weaver, K. A., \& Reynolds,
    C. S. 1998, ApJ, 503, L39

\bibitem[Wilson et al. (1985)]{Wilson85}Wilson, A. S., Baldwin, J. A.,
  \& Ulvestad, J. S. 1985, ApJ, 291, 627

\bibitem[Wilson \& Tsvetanov (1994)]{Wilson94}Wilson, A. S. \&
  Tsvetanov, Z. I. 1994, ApJ, 107, 1227  

\bibitem[Winge et al. (2000)]{winge}Winge, C., Storchi-Bergmann, T.,
  Ward, M. J., \& Wilson, A. S. 2000, MNRAS, 316, 1

\bibitem[Woo \& Urry (2002)]{woo}Woo, J.-H., \& Urry, C. M. 2002, ApJ,
  579, 530 

\bibitem[Young et al. (2007)]{young07}Young, S., Packham, C., Mason,
  R. E., Radomski, J. T., 
  \& Telesco, C. M. 2007, MNRAS, 378, 888
\end{thebibliography}
\end{document}